\pdfoutput=1
\documentclass[]{aa}   

\usepackage{amsmath}
\usepackage{graphicx}
\usepackage{array,multirow,makecell}
\usepackage{txfonts}
\usepackage{natbib}
\usepackage{ulem} 
\usepackage{booktabs}
\usepackage{arydshln}
\usepackage{gensymb}
\usepackage{color}
\usepackage{afterpage}
\usepackage[colorinlistoftodos]{todonotes}
\usepackage[caption = false]{subfig}
\usepackage{float}
\usepackage{pdflscape}
\usepackage{rotating}

\bibpunct{(}{)}{;}{a}{}{,} 

\newcommand{\Msol}{\ensuremath{M_{\odot}}}

\newcommand{\kms}{\ensuremath{\rm km\,s^{-1}}}

\usepackage[colorlinks=true,linkcolor=blue,citecolor=blue]{hyperref}
\begin{document} 

   \title{Episodic accretion constrained by a rich cluster of outflows
	}

  \author{T. Nony\inst{1}
          \and F. Motte\inst{1}
          \and F. Louvet\inst{2,3,4} 
          \and A. Plunkett\inst{5}
          \and A. Gusdorf\inst{4,6}
          \and S. Fechtenbaum\inst{1} 
          \and Y. Pouteau\inst{1}
          \and B. Lefloch\inst{1}
          \and S. Bontemps\inst{7}
          \and J. Molet\inst{7}
          \and J.-F. Robitaille\inst{1}
          }

   \institute{Univ. Grenoble Alpes, CNRS, IPAG, 38000 Grenoble, France
   \and Departmento de Astronomia de Chile, Universidad de Chile, Santiago, Chile
   \and AIM, CEA, CNRS, Université Paris-Saclay, Université Paris Diderot, Sorbonne Paris Cité, 91191 Gif-sur-Yvette, France
   \and Observatoire de Paris, PSL University, Sorbonne Université, LERMA, 75014 Paris, France
   \and National Radio Astronomy Observatory, Charlottesville, VA 22903, USA
   \and Laboratoire de Physique de l’École Normale Supérieure, ENS, Université PSL, CNRS, Sorbonne Université, Université de Paris, Paris, France
   \and Laboratoire d'astrophysique de Bordeaux, Univ. Bordeaux, CNRS, B18N, allée Geoffroy Saint-Hilaire, 33615 Pessac, France
   			}
             
   \date{}

  \abstract
  {The accretion history of protostars remains widely mysterious even though it represents one of the best ways to understand the protostellar collapse that leads to the formation of stars.}
  {Molecular outflows, which are easier to detect than the direct accretion onto the prostellar embryo, are here used to characterize the protostellar accretion phase in W43-MM1.}
  {The W43-MM1 protocluster host a sufficient number of protostars to statistically investigate molecular outflows in a single, homogeneous region. We used the CO(2-1) and SiO(5-4) line datacubes, taken as part of an ALMA mosaic with a 2000~AU resolution, to search for protostellar outflows, evaluate the influence that the environment has on these outflows' characteristics and put constraints on outflow variability in W43-MM1.} 
  {We discovered a rich cluster of 46 outflow lobes, driven by 27 protostars with masses of $1-100~\Msol$. The complex environment inside which these outflow lobes develop has a definite influence on their length, limiting the validity of using outflows' dynamical timescales as a proxy of the ejection timescale in clouds with high dynamics and varying conditions. 
  We performed a detailed study of Position-Velocity (PV) diagrams of outflows that revealed clear events of episodic ejection. The time variability of W43-MM1 outflows is a general trend and is more generally observed than in nearby, low- to intermediate-mass star-forming regions. The typical timescale found between two ejecta, $\sim$500~yr, is consistent with that found in nearby protostars.}
  {If ejection episodicity reflects variability in the accretion process, either protostellar accretion is more variable or episodicity is easier to detect in high-mass star-forming regions than in nearby clouds. 
 The timescale found between accretion events could be resulting from instabilities, associated with bursts of inflowing gas arising from the close dynamical environment of high-mass star-forming cores.}

               
   \maketitle

   \begin{figure*}[p]
\vskip -1.5cm 
\centerline{\hskip 2cm \includegraphics[width=1.02\hsize]{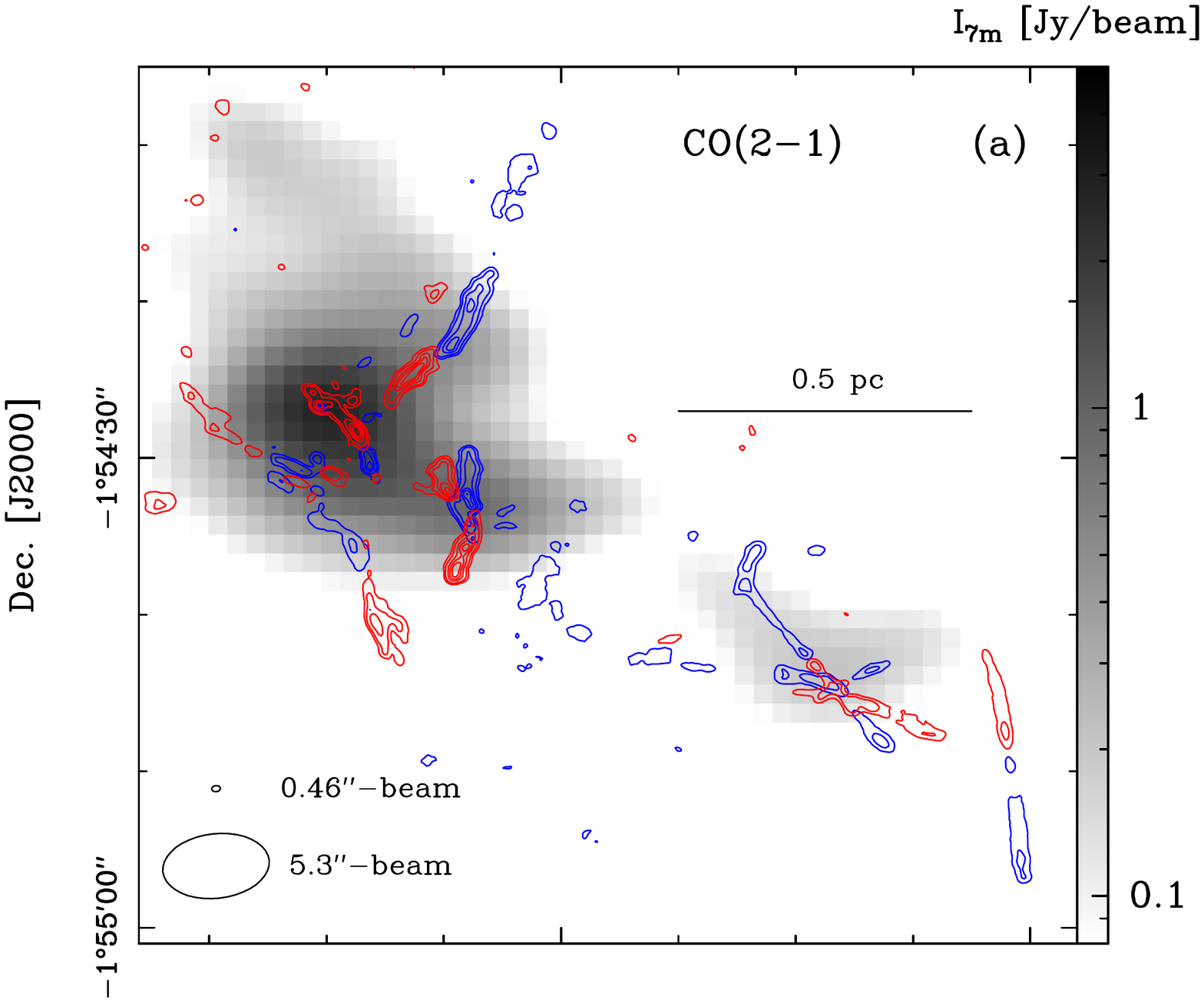}}
\vskip -4.cm 
\centerline{\hskip 2cm \includegraphics[width=1.02\hsize]{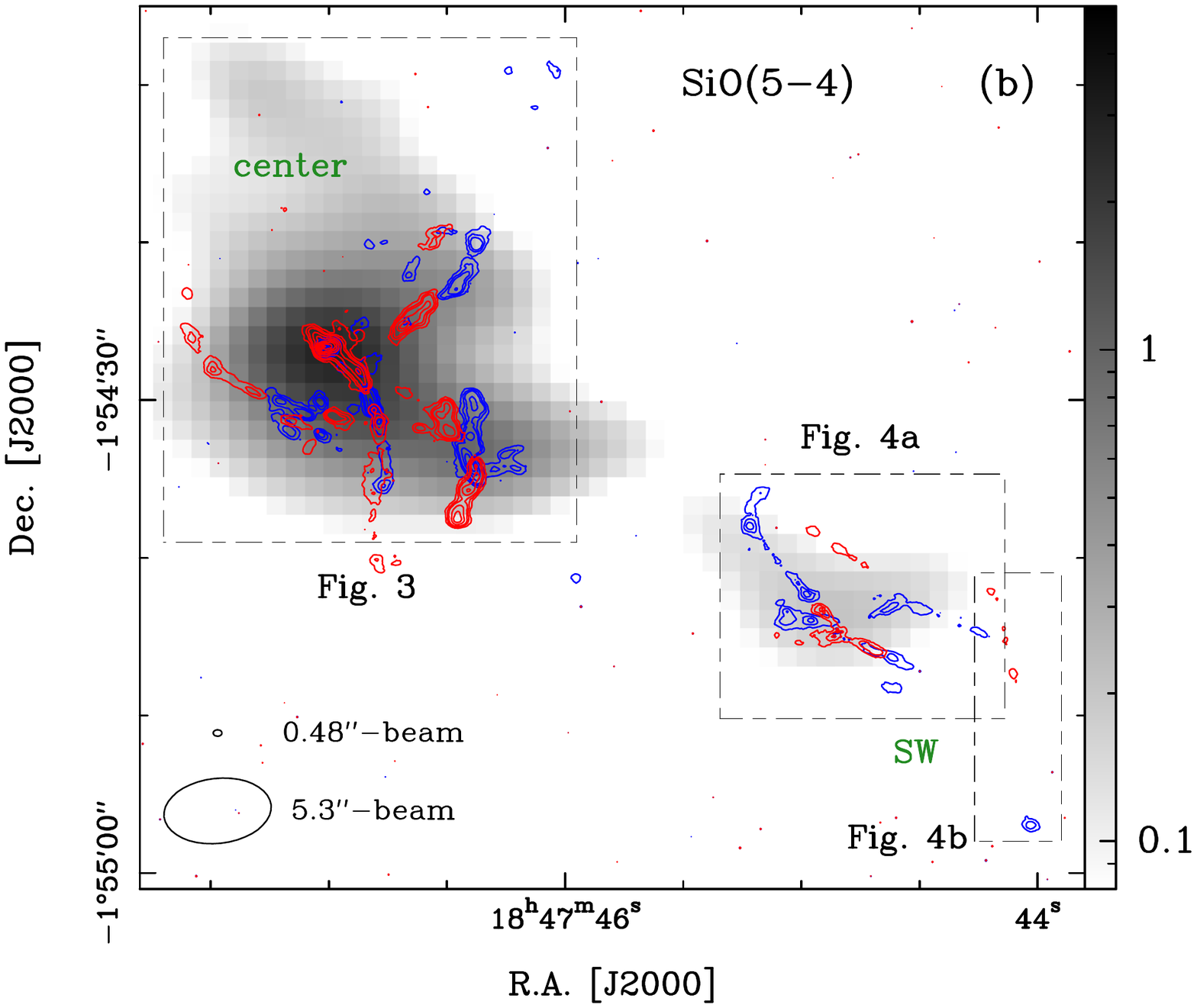}}
\vskip -1.5cm 
\caption{Cluster of molecular outflows in the W43-MM1 massive protocluster.
Contours of the integrated blue- and red-shifted emission of the CO(2-1) (in \textbf{a}) and SiO(5-4) (in \textbf{b}) line wings are overlaid on the 1.3~mm continuum emission of the 7~m array (gray scale, with a $5.3\arcsec$ beam and 10~mJy/beam rms). 
{\bf(a)} The CO line is integrated over $42-85~\kms$ for blue lobes and $111-154~\kms$ for red lobes. Contours are 15, 35 to 235 by steps of 50 in unit of $\sigma_{\rm blue}=33\,\rm mJy\,beam^{-1}\,\kms$ and $\sigma_{\rm red}=42\,\rm mJy\,beam^{-1}\,\kms$.
{\bf (b)} The SiO line is integrated over $43-93~\kms$ for blue lobes and $103-153~\kms$ for red lobes. Contours are 8, 18, 28 to 203 in unit of $\sigma=23\,\rm mJy\,beam^{-1}\,\kms$.
Ellipses in the lower-left corner represent the angular resolutions of each image and a scale bar is shown on the top panel to indicate the size in physical units. Boxes outline the zooms-ins of Figs.~\ref{f:co-ctr-high}-\ref{f:co-ext}.}
\label{f:general}
\end{figure*}

\section{Introduction}

It is now common knowledge that gas accretion during star formation is not a continuous process \citep[see, e.g.,][]{Audart14}. The temporal variability of accretion has recently been statistically studied for young low-mass stars in the final phase of star formation. Variability timescales from weeks to years were found in the light curves of young stars observed in the near-IR wavelength range \citep[e.g.,][]{Parks14, Cody18}. 
Events of much greater variability on larger timescales have also been found in the light curves of FU Orionis-type stars. 
Luminosity variability could probe a wide variety of processes with different timescales. When associated with variable accretion rates, this variability could trace instabilities developing within the disk, tidal interactions between  the disk and a companion, or even inflowing gas streams arising from the stellar environment \citep[e.g.,][]{Zhu09,Lodato04,Alves19}. 

Unfortunately, much less is known about the variability of accretion during the initial phase of star formation, whereas it is during this phase that most of the final stellar mass accumulates from the gas mass reservoir. Jets and outflows develop as protostellar accretion occurs and they possibly evacuate the angular momentum of the collapsing protostellar core. The morphology and kinematics of molecular outflows could thus provide fossil records of the accretion history of a protostar \citep[see, e.g., the reviews by][]{Arce06, Bally16}. The link between the outflow power and the accretion rate is far from direct, however. It could even be loose due to the still-unknown, physical mechanism that 
is at the origin of jets and outflows and because of the density and kinematic structure of the protostar's environment. Despite these caveats, outflow studies have already given very valuable constraints on the protostellar accretion mechanisms. They revealed that accretion rates are stronger at the beginning of the protostellar accretion phase of low-mass stars \citep[e.g.,][]{Bontemps96} and that they are stronger in high-mass ($> 8~\Msol$), compared to low-mass, protostars \citep[e.g.,][]{Beuther02, Duarte13CO, Maud15}. Detailed studies of molecular outflows driven by low-mass stars also suggested that the accretion could be episodic \citep[e.g.,][]{Cernicharo96, Arce01a,  Plunkett15, Jhan16}. 
A statistical study of bipolar outflows driven by high-mass protostars of Cygnus~X also favored a scenario with intermittent accretion rates \citep[e.g.,][]{Duarte13CO}. 

Several theoretical models have been proposed to explain how molecular outflows are formed and driven in the cloud \citep[see the review by][]{Frank14}. The two most compelling ones invoke either a wind launched from the whole surface of the protostellar disk \citep[disk-wind model by, e.g.,][]{Konigl00} or a jet $+$ wind initiated at the innermost region of this disk \citep[X-wind model by, e.g.,][]{Shu00}. Whatever the launching mechanism, these winds could easily recollimate along magnetic-field lines and become jet-like. Outflows could then develop from this collimated jet when it crosses the envelope, creates internal shocks and a terminal bow shock, and sweeps the ambient gas. A wide-angle wind would itself drive shells through the surrounding envelope. Given the large variety of length, opening angle, mass, and velocity structure observed for protostellar outflows, both X-winds and disk-winds physical mechanisms may be acting. 
Each model predicts different relations in diagrams, such as the Position-Velocity (PV), mass-velocity, and position-magnetic field diagrams, that thus could be used to differentiate between them \citep[e.g.,][]{Arce06}. A couple of models have shown PV diagrams associated with episodic ejection \citep[e.g.,][]{Rohde19,Vorobyov18} but they strongly depend on the structure adopted for the protostellar envelope.

From an observational point of view, young, low-mass protostars referred to as Class~0s have molecular outflows extending up to a few $\sim$0.1~pc, with collimation angles of a few degrees only, and velocities up to $100~\kms$ \citep{Bally16}. These molecular outflows are often composed of a collimated jet at high velocity, referred to as the  `extremely high-velocity' component,
plus a shell-like wind tracing gas at lower speeds, referred to as the `low-velocity' or sometimes `standard high-velocity' component.
Besides, collimated jets frequently appear as chains of knots interpreted as internal shocks produced by episodic variations in the protostar mass-loss rate and/or the ejection velocity.
PV diagrams of these jets often display velocity spurs referred to as `Hubble laws', since the gas velocity of these ejection phenomena increases with the distance to the protostar \citep[case of, e.g., the HH211 Class~0,][]{Gueth99}. A series of these velocity spurs composes a jagged profile sometimes called `Hubble wedge' \citep{Arce01a}. These complex PV diagrams have been investigated in about a dozen of low-mass Class~0 protostars so far including L1448-C(N) \citep{Bachiller90}, HH111 \citep{Cernicharo96}, L1551 \citep{Bachiller94}, L1157 \citep{Gueth98}, IRAS 04166 \citep{Tafalla04}, HH212 \citep{Lee06}, HH46-47 \citep{Arce13}, and CARMA-7 in Serpens South \citep{Plunkett15}. 

As for high-mass protostars, current outflow studies often are limited by their angular resolution \citep{Beuther07, Maud15, Cunningham16} or are focused on a handful of objects only \citep{Duarte13CO, Cheng19}. Surveys with high-angular resolution in high-mass star-forming regions are thus unavoidable to characterize high-mass protostellar outflows and finally constrain the mass-loss or even accretion history of high-mass protostars. Interferometric mosaics at (sub)millimeter wavelengths currently are the best way to image a large and homogeneous sample of outflows driven by protostars forming within the same cloud.
W43-MM1 has been imaged with ALMA and represents one of the richest protocluster forming massive stars found to date \citep{motte18b}. It is located at 5.5\,kpc from the Sun \citep{zhang14} at the tip of the Galactic bar \citep{nguyen11b}. This $\rm 6\,pc^2$ cloud ridge has a $2\times10^4\,\Msol$ mass and qualifies as `mini-starburst' because its star-formation activity is reminiscent of that of starburst galaxies \citep[$\rm SFR\,\sim\,6000\,\Msol\,Myr^{-1}$,][]{motte03, louvet14}.  
\cite{motte18b} detected 131 cores in W43-MM1, and among them 13 high-mass cores. With masses from 16 to $100~\Msol$ within their $\sim$2000~AU diameters, these cores are expected to form high-mass stars in the near future.

Here, we present an ALMA mosaic of the W43-MM1 cloud at $\sim$230~GHz, which reveals a cluster of molecular ouflows driven by high- to intermediate-mass protostars. From observations presented in Sect.~\ref{s:obs}, we identify 46 molecular outflow lobes and characterize their morphology (see Sect.~\ref{s:res}). In Sect.~\ref{s:disc}, we discuss the ejection events found in molecular jets and interpret them as reflecting some accretion bursts, and in Sect.~\ref{s:conc} we summarize our findings.

\section{Observations}
\label{s:obs}
 
 Observations were carried out in Cycle 2 between July 2014 and June 2015 (project \#2013.1.01365.S), with ALMA 12~m and 7~m (ACA) arrays and baselines ranging from 7.6~m to 1045~m. W43-MM1 was imaged with a $78\arcsec\times$ 53$\arcsec$ (2.1 pc $\times$ 1.4 pc) mosaic composed of 33 fields with the 12~m array and 11 fields with ACA. In the 12~m configuration, the primary beam is of 26.7$\arcsec$ (45.8$\arcsec$ with ACA) and the maximum detectable scale is of $\sim 12\arcsec$ ($\sim 21\arcsec$ with ACA). Data were reduced in CASA~4.7, applying manual and selfcalibration scripts. We ran the CLEAN algorithm with CASA~5.4 using a robust weight of 0.5 and the \textit{multiscale} (with 0, 1, 3, 9 and 17 beams) option in order to minimize interferometric artefacts due to missing short spacings. Line data were imaged using the merged (12~m + 7~m) configurations while the continuum was imaged with the 12~m and 7~m arrays separately. The 12~m-only data  provide the best sensitivity image chosen to identify compact cores \citep{motte18b} while the 7~m only map is the best to probe the column density background of these cores. Parameters of the four line spectral windows used in this study are presented in Table~\ref{tab:spw}. The 1.3~mm continuum images themselves have $0.44\arcsec$ and $5.3\arcsec$ angular resolutions and $1\sigma$ rms noise levels of $\sim$0.13~mJy\,beam$^{-1}$ and $\sim$9.1~mJy\,beam$^{-1}$ (0.015 K and 0.007 K).

 \begin{table}[!hb]
   \caption[]{Parameters of the merged data for the spectral windows used here.}
   \label{tab:spw}
    \begin{tabular}{ c c c c c c c }
    \hline
    \hline
    Spectral     &   $\nu_{\rm obs}$  & Bandwidth  & \multicolumn{2}{c}{Resolution}  & rms \\
    window           &  [GHz]        &   [MHz]    & \multicolumn{2}{c}{[$\arcsec$] ~ [$\kms$] } & \tablefootmark{a} \\
     \hline
       SiO(5-4)    &  217.033  & 234  & 0.48 &  0.5 & 3.3 \\
     O$^{13}$CS(18-17) & 218.125 & 234 & 0.46 &  0.25 & 3.2 \\
     CO(2-1)    &  230.462   & 469  & 0.46 & 1.3 & 2.3 \\
     $^{13}$CS(5-4)  &  231.144  & 469 & 0.46 & 0.35  & 2.9 \\
    \hline
    \end{tabular}\\
    \tablefoottext{a}{$1\sigma$ rms in unit of mJy\,beam$^{-1}$. 1 mJy\,beam$^{-1}$ corresponds to 0.12~K at 230 GHz.} 
   \end{table}
 
\begin{figure}
\label{f:spec}
\includegraphics[width=0.95\hsize]{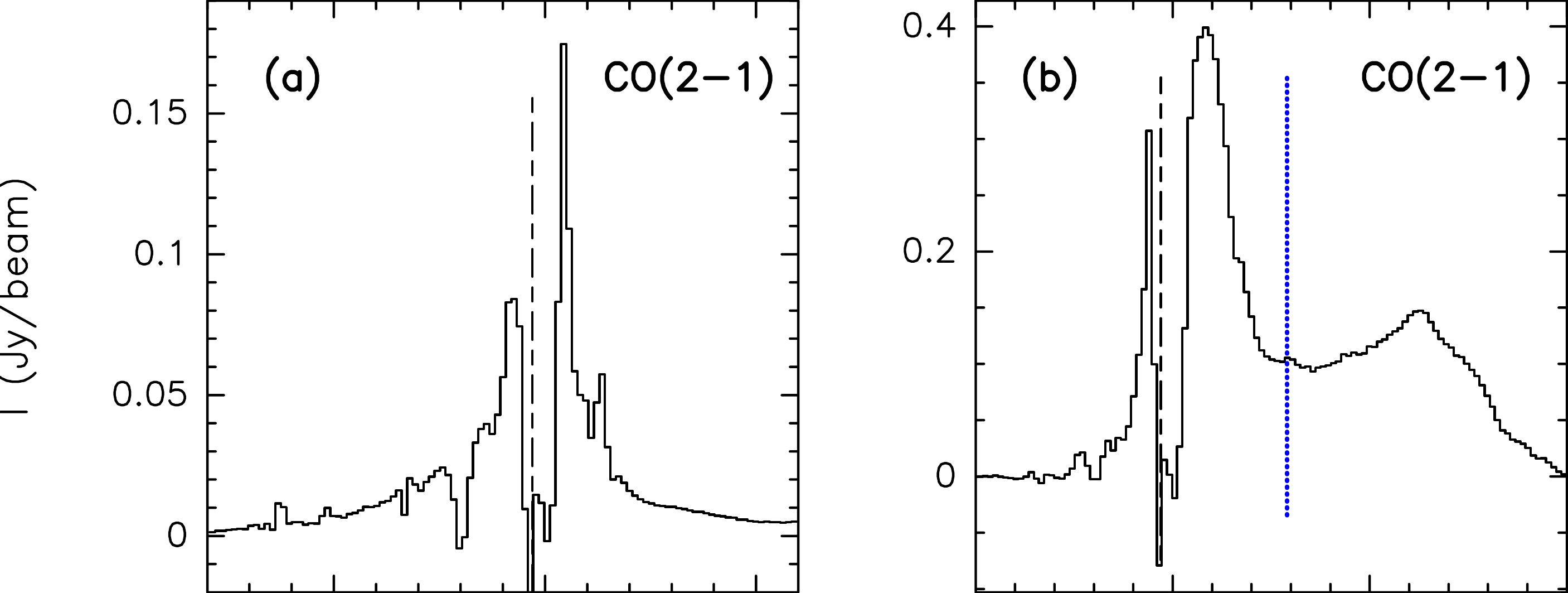}
\vskip 0.3cm
\includegraphics[width=0.965\hsize]{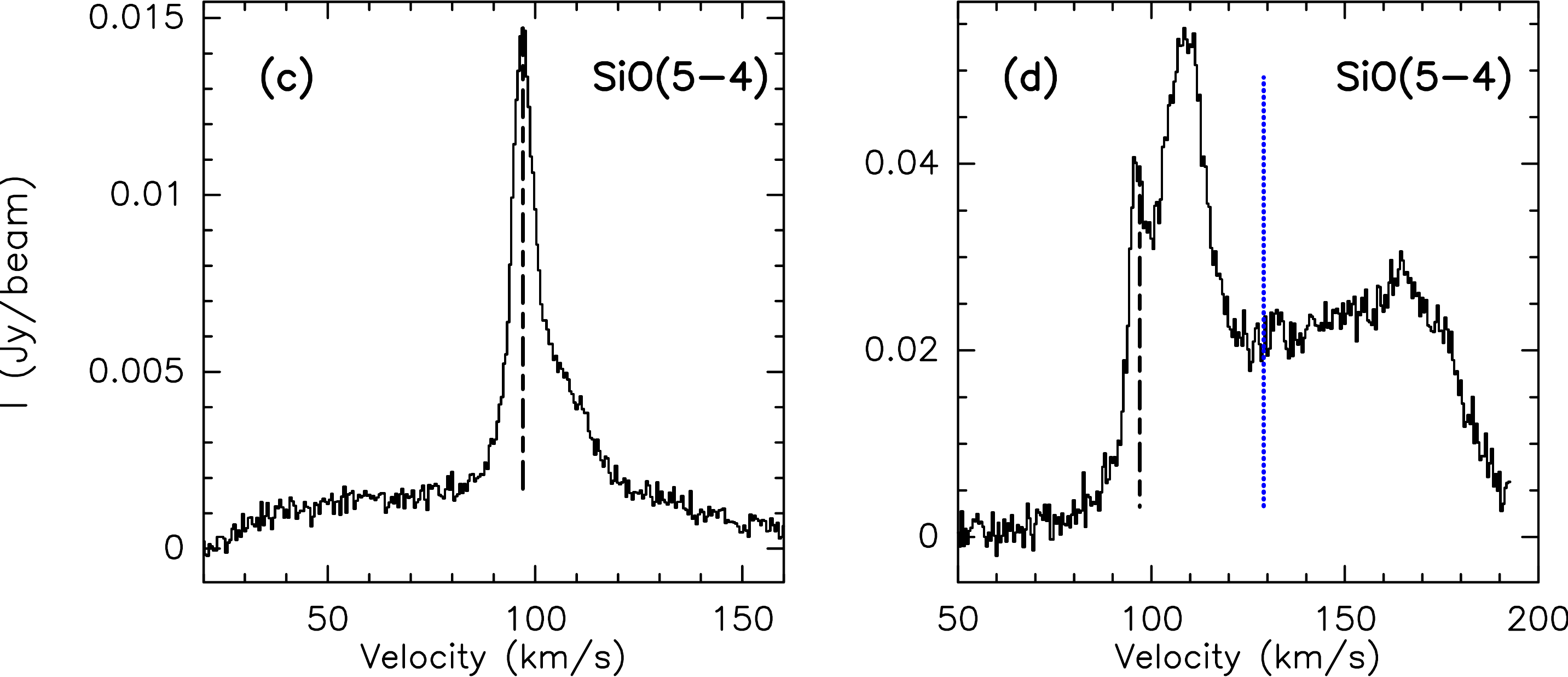}
\caption{Typical CO(2-1) (in {\bf a} and {\bf b}) and SiO(5-4) (in {\bf c} and {\bf d}) lines toward molecular outflows, here that of core \#8. {(\bf a, c)} Lines averaged over the complete extent of the outflow, thus including both the blue and red lobes.
{(\bf b, d)} Lines averaged over one high-velocity knot, here knot R2 of core \#8. Line absorption around the velocity at rest, indicated by a dashed line, is due to missing short spacings and optical thickness. The transition between the low- and high-velocity components of the molecular outflows is indicated in panels {\bf b} and {\bf d} with a blue dotted line.}
\end{figure}

\begin{figure*}[p]
\subfloat{\includegraphics[width=0.5\hsize]{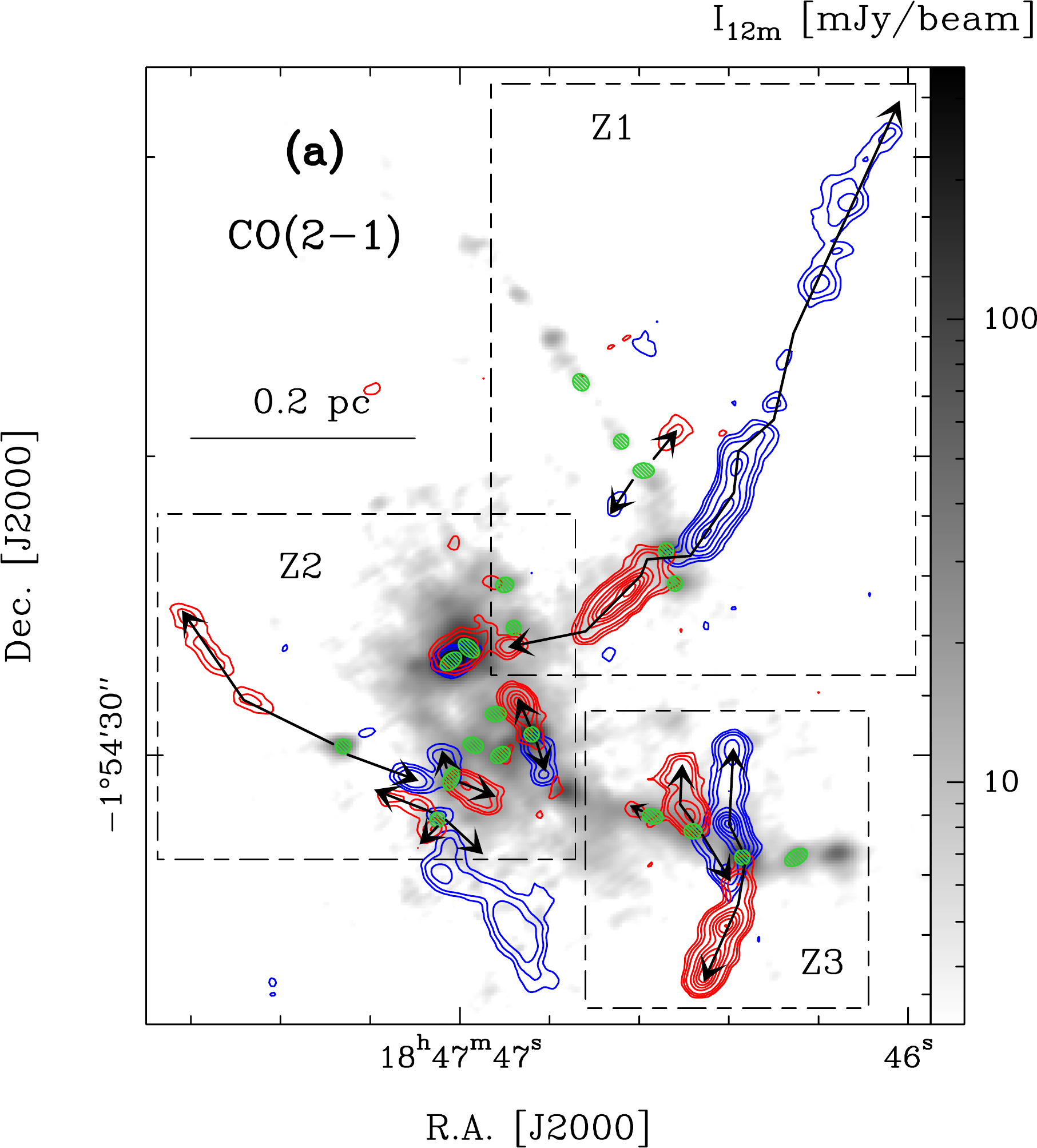}}
\hskip 0.8cm \vspace{0.5cm}
\subfloat{\includegraphics[width=0.47\hsize]{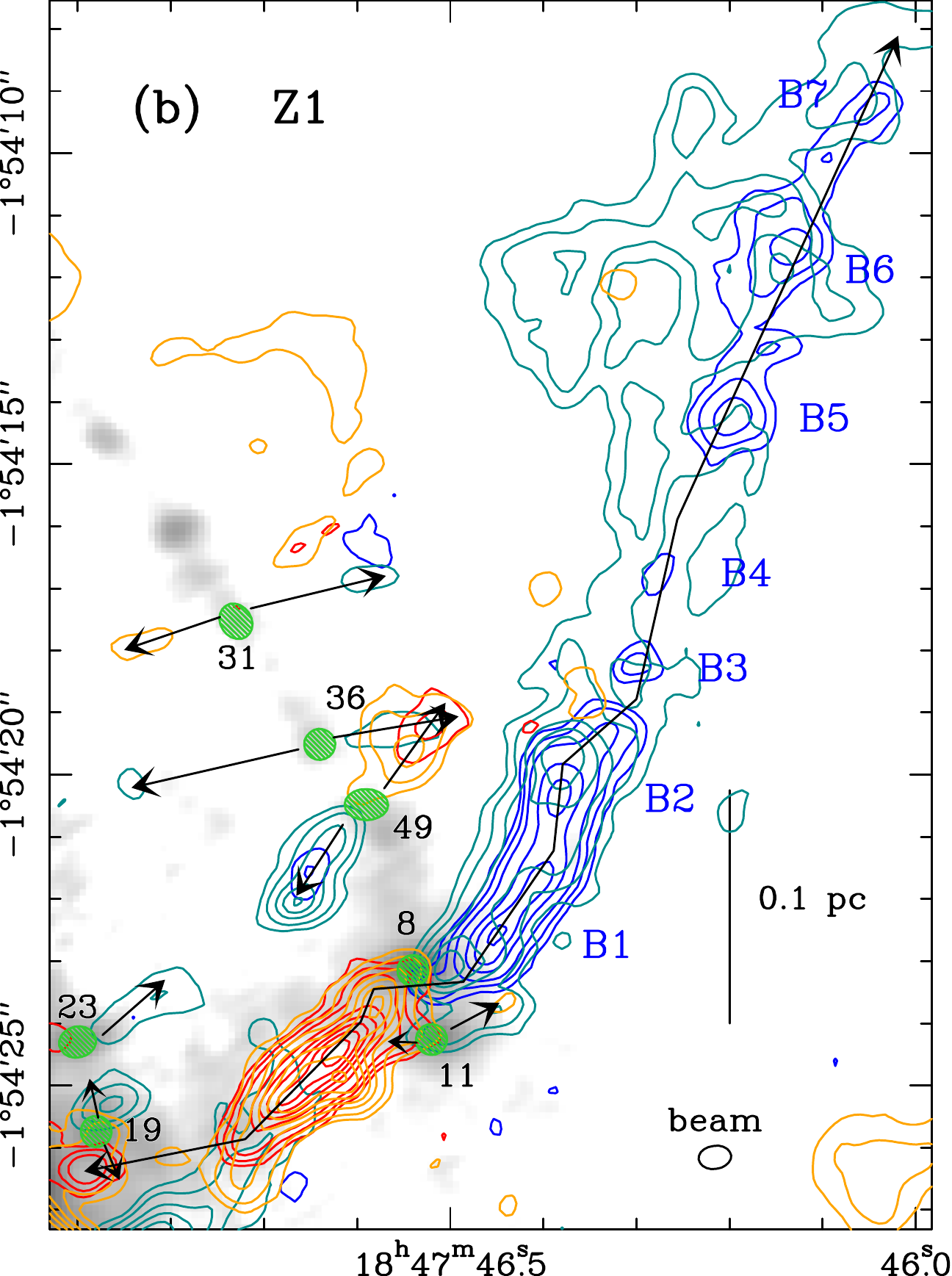}} \\
\subfloat{\includegraphics[width=0.57\hsize]{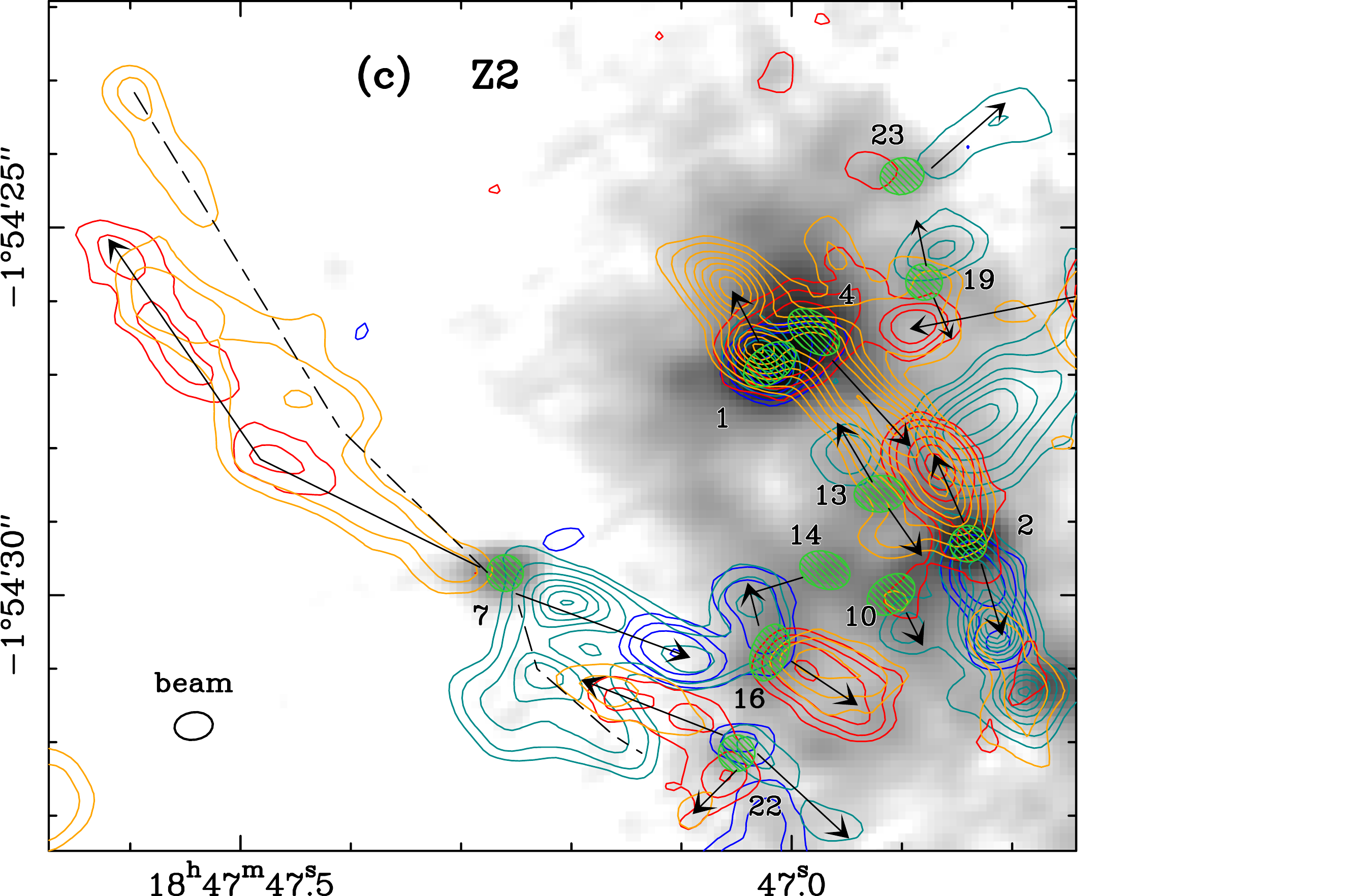}} \hskip -0.3cm
\subfloat{\includegraphics[width=0.31\hsize]{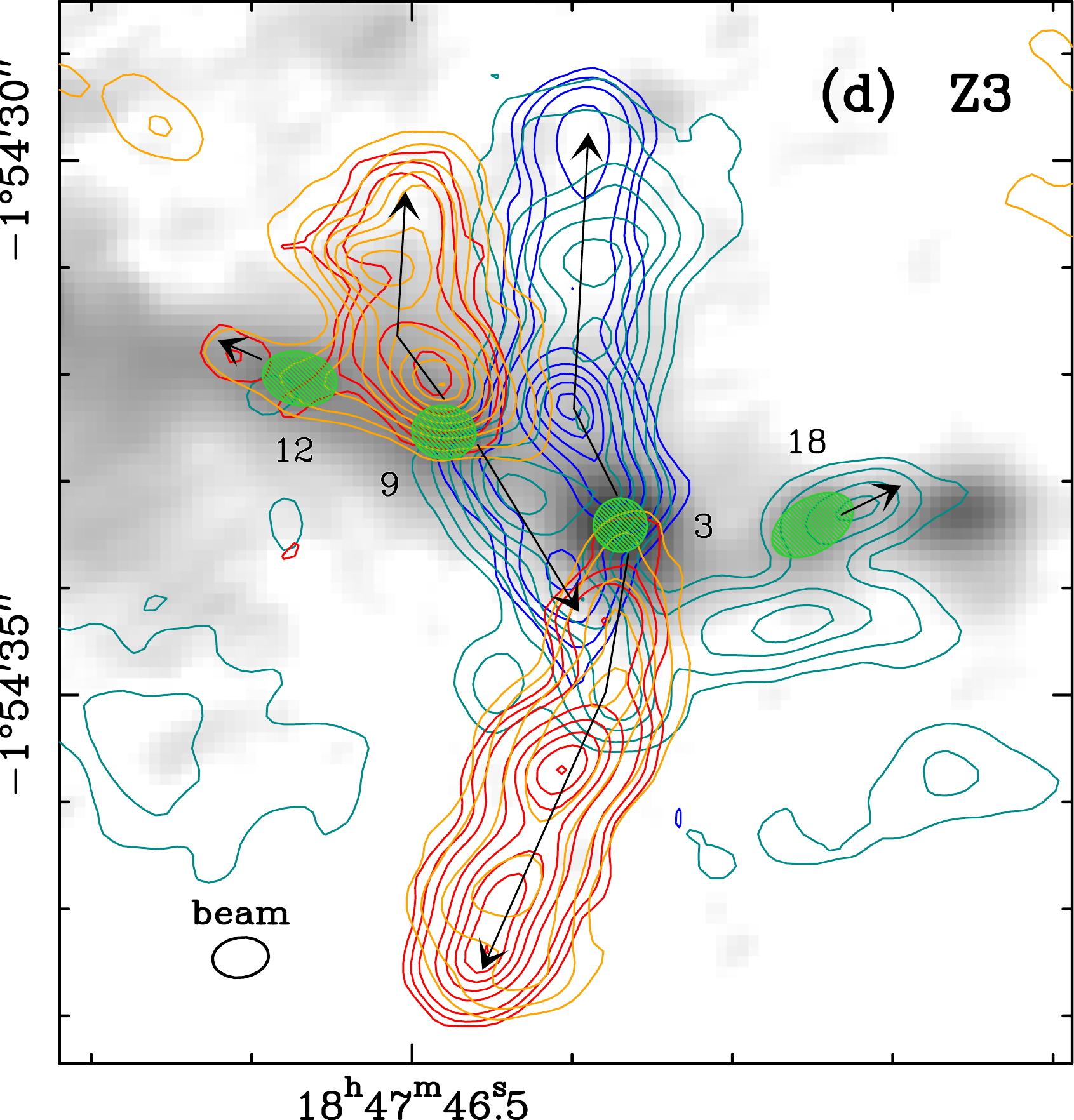}}
\vskip 0.3cm

\caption{Molecular outflows driven by cores in the central region of the W43-MM1 ridge. Outflows detected at high velocity (in {\bf a}) are shown in detail with their low- and high-velocity components in zooms {\bf b}, {\bf c}, and {\bf d}. Contours of the integrated blue- and red-shifted emission of the CO(2-1) line wings are overlaid on the 1.3~mm continuum emission of the 12~m array (gray scale). For blue lobes, emission is integrated over $42-64~\kms$ (HV, blue contours) and $82-88~\kms$ (LV, cyan contours). For red lobes, emission is integrated over $108-119~\kms$ (LV, orange contours) and $128-158~\kms$ (HV, red contours).  
For blue lobes, contours are 7, 15, 30 to 230 by steps of 40 (HV) and 7, 15 to 120 by steps of 15 (LV), in unit of $\sigma=20\,\rm mJy\,beam^{-1}\,\kms$. 
For red lobes, contours are 10, 20 to 160 by steps of 20, in unit of $\sigma_{\rm LV,\, R}=37\,\rm mJy\,beam^{-1}\,\kms$ (LV) and 7, 15, 30 to 280 by steps of 50, in unit of $\sigma$ (HV). Green ellipses locate the W43-MM1 cores, arrows (sometimes broken) indicate the direction of their outflows, dashed line outline the best developed outflow cavities. Knots of the blue lobe of core~\#8 are labelled B1 to B7 in zoom {\bf b} (the first knot B0 is not visible on this integrated map). Ellipses in the lower corners represent the angular resolutions of the images and a scale bar is shown in \textbf{a} and {\bf b} to indicate the size in physical units (same physical units in \textbf{c}-\textbf{d} and Fig.~\ref{f:co-ext} than in \textbf{b}). Some isolated features, like those outlined by the blue contours below the Z2 box in \textbf{a}, correspond to diffuse gas not associated with the W43-MM1 outflows.}
\label{f:co-ctr-high}
\end{figure*}

\begin{figure*}
\subfloat{\includegraphics[width=0.69\hsize]{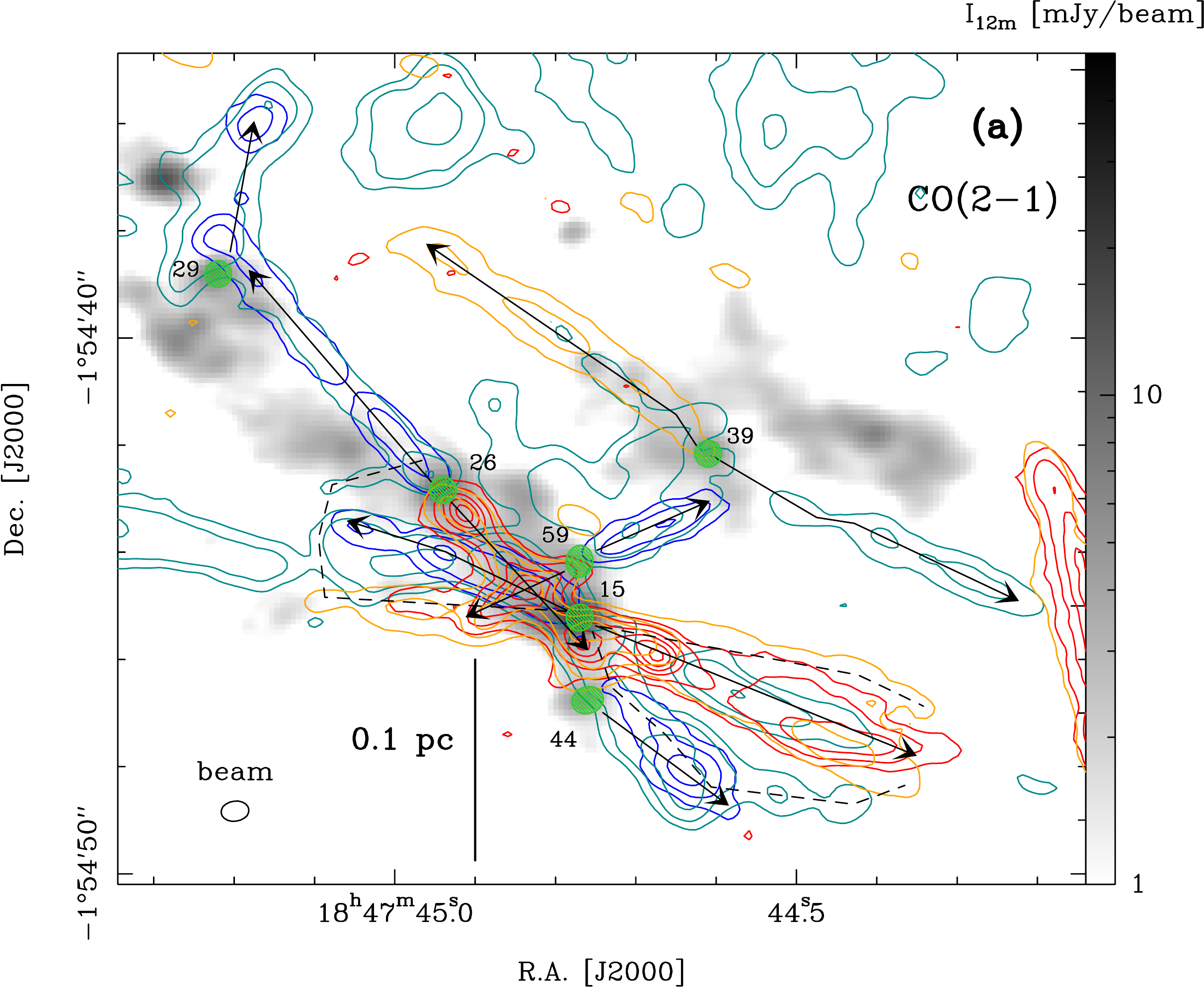}}
\hskip -1.5cm
\subfloat{\includegraphics[width=0.41\hsize]{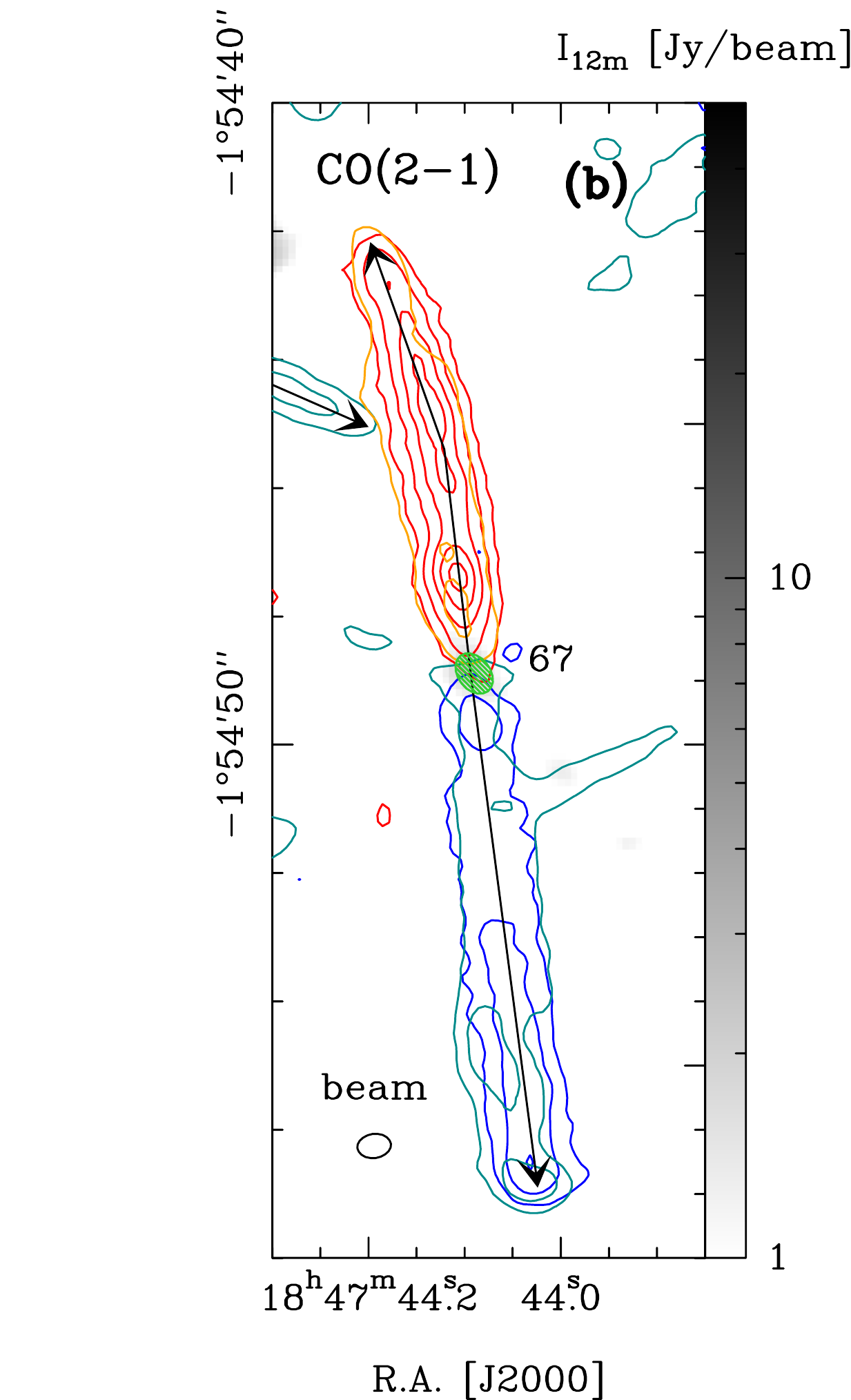}}

\caption{Molecular outflows driven by cores of the south-western part of W43-MM1. Contours correspond to the integrated blue- and red-shifted emission of the CO(2-1) line shown with the same color code as in Fig.~\ref{f:co-ctr-high}. They are overlaid on the 1.3~mm continuum emission of the 12~m array (gray scale). 
For blue lobes, contours are 5, 15, 30 by steps of 15 (HV) and 10, 20, 40, 60 (LV), in unit of $\sigma= 20\,\rm mJy\,beam^{-1}\,\kms$.
For red lobes, contours are 5, 15, 25, 35, in unit of $\sigma_{\rm LV,\, R}= 37\,\rm mJy\,beam^{-1}\,\kms$ (LV) and 5, 15 to 75 by steps of 15, in unit of $\sigma$ (HV). 
Same physical units and same convention as in Figs.~\ref{f:co-ctr-high} for lines, arrows, and ellipses. Isolated features, like those outlined by the cyan contours at the top of panel \textbf{a}, correspond to diffuse gas not associated with the W43-MM1 outflows.}

\label{f:co-ext}
\end{figure*}

   \section{Results and analysis}
\label{s:res} 
Using the CO(2-1) and SiO(5-4) lines, we have conducted a systematic search for outflows driven by the cores that have been detected at $0.44\arcsec$ angular resolution (or 2400~AU at 5.5 kpc) in the 1.3~mm continuum image by \cite{motte18b}.

\subsection{Discovery of a rich cluster of outflows} 
\label{s:pres}
Figures~\ref{f:general}a-b provide an overview of the cluster of molecular outflows found in W43-MM1. The blue- and red-shifted wings of the CO(2-1) and SiO(5-4) lines centered at the cloud velocity at rest, $V_{\rm LSR} \sim 98~\kms$, were there integrated over the entire velocity range that we observed for the line wings.
In the following, we took the same velocity range for all outflows: $42-85~\kms$ and $111-154~\kms$ for CO(2-1); $43-93~\kms$ and $103-153~\kms$ for SiO(5-4) (see Fig.~\ref{f:spec}). 
These velocity ranges are different because CO lines present effects of self-absorption and missing short spacings close to their central velocity (see Fig.~\ref{f:spec}a). The CO(2-1) line also exhibits emission and absorption features associated with low-density clouds located along the line of sight of W43, which have been identified by \cite{nguyen11b} and \cite{Carlhoff13}. These features have a negligible impact on our wide-band integrations except for the absorption observed at $\sim$80~$\kms$, which we excluded by ignoring a handful of channels.

In Figs.~\ref{f:general}a-b, we discovered several tens of outflow lobes, whose existence was confirmed by investigating the velocity cubes of both CO(2-1) and SiO(5-4) lines (see Sect.~\ref{s:flow}). 
Table~\ref{tab:outflow} lists the 27 cores that drive these outflows, gives their core masses, $M_{\rm core}$, their velocity at rest, $V_{\rm LSR}$, and characterize each of their associated outflow lobes. Table~\ref{tab:outflow} separates cores according to their location in W43-MM1: 20 are located within its center and 7 in its south-western part (Center and SW in Fig.~\ref{f:general}b).

A channel map analysis revealed that most of the CO outflows of W43-MM1 protostars present two distinct velocity components: a collimated \textit{jet}-like structure at high velocity ($|\Delta V| > 30~\kms$ from the cloud velocity at rest, $V_{\rm LSR} \sim 98~\kms$) plus a broader and more intense \textit{outflow} at lower velocity. Molecular jets generally exhibit a series of knots that correspond to bumps in the CO(2-1) and SiO(5-4) spectra (see Figs.~\ref{f:spec}b-d). Similar features have been observed in the spectra of outflows driven by low-mass protostars \citep[e.g.,][]{Santiago09,Lefloch15}. 
We therefore integrated the wings of the CO line in two separated ranges, hereafter called the low- and high-velocity (LV and HV) ranges, recalling the standard and extremely high-velocity ranges used by, e.g., \cite{Tafalla04}.
For the blue- and red-shifted wings, they correspond to $\Delta V=-16\,;\,-10~\kms$ (resp. $10\,;\,21~\kms$) and $\Delta V =-56\,;\,-34~\kms$ (resp. $30\,;\,60~\kms$) from the $V_{\rm LSR}$. Figures~\ref{f:co-ctr-high}-\ref{f:co-ext} present the outflows driven by the protostars listed in Table~\ref{tab:outflow} and highlight their ejection direction and length.
Isolated contours in Figs.~\ref{f:co-ctr-high}-\ref{f:co-ext} outline diffuse gas observed along the same line of sight as W43-MM1 but not strictly associated to it. These diffuse structures could lie within the outskirt of the W43 complex \citep{nguyen11b}.

\subsection{Association with dust cores}
\label{s:assoc}

The $^{13}$CS(5-4) line was chosen to estimate the cores' velocity because it traces high-density gas (critical density of $\sim 10^6 \rm ~cm^{-3}$). The $^{13}$CS emission is strong in the central region of W43-MM1 but weaker in its less dense regions like its southwestern and northern parts. 
It is clearly detected toward 
cores or their close surrounding in the central region (see Fig.~\ref{f:cs-velo}a), while only 
half of the cores are detected in the south-western part of W43-MM1 (see Fig.~\ref{f:cs-velo}b). We used the velocities of a Gaussian fitted to the $^{13}$CS(5-4) line integrated over the core, except for cores \#1, \#2, and \#4 for which we used the O$^{13}$CS(18-17) line. This line traces extremely-dense and warm gas and is thus less confused than $^{13}$CS(5-4) by the complex environment of the most central part of W43-MM1, while still being strong enough for a fit (Pouteau et al. in prep.). For the cores neither detected in $^{13}$CS nor in O$^{13}$CS(18-17), we assumed that their velocity at rest would be close to that of their neighboring cores and used $95~\kms$ for undetected cores in the south-westernmost part and $96~\kms$ for core \#31. 
In W43-MM1, the cores' velocities at rest, listed in Table~\ref{tab:outflow}, range from 95 to $102~\kms$ with a gradient from north-east to south-west and a median of $\sim$97.7~$\kms$ over the whole cloud and $\sim$98.6~$\kms$ in the central region (see Fig.~\ref{f:cs-velo}). Given that the $^{13}$CS(5-4) lines were fitted without subtracting any emission from the core's foreground and background, their velocities at rest, $V_{\rm LSR}$, are estimated to be uncertain by up to 1~km\,s$^{-1}$. These $V_{\rm LSR}$ values and gradient are consistent with the ones measured at large scales ($28\arcsec$ or $\sim$0.7~pc) by \cite{nguyen13}.

The association between cores and molecular outflows was mostly done using the high-velocity jet component of CO(2-1) which is the most collimated and the easiest to distinguish from the cloud emission (see Figs.~\ref{f:co-ctr-high}a-d). However the low-velocity component was necessary to identify one third of the 46 outflow lobes because their high-velocity component is either missing or too weak.
SiO(5-4) was used as a secondary outflow tracer, as in \cite{Nony18}, since it is detected along each of the CO outflow lobes (compare Figs.~\ref{f:general}a and \ref{f:general}b, Figs.~\ref{f:co-ctr-high} and \ref{fig:co-sio-main},  Figs.~\ref{f:co-ext} and \ref{fig:co-sio-app}), but four (from cores \#11, \#23, \#31, see Table~\ref{tab:outflow}). 
The association between cores and outflows is considered reliable for 19 cores and more tentative for 8 cores (see Table~\ref{tab:outflow}). The latter consist of outflows developing in confused environments, with several lobes overlapping (cores \#4, \#10, \#13, \#14, red lobes of cores \#11 and \#14); outflows only detected in CO (cores \#23, \#11, red lobe of core \#31); and/or outflows only detected at low velocities ($\Delta V_{\rm max} \leq 20~\kms$, cores \#4, \#13, \#31, blue lobes of cores \#11 and \#36). 
 
\subsection{Main characteristics of outflow lobes}
\label{s:flow}

Among the main characteristics of outflows, one can estimate their maximum velocities and lengths. For this purpose, we investigated the channel maps of CO(2-1) using a $5\sigma=12.5$~mJy\,beam$^{-1}$ threshold. For each outflow, we measured its maximum velocity projected along the line of sight and relative to the core velocity at rest, $\Delta V_{\rm max}=|V_{\rm max}-V_{\rm LSR}|$, and its maximum length projected on the plane of the sky, $L_{\rm max}$ (see Table~\ref{tab:outflow}). In a handful of cases, these values are lower limits because their estimates are confused by other outflows or line-of-sight features that dominate over the CO wings (see Figs.~\ref{f:co-ctr-high}-\ref{f:co-ext} and Fig.~\ref{f:spec}a). We estimated that these velocity and size estimates are uncertain by up to $2~\kms$ (twice the precision of our velocity measurements) and 0.2$\arcsec$ 
(about half our resolution element, 
0.005~pc at 5.5~kpc), respectively. 
Measured velocities range from 11 to $101~\kms$, with a median value of $\sim$47~$\kms$, and measured lengths range from 0.02~pc to 0.4~pc. 

Figures~\ref{f:co-ctr-high}b-d and \ref{f:co-ext}a-b display the directions of the molecular jets associated with each lobe of the 28 detected outflows. Molecular jets are generally well collimated, keeping a constant width close to our resolution limit, of about 0.04~pc (or 8400~AU) all along their length.
The longest molecular jets, $\gtrsim$0.1~pc without any correction for the plane-of-the-sky projection, frequently show direction variations, outlined by broken segments and arrows in Figs.~\ref{f:co-ctr-high}-\ref{f:co-ext}.
Table~\ref{tab:outflow} lists the position angles of molecular jets, defined as the angle of the line passing through the furthest knot of the molecular jet before it undergoes any jet deflection. We thus characterized the closest, and thus youngest, ejections from the core, connecting at best the blue and red lobes. In most of the cases this comes down to taking the core center as reference. However a couple of jets are offset by $\sim$0.3$\arcsec$ with respect to the core center (cores \#8 and \#9 especially, see Figs.~\ref{f:co-ctr-high}b and \ref{f:co-ctr-high}d). The angle uncertainty relies on the precision of the knots' location, corresponding to our ability to connect one beam element at a typical distance of $\sim$ $2\arcsec$.
It generally is $0.5\degree-1\degree$ but reaches $4\degree$ for the shortest lobes. 

The majority (18 out of 28) of the molecular outflows we discovered are bipolar (see Table~\ref{tab:outflow}). Three of the monopolar outflows could have their blue or red counterparts confused by other outflows (\#10 and \#14) or disregarded at low velocity (\#4). In total, we thus evaluate the number of monopolar outflows to correspond to about $25\%$ of the complete sample of outflows in W43-MM1.

\begin{landscape}
   \begin{table}[p]
   \caption[]{Main characteristics of each outflow lobe driven by W43-MM1 protostars located in its center (top rows) and south-western (bottom rows) part.}
   \label{tab:outflow}
    \small
 \makebox[1.25\textwidth][c]{
    \begin{tabular}{ c  c  c  c  c | c  c  c  c | c  c  c  c  | c }
    \hline
    \hline
       &  & Associated core & &  & \multicolumn{4}{c}{Blue lobe} & \multicolumn{4}{c}{Red lobe} &   \\
     \hline  
      Name   &  RA\tablefootmark{a}  &     Dec.\tablefootmark{a}   & $M_{\rm core}$\tablefootmark{a} & $V_{\rm LSR}$\tablefootmark{b}   & ${\Delta V}_{\rm max}$\tablefootmark{c} & $L_{\rm max}$\tablefootmark{c} & 
      PA\tablefootmark{d} & $A_{\rm V}$\tablefootmark{e} & ${\Delta V}_{\rm max}$\tablefootmark{c} & $L_{\rm max}$\tablefootmark{c} & 
      PA\tablefootmark{d} & $A_{\rm V}$\tablefootmark{e} & Comment\tablefootmark{f} \\
          &  [J2000] &  [J2000]  & [$\Msol$]  & [$\kms$]   & [$\kms$] & [$\times 10^{-2}$ pc] & [deg] & [$\times 10^{3}$ mag] & [$\kms$] &[$\times 10^{-2}$ pc]  & [deg] & [$\times 10^{3}$ mag] &     \\
     \hline
      \#1   &  18:47:47.02  &  -1:54:26.86 & $102 \pm 5$ & 98  &- &- &- &- & 34 & 5.3 & $121.5 \pm 0.5$ & 2.5 &    \\
      \#2   &  18:47:46.84  & -1:54:29.30 & $55 \pm 5$ & 99 & 57 & 5.0 & $-74.5 \pm 0.8$ & 1.7 & 48 & 6.1 & $114.6 \pm 0.9$ & 2.4 &   \\
      \#3   & 18:47:46.37   &  -1:54:33.41  & $59 \pm 4$ & 98 & 83$^*$-91 & 11 & $115 \pm 0.5$ & 0.78 (0.53) & 60 & 13 & $-105.5 \pm 0.9$ & 0.52 (0.28) &  \\
      \#4   & 18:47:46.98   &  -1:54:26.42  & $36 \pm 3$ & 102  &- &- &- &- & 19$^*$-25 & 6.9$^*$-9.6 & $-50 \pm 1$  & 2.4 & Conf  \\
      \#7   & 18:47:47.26   &  -1:54:29.70  & $23 \pm 2$ & $99$ & 50 & 8.3 & $-25.1^* \pm 0.5$ & 1.2 & 43 & 18 & $154^* \pm 1$ & 0.39 (0.25) &   \\ 
      \#8   & 18:47:46.54  & -1:54:23.15  &  $14 \pm 2$ & $97$ & 82$^*$-90 & 44 & $18^* \pm 3$ & 0.22 (0.13) & 101 & 13 & $-171^* \pm 2$ & 1.4 (2.5) &  \\
      \#9   & 18:47:46.48   & -1:54:32.54  & $17.8 \pm 0.9$ & $97$ & 61 & 9.6 & $-56^* \pm 1$ & 0.83 (0.68) & 64 & 7.2 & $127.0^* \pm 0,5$ & 0.99 &  \\
      \#10  & 18:47:46.91   &  -1:54:29.99  &  $16 \pm 1$ & $101$ & 29 & 2.7 & $-66.2 \pm 0.5$ & 1.7 &-&-&- &- & Conf  \\
      \#11 & 18:47:46.52   &  -1:54:24.26   &  $2.1 \pm 0.3$ & $95$   & 20  & 4.8  & $-25.9 \pm 0.5$ & 0.55 & 52$^*$-65 & 1.9$^*$-6.1 &$173 \pm 4$ & 0.63 & CO, Conf(R) \\ 
      \#12  & 18:47:46.57   &   -1:54:32.04  &  $ 31 \pm 4$ & $98$ &- &- &- & - & 42 & 2.9 & $157.5 \pm 0.5$ & 1.2 &  \\
      \#13 &  18:47:46.92   &   -1:54:28.62  &  $11 \pm 3$ &  99 & 27 & 2.9 & $131 \pm 4$ & 2.6 & 18 & 2.9 & $-52 \pm 3$ & 2.0 & Conf \\ 
      \#14  & 18:47:46.97   &   -1:54:29.66  & $ 19 \pm 4$ & $100$ & 45 & 4.2 & $-165.8 \pm 0.5$ & 1.8 &- &- &- &- & Conf \\
      \#16  & 18:47:47.02   &   -1:54:30.78   &  $ 36 \pm 6$ & $100$ & 60 & 3.7 & $110.2 \pm 0.5$ & 1.8 & 62 & 4.8 & $-33 \pm 1$ & 1.5 & \\
      \#18  & 18:47:46.25   &   -1:54:33.41   & $ 28 \pm 4$ & $96$ & 25 & 2.9 & $24.7 \pm 0.5$ & 0.69 &- &- &- & -  \\
      \#19  & 18:47:46.88   &   -1:54:25.74   &   $ 9 \pm 3$ & $100$ & 27 & 2.1 & $107 \pm 4$ & 1.5 & 30 & 2.9 & $-68 \pm 1$ & 2.0 &  \\
      \#22-a  &  18:47:47.05  &   -1:54:32.15  &  $ 8 \pm 2$ & $100$ & 45 & 6.1 &$-35.4 \pm 0.5$ & 0.91 & 46 & 7.2 & $159 \pm 2$ & 0.83 &  \\
      \#22-b  &   &  &   &     &- &- &- &- & 58 & 4.0 & $-127.5 \pm 0.5$ & 0.70 &  \\
      \#23  &  18:47:46.90  &   -1:54:24.30  &  $ 11 \pm 3$ & $101$ & 39 & 5.1 & $22 \pm 1$ & 1.1 &- &- &- &-  & CO  \\ 
      \#31  &  18:47:46.73   &  -1:54:17.53  & $4.3 \pm 0.7$  & 96$^*$  & 11$^*$-14 & 7.7 & $17.2^* \pm 1$ & 0.31 & 16$^*$-20 & 5.9 & $-169.3^* \pm 1$ & 0.36 &  CO(R)\\ 
      \#36  & 18:47:46.64   &   -1:54:19.51  &  $3.1 \pm 0.5$  & $96$ & 11$^*$-14 & 8.5 & $-165 \pm 2$ & 0.45 & 27 & 6.3 & $14.6 \pm 0.6$ & 0.32 & Conf(R) \\
      \#49  & 18:47:46.59   &   -1:54:20.48  &  $ 3.7 \pm 0.8$ & $96$ & 49 & 6.1 & $-120.1^* \pm 0.5$ & 0.85 & 50 & 5.9 & $44.8^* \pm 0.5$ & 0.33 &   \\ 
      
      \hline
      \#15 & 18:47:44.77   &   -1:54:45.22   &   $3.9 \pm 0.3$ & 95 & 78 & 13 & $152 \pm 2$ & 0.31 & 68 & 18 & $-24.6 \pm 0.5$ & 0.17 (0.14) &   \\
      \#26 & 18:47:44.94   &   -1:54:42.84  &  $5.2 \pm 0.7$ & 95$^*$ & 45 & 18 & $131.1 \pm 0.5$ & 0.24 & 71 & 12 & $-48.5 \pm 0.5$ & 0.31 (0.23) &  \\ 
      \#29 & 18:47:45.22   &   -1:54:38.81  & $3.1 \pm 0.5$ & 95 & 39 & 9.3 & $78 \pm 1$ & 0.24 &- &- &- &- & \\ 
      \#39 & 18:47:44.61   &   -1:54:42.16  & $2.3 \pm 0.4$ & 95$^*$ & 23 & 19 & $-30.1\pm 0.9$ & 0.21 & 24 & 19 & $133 \pm 4$ & 0.28 & \\
      \#44 & 18:47:44.76   &   -1:54:46.76  & $2.5 \pm 0.3$ & 97 & 50 & 9.3 & $-36.2 \pm 0.5$ & 0.19 &- &- &- &- & \\ 
      \#59 & 18:47:44.77   &  -1:54:44.17  & $1.4 \pm 0.5$ & 95$^*$ & 62 & 7.5 & $24.9 \pm 0.5$ & 0.32 & 64 & 6.7 & $-150.8 \pm 0.5$ & 0.31 & \\ 
      \#67 & 18:47:44.09  &  -1:54:48.89   &  $2.2 \pm 0.3$ & 95$^*$ & 74 & 22 & $-83.8 \pm 0.5$ & 0.14 & 67 & 19 & $101.5 \pm 1.5$ & 0.16 & \\ 
    \hline 
    \end{tabular} 
    } \\
    \tablefoottext{a}{Values taken from \cite{motte18b}.}\\
   \tablefoottext{b}{Velocity at rest estimated with the $^{13}$CS(5-4) line, except for cores \#1, \#2, and \#4, for which the O$^{13}$CS(18-17) line was used instead. A star marker indicates the velocity of the neighboring core that is used for cores neither detected in $^{13}$CS(5-4) nor in O$^{13}$CS(18-17).} \\
   \tablefoottext{c}{Maximum velocity offset and length, uncertain by $2~\kms$ and 0.2$\arcsec$ (0.005~pc), respectively. A star marker indicates lower limits used in our calculations when measurements are confused by line absorption features or neighboring outflows.} \\
   \tablefoottext{d}{Position angle of the molecular jet, measured from West to North. A star marker indicates molecular jets that are offset by more than $0.2\arcsec$ from their core center.} \\
   \tablefoottext{e}{Visual extinction measured over the complete extent of the outflow lobe, uncertain by 30\%. A second measure at the tip of the molecular jet is given in parenthesis for lobes covered by more than 3 pixels in the 7~m array continuum map (i.e., with $L_{\rm max} \geq 9.6 \times 10^{-2}$~pc) and when it differs by more than 20\%.} \\
   \tablefoottext{f}{Less reliable outflow lobes are pinpointed by mentioning their confused environments (denoted with `Conf') or their non-detection in SiO(5-4) but in CO(2-1) (denoted with `CO'). `(R)' means that the note applies to the red lobe only.
   }
   \end{table}
\end{landscape}

For most of the molecular outflow lobes, the low-velocity ($|\Delta V| \sim 10-20~\kms$) emission presents a wider morphology but is similar to that of the high-velocity jet (see flows driven by cores \#3, \#7, \#8, \#9, and \#26 in Figs.~\ref{f:co-ctr-high}b-d and \ref{f:co-ext}a). 
For the longest outflows the observed cavity however tends not to be symmetric around the jet axis (case of cores  \#7 and \#8, see Figs.~\ref{f:co-ctr-high}b-c).
In marked contrast, core \#15 developed outflow lobes with wider-open angle, $30\degree$, cavities, and varying velocities when projected on the line-of-sight (see  Fig.~\ref{f:co-ext}a). To give more general trends is however difficult because our observation limitations (effect of self-absorption and missing short spacings) preclude the detection of the outflow cavities developing close to the systemic velocity ($|\Delta V| < 10~\kms$).

\subsection{Cloud environment crossed by outflow lobes}
\label{s:env}

We characterized the cloud density within which each outflow lobe develops by estimating the mean column density, or equivalently the visual extinction, of the cloud over the complete extent of each outflow lobe. The (filtered) column density along any given line of sight may be derived from $I_{\rm 7m}$, the 1.3~mm flux density measured in the 7~m-array continuum image in a $5.3\arcsec$-beam, by:
\begin{equation}
\begin{split}
N_{\rm H_2, f}\: &= \frac{I_{\rm 7m}}{ \Omega_{\rm beam}\;\mu\,m_{\rm H} \: \kappa_{\rm 1.3mm}\: B_{\rm \tiny 1.3mm}(T_{\rm dust})} \\
 &=\:  9.4\,10^{23}\,\text{cm}^{-2}\left(\frac{I_{\rm 7m}}{\rm Jy\,beam^{-1}}\right),   
\end{split}
\label{eq-Av-filter} 
\end{equation}
where $\Omega_{\rm beam}$ is the beam solid angle, $\mu = 2.8$ is the mean molecular weight, $m_{\rm H}$ is the mass of atomic hydrogen, $\kappa_{\rm 1.3mm}$ is the dust opacity per unit mass column density at 1.3~mm, and $B_{\rm \tiny 1.3mm}(T_{\rm dust})$ is the Planck function for a dust temperature $T_{\rm dust}$. Because W43-MM1 is a warm and dense cloud, we assumed $T_{\rm dust}=23$~K and $\kappa_{\rm 1.3mm}=0.01\,\rm cm\,g^{-1}$ \citep[see arguments in][]{motte18b}.
Since, due to interferometer filtering, scales larger than $\sim$21$\arcsec$ are missing, we added a Av=130~mag level measured at the outskirt of the W43-MM1 cloud ridge on the \textit{Herschel} column density image, with $25\arcsec$ resolution, by \cite{nguyen13}. 

Equation~\ref{eq-Av-filter} then leads to the following equation for the visual extinction, $A_{\rm V}$:
\begin{eqnarray}
A_{\rm V} = \left(\frac {N_{\rm H_2}}{10^{21}\,\rm cm^{-2}}\right) \approx 130~\rm mag + 940~mag \times  \left(\frac {I_{\rm 7m}}{\rm Jy\,beam^{-1}}\right).
\end{eqnarray}

$A_{\rm V}$ values span more than one order of magnitude, from 140 to 2600 mag (see Table~\ref{tab:outflow}). 
Since the median core environments should have rather homogeneous temperature and dust opacity, we estimate the relative uncertainty of our $A_{\rm V}$ values to be about 30\% or less.
Similarly, the large-scale filtered cloud structures have $A_{\rm V}$ values uncertain by up to 30\%, $A_{\rm V}=130\pm 40$~mag. As for the absolute values of the visual extinction, they could be uncertain by up to a factor of two due to opacities which are generally badly constrained.

\subsection{Knots in jets}
\label{s:knot}

\begin{figure*}[]
\centerline{\includegraphics[scale=1]{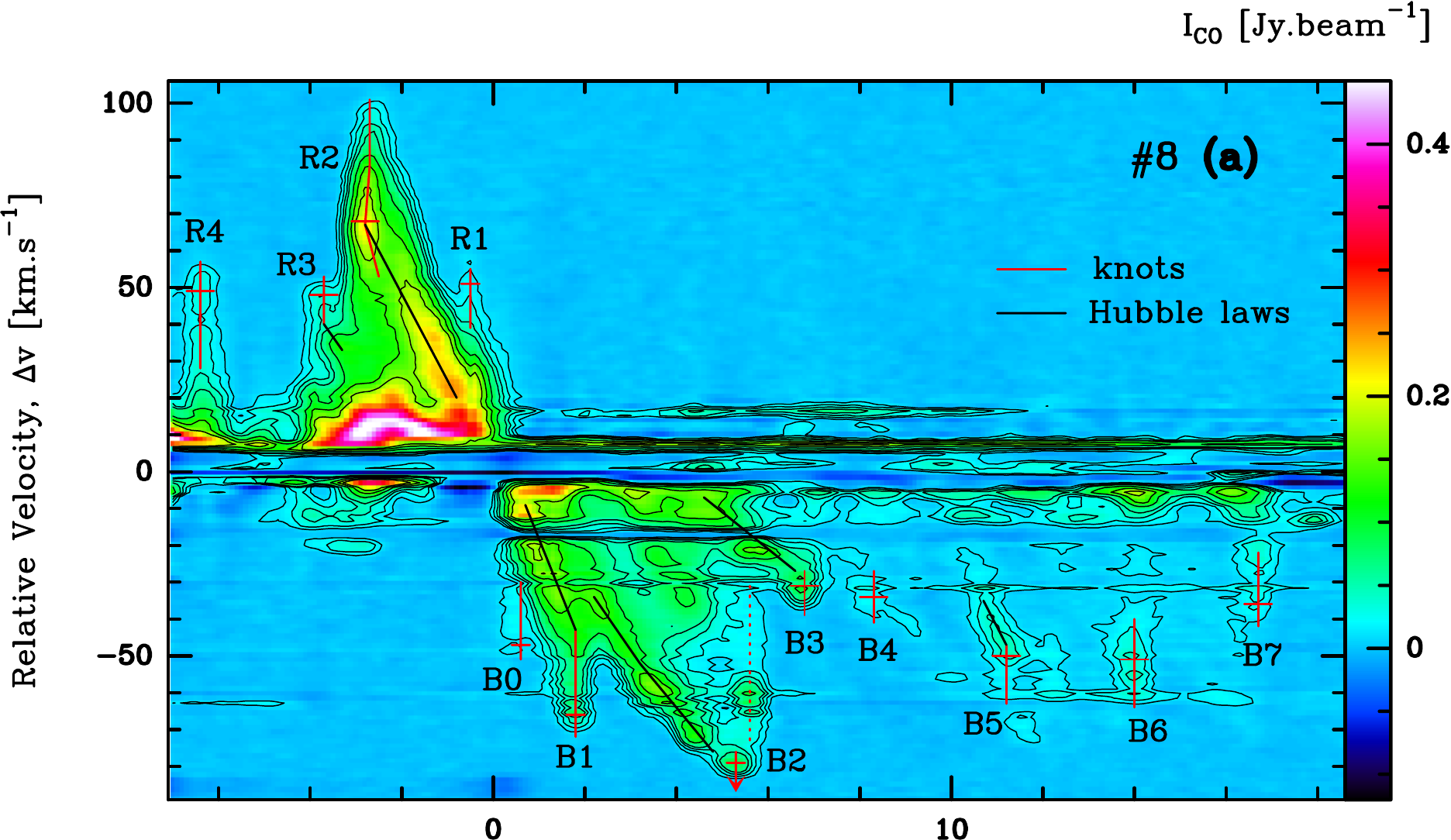}}
\vskip 1cm
\begin{minipage}{0.5\textwidth}
    \centering
    \includegraphics[scale=1]{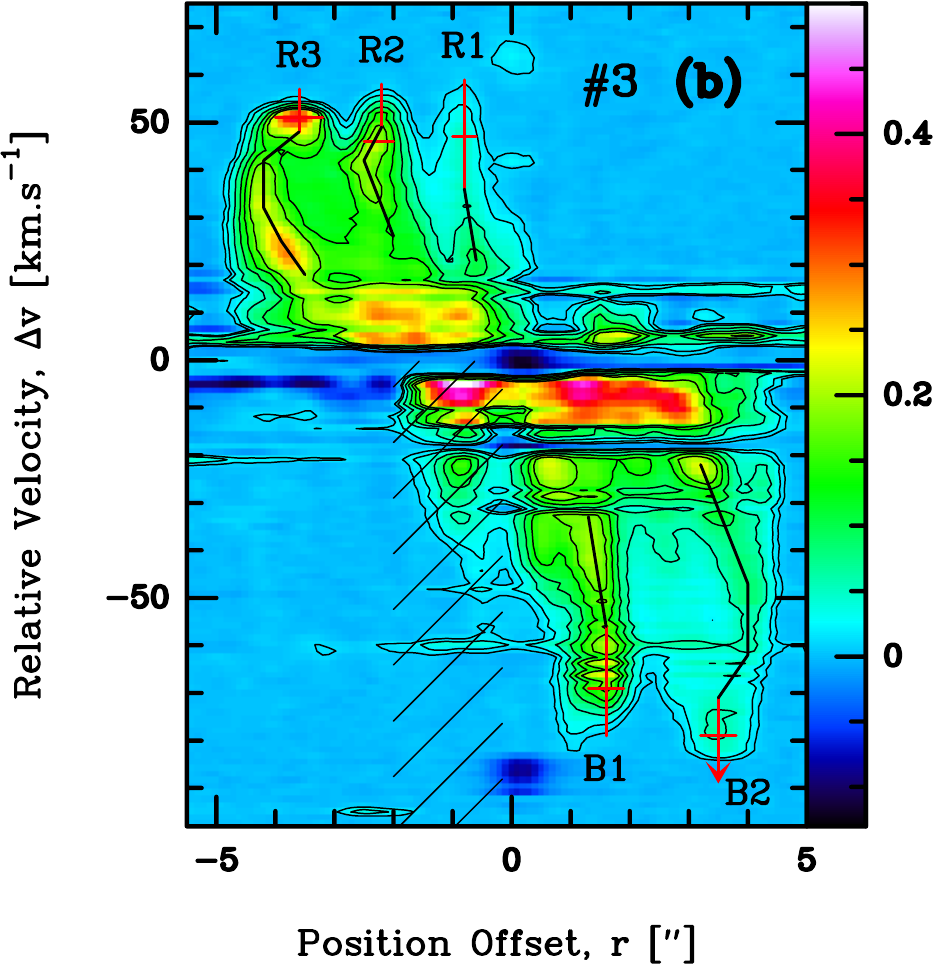}
\end{minipage}
\begin{minipage}{0.5\textwidth}
    \centering
    \includegraphics[scale=1]{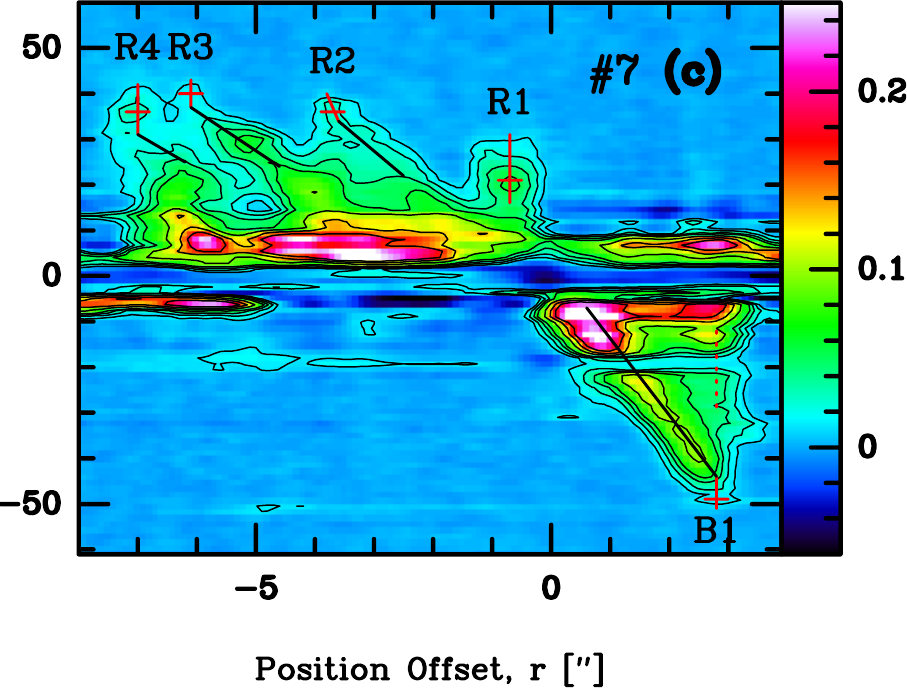}
\end{minipage}
\caption{PV diagrams along the molecular outflows of cores \#8 (in \textbf{a}), \#3 (in \textbf{b}), and \#7 (in \textbf{c}). Position offsets, $r$ on the X axis, are given with respect to the core center or the connecting point between the red and blue lobes. Negative and positive position offsets correspond to the red and blue lobes, respectively. Each figure has the same physical scales on the X axis and on the Y axis.
Contours are 5, 10, 22 to 67 by 15 steps in unit of $\sigma_{\rm CO}=2.5\,\rm mJy\,beam^{-1}$. Additional 16$\sigma_{\rm CO}$ contours are drawn in \textbf{a} and \textbf{c}. Vertical and horizontal segments in red locate the detected knots (at $r_{\rm knot}$ in Table~\ref{tab:tdyn}) and outline their associated velocity range from which $V_{\rm knot}$ was measured. Segments with arrows in \textbf{a} and \textbf{b} indicate the knots whose velocity range is limited by absorption features.
Fingers, with Hubble-law gas distributions connecting the knots with lower-velocity structures, are indicated with continuous black lines, gas layers possibly associated with lateral forward shocks are indicated with dotted red lines. The PV area confused by another outflow is hatched in \textbf{b}.} 

\label{f:PV}
\end{figure*}

\begin{figure*}[]
\centering
\begin{minipage}{\textwidth}
    \centering
    \includegraphics[scale=1]{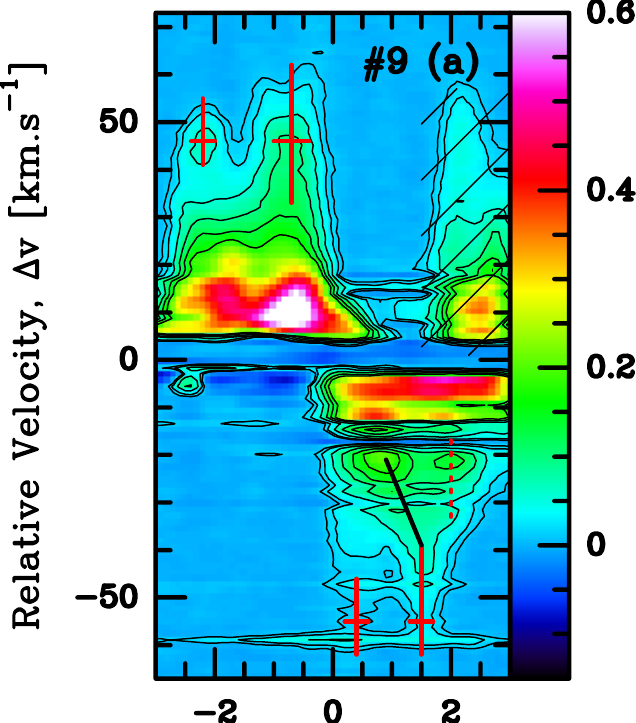}
\includegraphics[scale=1]{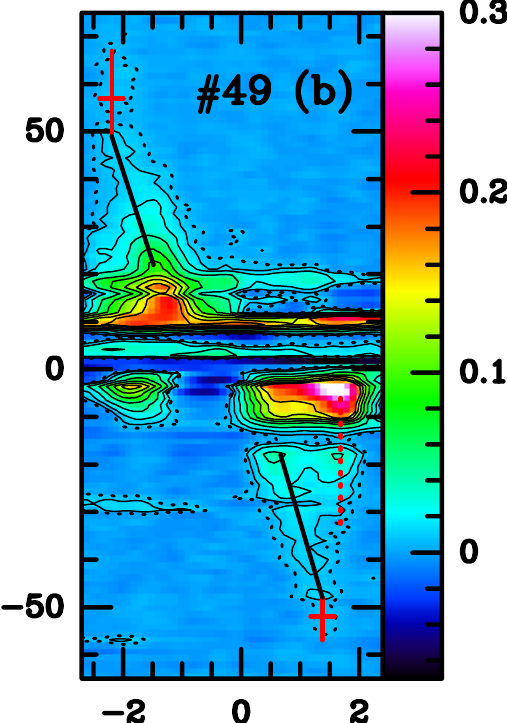}
\includegraphics[scale=1]{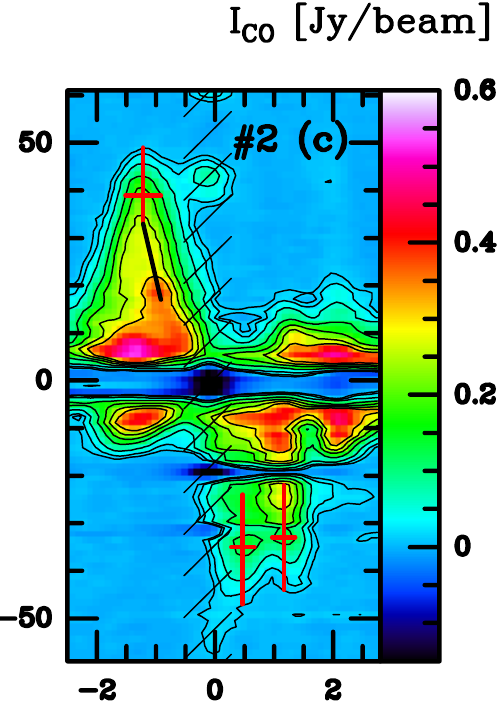} 
\end{minipage}
\vskip 0.5cm
\includegraphics[scale=1]{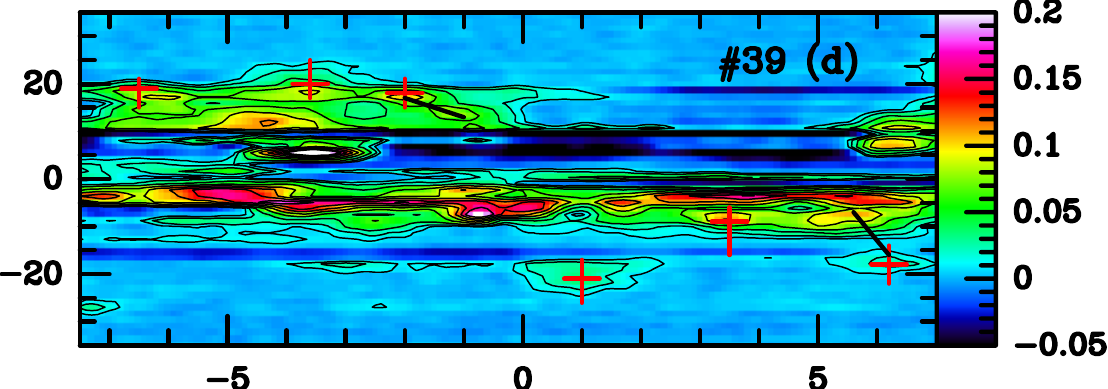}
\includegraphics[scale=1]{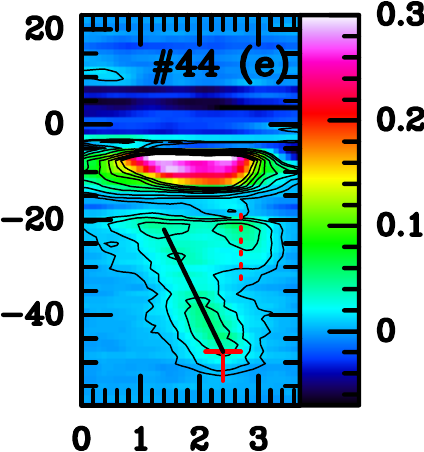}
\vskip 0.5cm
\includegraphics[scale=1]{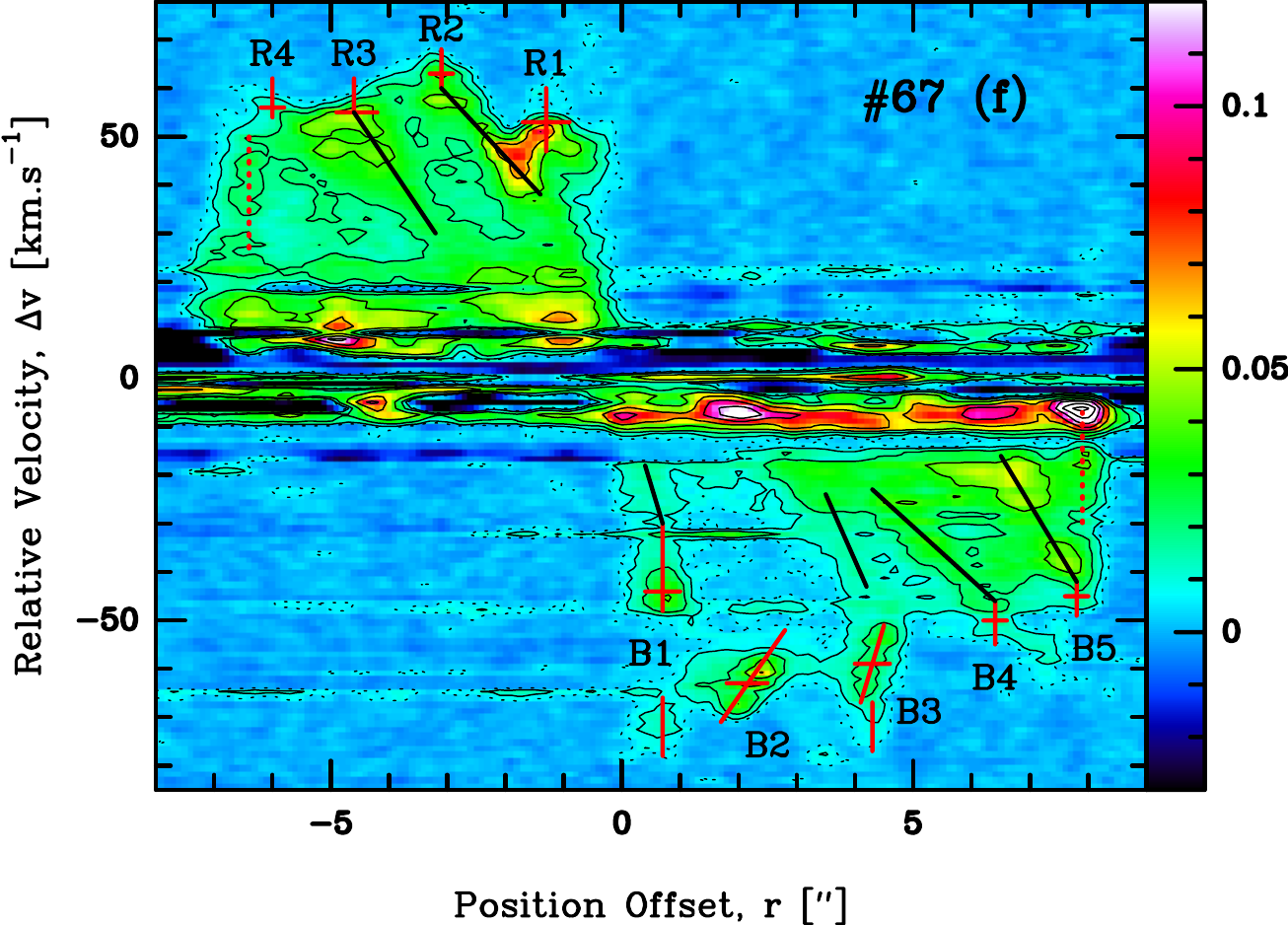}
\vskip 0.3cm
\caption{PV diagrams along the molecular outflows of cores \#9 (in \textbf{a}), \#49 (in \textbf{b}), \#2 (in \textbf{c}), \#39 (in \textbf{d}), \#44 (in \textbf{e}),  and \#67 (in \textbf{f}). 
Same convention of position offsets, segments, and lines as in Fig.~\ref{f:PV}.
Contours are 5, 10, 22 to 67 by steps of 15 in \textbf{a}, 4, 8, 15 to 65 by steps of 10 in \textbf{b} and \textbf{d}--\textbf{f}, 5, 10, 22 to 122 by steps of 25 in \textbf{c}, in unit of $\sigma_{\rm CO}=2.5\,\rm mJy\,beam^{-1}$. Additional 2$\sigma_{\rm CO}$ contours are drawn in dotted line in \textbf{b} and \textbf{f} to highlight weaker emission. PV areas confused by another outflow in \textbf{a} and hot core lines in \textbf{c} are hatched. 
}
\label{f:PV-49}
\end{figure*}

\begin{figure*}[]
\centering
\includegraphics[scale=0.95]{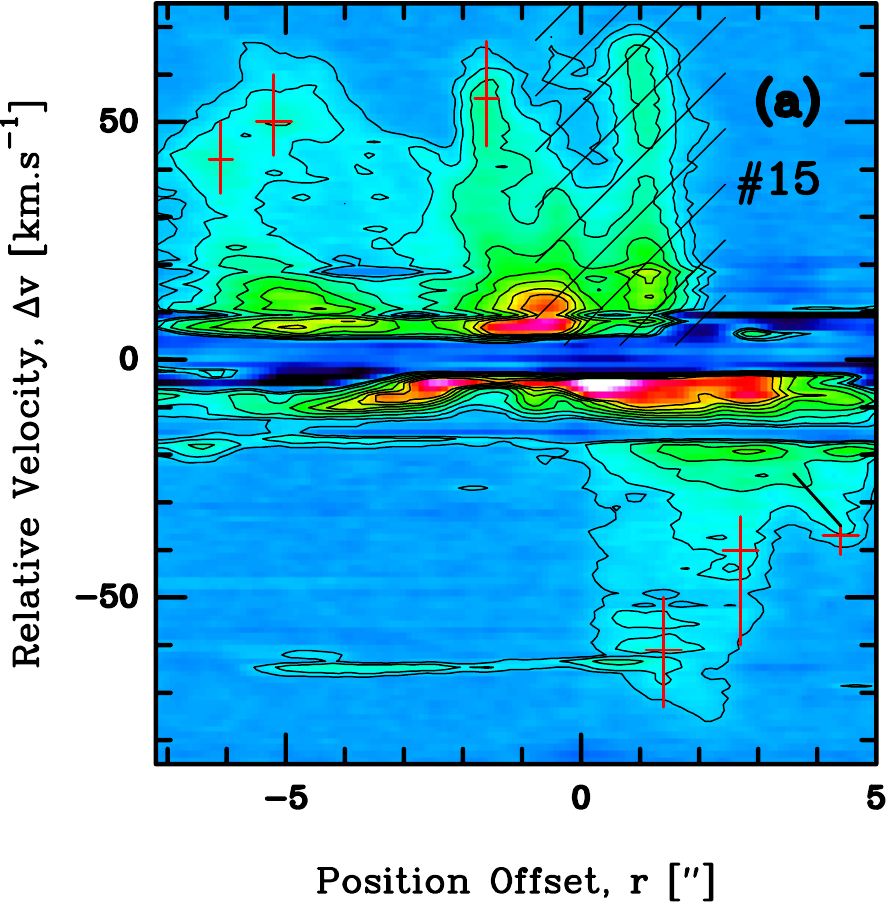}
\includegraphics[scale=0.95]{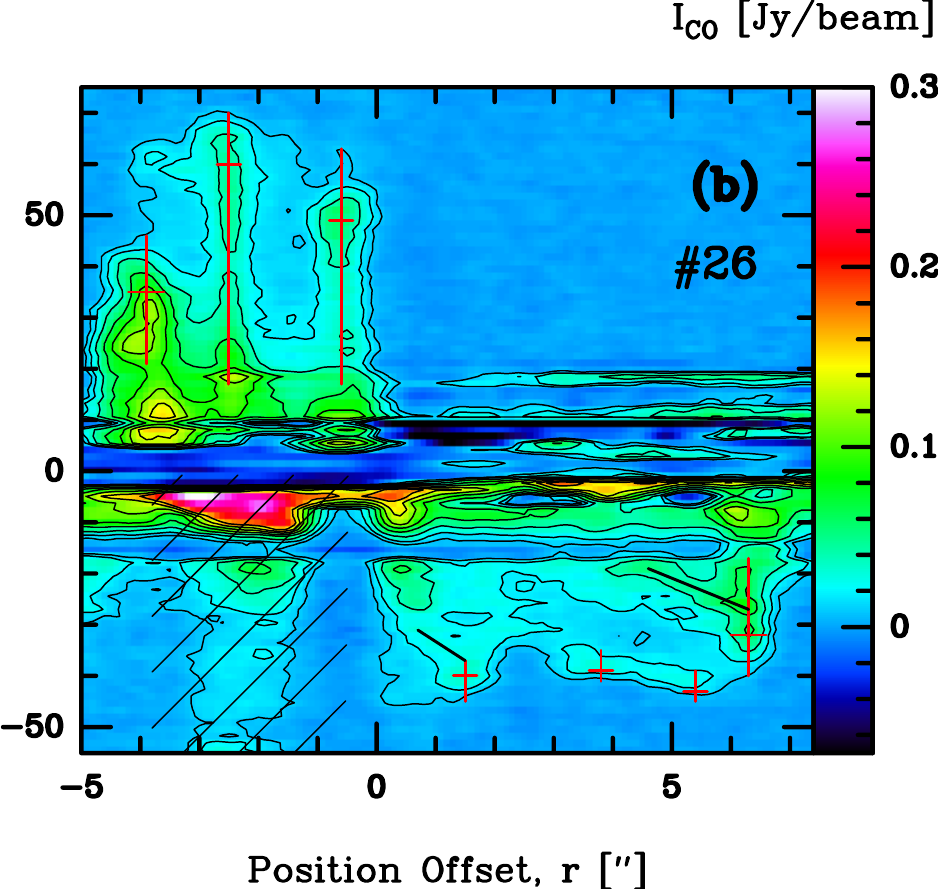} 
\vskip 0.3cm
\caption{PV diagrams along the outflows of cores \#15 (in \textbf{a}) and \#26 (in \textbf{b}). Same convention of position offsets, segments, and lines as in Fig.~\ref{f:PV}.
Contours are 4, 8, 15 to 65 by steps of 10 in unit of $\sigma_{\rm CO}=2.5\,\rm mJy\,beam^{-1}$. PV areas with confusion between the blue lobe of core~\#15 and the red lobe of core~\#26 are hatched.}
\label{f:PV-15}
\end{figure*}

Most of the molecular jets observed both in CO(2-1) and SiO(5-4) integrated maps consist of a continuous emission plus a succession of knots, which are emission peaks aligned along the outflow axis (see Figs.~\ref{f:co-ctr-high}b-d, \ref{f:co-ext}a-b and Figs.~\ref{fig:co-sio-main}a-b, \ref{fig:co-sio-app}a-c).
In order to validate the velocity coherence of these knots initially identified in space (see, e.g., Figs.~\ref{f:co-ctr-high}b-d and \ref{f:co-ext}a-b) and to characterize their velocity and intensity structure, we built Position-Velocity (PV) diagrams along the axis of each jet as defined in Sect.~\ref{s:flow} and displayed in Figs.~\ref{f:co-ctr-high}-\ref{f:co-ext}. In order to smooth local variations and gain in signal-to-noise level, we averaged the PV diagrams over $0.6\arcsec$, about half of the mean width of molecular jets detected in W43-MM1. A smaller averaging width was taken when needed to avoid confusion from neighboring outflows. 
In Figs.~\ref{f:co-ctr-high}-\ref{f:co-ext}, directions of molecular jets are indicated by arrows which follow the jet deflections when they exist. For the few molecular jets with strong deflection, the PV diagram is the juxtaposition of the ones built along each segment of the jet. 

Figures~\ref{f:PV}a-c, \ref{f:PV-49}a-f and \ref{f:PV-15}a-b show the PV diagrams of the outflows that display several knots and are not too confused by other outflows.
For a given core with a bipolar outflow, we chose to give negative position offsets to its red lobe and positive offsets to its blue lobe. PV diagrams present a very complex structure that varies from core to core and often, for a single core, from its red lobe to its blue lobe. As a general trend, the PV diagram of an outflow lobe is constituted of an intense emission at very low velocity plus several linear features/fingers that develop from these low velocities up to the highest velocities of the outflow. 
The low-velocity part of the outflow is not properly constrained by present observations as it corresponds to absorbed zones at the cloud velocity at rest (see Sect.~\ref{s:pres}).
As for the fingers, most of them have velocities that increase approximately linearly with their positional offsets, and thus their distance, from the core.
This distance-to-velocity relation, generally qualified as `Hubble law' 
\cite[see, e.g., review by][]{Arce06}, is outlined by segments in Figs.~\ref{f:PV}--\ref{f:PV-15}. 
The best examples of fingers with Hubble law gas distribution can be found in the outflow lobes of core \#8 and core \#7 (see Figs.~\ref{f:PV}a and \ref{f:PV}c).
The tip, at high velocity, of these linear features are labelled with `R' or `B' plus their ordered number, because they correspond to the spatial knots observed along the jets (see Figs.~\ref{f:co-ctr-high}b-c and \ref{f:co-ext}a-b). 
We stress that this numbering does not indicate the pairs of (red and blue) knots that would have been simultaneously ejected by the protostar.

In total, we discovered 86 knots in the integrated maps and PV diagrams of CO(2-1), contributing to 38 out of 46 outflow lobes driven by 22 cores in W43-MM1. The mean number of jet knots per outflow lobe with at least one knot thus is $\sim$2.3.
Interestingly enough, outflows that do not display any jet knot are low-velocity (<$15~\kms$, for cores \#31 and \#36) or are confused by the overlap with other outflows (cores \#4, \#11, and \#14) (see Table~\ref{tab:outflow} and Figs.~\ref{f:co-ctr-high}b-c). 

Table~\ref{tab:tdyn} lists each of the detected knots with the distance to their associated protostellar core or the connecting point between red and blue lobes, $r_{\rm knot}$, and their velocity relative to the core velocity at rest, $\Delta V_{\rm knot}=|V_{\rm knot}-V_{\rm LSR}|$.
Velocities are measured on the high-velocity vertical segments associated to every knot, either at its local maximum or at half the segment when no clear maximum is observed (see Figs.~\ref{f:PV}--\ref{f:PV-15}). The uncertainties of knot velocities and positions were estimated to be $\sim$3~$\kms$ and $\sim$0.2$\arcsec$ ($\sim$0.005~pc), respectively. 
Table~\ref{tab:tdyn} also lists the mean velocity offset of knots detected in each lobe, calculated as $\Delta V_{\rm lobe} = \overline{\Delta V_{\rm knot}}^{\rm lobe}$. 
The uncertainty on $\Delta V_{\rm lobe}$ is taken to be the dispersion between all the (1 to 8) measurements of $V_{\rm knot}$ with a minimum value of 3~$\kms$. While the dispersion is relatively small, typically $\sim$4$-5~\kms$ (see, e.g, Figs.~\ref{f:PV}b or \ref{f:PV-49}a), four lobes have a dispersion larger than $10~\kms$, among which the blue lobe of cores \#8 and \#15. The first one is the longest lobe in the sample and could display velocity wiggling like those found by \cite{Lee15} or \cite{Choi17}. 
The second atypical lobe propagates in a complex environment. The velocity offsets measured for the blue and red lobes of the 15 bipolar outflows of Table~\ref{tab:tdyn} are similar to within $\sim$15\% or $\sim$5~$\kms$, with the highest asymmetry measured for the high-mass cores \#3, \#7, and \#16. Over the complete sample of 46 lobes, the blue and red lobes have the same median value, $\overline{\Delta V_{\rm knot}}^{\rm blue\;lobe} \simeq \overline{\Delta V_{\rm knot}}^{\rm red\;lobe}\sim 42~\kms$.

\begin{table*}[]
    \caption{Characteristics of the series of knots detected in the molecular jets driven by W43-MM1 protostars located in its center (top rows) and south-western (bottom rows) part.}
    \label{tab:tdyn}
    \begin{tabular}{c | c c c c c | c c c c c }
    \hline
    \hline
     Cores  & \multicolumn{5}{c}{Blue lobe} &  \multicolumn{5}{c}{Red lobe}    \\ \cmidrule{2-6} \cmidrule{7-11}
            &  $r_{\rm knot}$\tablefootmark{a}  &  $\Delta V_{\rm knot}$\tablefootmark{a} & $\Delta V_{\rm lobe}$ &  $t_{\rm knot}$ & $\Delta t$\tablefootmark{b}  &  $r_{\rm knot}$\tablefootmark{a}  &  $\Delta V_{\rm knot}$\tablefootmark{a}  & $\Delta V_{\rm lobe}$ &  $t_{\rm knot}$ & $\Delta t$\tablefootmark{b} \\
            & [$\arcsec$] &  [$\kms$] &  [$\kms$]  & [kyr] & [kyr] & [$\arcsec$] &  [$\kms$]  &  [$\kms$] & [kyr] & [kyr] \\
    \hline       
    \#1   &  &  & & &  & 0.8 & 20 & $20 \pm 3$ & $1.0 \pm 0.4 $ & \\
    \hdashline
    \#2   & 0.4 & 35 & & $0.3 \pm 0.2$ & & & & & & \\
          & 1.4 & 33 & $34 \pm 3$ & $1.0 \pm 0.2$ & $0.7 \pm 0.4$ & 1.2 & 39 & $39 \pm 3$ & $0.8 \pm 0.2$ & \\
    \hdashline
    \#3   & 1.6 & 69 & & $0.5 \pm 0.1$ &  & 0.8 & 47 & & $0.4 \pm 0.1$ & \\
          & 3.5 & 79 & $74 \pm 5$ & $1.2 \pm 0.2$ & $0.7 \pm 0.3$ & 2.2 & 46 & $48 \pm 3$ & $1.2 \pm 0.2$ & $0.7 \pm 0.3$ \\  
          &  &  & &  &  & 3.6 & 51 & & $1.9 \pm 0.2$ & $0.7 \pm 0.3$\\
    \hdashline
    \#7   & & & & & & 0.7 & 21 & & $0.5 \pm 0.3$ & \\
          & 2.8 & 49 & $49 \pm 3$ & $1.5 \pm 200$ & & 3.8 & 36 & $33 \pm 7$ & $2.9 \pm 0.8$ &$2.4 \pm 0.8$ \\ 
          & & & & & & 6.1 & 40 &  & $4.7 \pm 1.2$ & $1.8 \pm 0.7$ \\
          & & & & & & 7.0 & 36 & & $5.4 \pm 1.3$ & $0.7 \pm 0.5$ \\
    \hdashline
    \#8   & 0.6 & 47 & & $0.3 \pm 0.2$ & & 0.8 & 51 & & $0.4 \pm 0.1$ & \\
          & 1.8 & 66 & & $0.9 \pm 0.4$ & $0.6 \pm 0.4$ & 3.0 & 68 & & $1.5 \pm 0.2$ & $1.1 \pm 0.3$ \\
          & 5.3 & 79 & & $2.7 \pm 1.0$ & $1.8 \pm 0.8$  & 3.9 & 48 &  & $2.0 \pm 0.2$ & $0.5 \pm 0.2$ \\
          & 6.8 & 31 & $49 \pm 16$ & $3.5\pm 1.2$ & $0.8 \pm 0.5$ & 6.2 & 49 &  $49 \pm 3$ & $3.2 \pm 0.3$ & $1.2 \pm 0.3$ \\
          & 8.3 & 34 & & $4.3 \pm 1.4$ & $0.8 \pm 0.5$ & & & & \\
          & 11.2 & 50 & & $5.8 \pm 1.9$ & $1.5 \pm 0.7$ & & & & \\
          & 14.0 & 51 & & $7.2 \pm 2.4$ & $1.4 \pm 0.7$ & & & & \\  
          & 16.7 & 36 & & $8.6 \pm 2.8$ & $1.4 \pm 0.6$ & & & & \\  
    \hdashline
    \#9   & 0.4 & 55 & &$0.2 \pm 0.1$ & & 0.7 & 46 & & $0.4 \pm 0.1$ & \\
          & 1.5 & 55 & $55 \pm 3 $ & $0.7 \pm 0.1$ & $0.5 \pm 0.2$ & 2.2 & 46 & $46 \pm 3$ & $1.2 \pm 0.2$ & $0.8 \pm 0.3$ \\
   \hdashline
   \#10   & 0.6 & 27 & $27 \pm 3$ & $0.6 \pm 0.3$ & & & & & & \\
   \hdashline
   \#12   & & & & & & 0.7 & 39 & $39 \pm 3$ & $0.5 \pm 0.2$ & \\
   \#13   & 0.7 & 26 & $26 \pm 3$ & $0.7 \pm 0.3$ & & 0.5 & 22 & $22 \pm 3$ & $0.6 \pm 0.3$ & \\
   \hdashline
   \#16   & 0.4 & 56 & & $0.2 \pm 0.2$ & & & & & & \\
          & 0.9 & 33 & $45 \pm 12$& $0.5 \pm 0.3$ & $0.3 \pm 0.3$ & 0.9 & 58 & $58 \pm 3$ & $0.4 \pm 0.1$ & \\
   \hdashline
   \#18   & 0.8 & 26 & $26 \pm 3$ & $0.8 \pm 0.3$ & & & & & & \\   
   \#19   & 0.4 & 23 & $23 \pm 3$ & $0.4 \pm 0.3$ & & 0.3 & 22 & $22 \pm 3$ & $0.4 \pm 0.3$ & \\   
   \hdashline
   \#22-a   & 0.3 & 36 & $36 \pm 3$ & $0.2 \pm 0.2$ & & 0.6 & 36 & $40 \pm 4$ & $0.4 \pm 0.2$ & \\
            & & & & & & 1.8 & 44 &  & $1.1 \pm 0.2$ & $0.8 \pm 0.3$  \\
   \#22-b   & & & & & & 0.3 & 51 & $41 \pm 11$ & $0.2 \pm 0.2$ & \\      
             & & & & & & 1.2 & 30 &   & $0.8 \pm 0.3$ & $0.6 \pm 0.4$  \\
    \hdashline
    \#23   & 0.9 & 36 & $36 \pm 3$ & $0.6 \pm 0.2$ & & & & & & \\ 
    \#49  & 1.4 & 53 & $53 \pm 3$ & $0.7 \pm 0.1$ & & 2.1 & 56 & $56 \pm 3$ & $1.0 \pm 0.1$ & \\
    \hline 
    \#15  & 1.4 & 61 & & $0.8 \pm 0.3$ & & 1.6 & 55 & & $0.8 \pm 0.2$ & \\
          & 2.7 & 40 & $47 \pm 11$ & $1.5 \pm 0.5$ & $0.7 \pm 0.4$ & 5.2 & 50 & $49 \pm 6$ & $2.7 \pm 0.5$ & $1.9 \pm 0.4$ \\
          & 4.4 & 37 & & $2.4 \pm 0.7$ & $0.9 \pm 0.4$ & 6.1 & 42 & & $3.2 \pm 0.5$ & $0.5 \pm 0.3$ \\
    \hdashline      
    \#26  & 1.5 & 40 & & $1.0 \pm 0.2$ & & 0.6 & 49 & & $0.4 \pm 0.2$  & \\
          & 3.8 & 39 & & $2.5 \pm 0.4$ & $1.5 \pm 0.4$ & 2.5 & 60\tablefootmark{b} & & $1.5 \pm 0.4$ & $1.1 \pm 0.4$ \\
          & 5.4 & 43 & $39 \pm 4$ & $3.6 \pm 0.5$ & $1.1 \pm 0.4$ & 3.9 & 35 & $42 \pm 7$ & $2.4 \pm 0.4$ & $0.8 \pm 0.4$ \\
          & 6.3 & 32 & & $4.1 \pm 0.6$ & $0.6 \pm 0.2$ & & & &  & \\
    \hdashline
    \#29  & 1.5 & 31 & & $1.2 \pm 0.3$ & & & & & \\
          & 2.9 & 35 & $33 \pm 3$ & $2.2 \pm 0.4$ & $1.1 \pm 0.4$ & & & & \\
    \hdashline
    \#44  & 2.4 & 46 & $46 \pm 3$ & $1.3 \pm 0.2$ & & & & & & \\
    \hdashline   
    \#39  & 1.0 & 21 & & $1.6 \pm 0.8$ & & 2.0 & 113 & & $2.7 \pm 0.7$  & \\
          & 3.5 & 9 & $16 \pm 5$ & $5.5 \pm 2.1$ & $4 \pm 1$ & 3.6 & 115 & $19 \pm 3$  & $4.8 \pm 1.0$ & $2.1 \pm 0.9$ \\
          & 6.2 & 18 & & $9.8 \pm 3.4$ & $4 \pm 1$ & 6.5 & 114 & & $8.7 \pm 1.6$ & $4 \pm 1$ \\
    \hdashline
    \#59  & 0.8 & 50 & & $0.5 \pm 0.1$ & & 0.5 & 50 & & $0.3 \pm 0.2$ & \\
          & 2.0 & 45 & $45 \pm 4$ & $1.1 \pm 0.2$ & $0.7 \pm 0.3$ & 1.1 & 60\tablefootmark{c} & $43 \pm 7$ & $0.6 \pm 0.3$ & $0.4 \pm 0.3$ \\
          & 2.5 & 41 & & $1.4 \pm 0.2$ & $0.3 \pm 0.3$ & 1.7 & 45 &  & $1.0 \pm 0.3$ & $0.4 \pm 0.3$\\
          & & & & & & 2.3 & 33 & & $1.4 \pm 0.3$ & $0.4 \pm 0.3$ \\
    \hdashline      
    \#67  & 0.8 & 44 & & $0.4 \pm 0.2$ & & 1.3 & 53 & & $0.6 \pm 0.1$ & \\
          & 2.3 & 63 & & $1.1 \pm 0.3$ & $0.7 \pm 0.3$ & 3.1 & 63 &  & $1.3 \pm 0.2$ & $0.8 \pm 0.2$ \\
          & 4.3 & 59 & $52 \pm 8$ & $2.1 \pm 0.4$ & $1.0 \pm 0.3$ & 4.6 & 55 & $57 \pm 4$ & $2.1 \pm 0.2$ & $0.7 \pm 0.2$ \\ 
          & 6.4 & 50 & & $3.1 \pm 0.6$ & $1.0 \pm 0.3$ & 6.0 & 56 & & $2.7 \pm 0.3$ & $0.6 \pm 0.2$\\ 
          & 7.8 & 45 & & $3.8 \pm 0.7$ & $0.7 \pm 0.3$ & & & & & \\
    \hline
    \end{tabular} \\
    \tablefoottext{a}{Distance and velocity offset of knots with respect to the core, uncertain by up to $0.2\arcsec$ (0.005~pc) and 3~$\kms$ respectively.} \\   
    \tablefoottext{b}{Difference between dynamical timescales $t_{\rm knot}$ of successive knots.} \\
    \tablefoottext{c}{This knot is part of the blue lobe of both core \#26 and core \#59. Because of this confusion, it is not used for $\Delta V_{\rm lobe}$ calculations.} 
\end{table*}


\section{Discussion}
\label{s:disc}

With 46 outflow lobes detected, corresponding to 36 bipolar and 10 monopolar outflows, W43-MM1 is one of the richest protocluster found to date, in terms of outflows per area. In particular, it is twice richer than the nearby protocluster called NGC~1333 \citep{Plunkett13}, over a similar spatial area, $\sim$1~pc. Our discovery is due to a combination of cluster characteristics -- with a close-packed cluster of cores  \citep{motte18b} --, suitable angular resolution ($2000$~AU) to disentangle outflow lobes, and the detailed study in velocity cubes (see Sect.~\ref{s:pres}) allowed by the 1.3~$\kms$ resolution.
This sample of 46 outflow lobes, ejected by 27 individual cores from a single protocluster at an early stage of evolution, uniquely provides us with the opportunity to make meaningful statistical studies of outflow properties. We hereafter investigate the influence of the core mass and cloud environment on the main characteristics of outflows (see Sect.~\ref{s:d-core+envi}) and discuss the history of ejection events in W43-MM1 (see Sect.~\ref{s:d-history}).

\subsection{Outflows characteristics in a dynamical environment}
\label{s:d-core+envi}

The protostars that drive the outflow lobes listed in Table~\ref{tab:outflow} have core masses spanning 2 orders of magnitude, $1.4~\Msol$ to $102~\Msol$ \citep{motte18b}. They could be the precursors of stars with masses ranging from $\sim$0.5~$\Msol$ to $\sim$80~$\Msol$, assuming a gas-to-star conversion factor of $30\%-80\%$ \citep{Alves07, Bontemps2010, louvet14}. Since the W43-MM1 protostellar population is young \citep{motte03, motte18b}, in agreement with the strong outflow collimation found in Sect.~\ref{s:flow}, we expect protostellar outflows to develop in a relatively pristine cloud, devoid of stellar feedback effects. The cloud regions crossed by outflows have different densities, with visual extinctions varying from 140~mag to 2600~mag (see Table~\ref{tab:outflow}).

\subsubsection{Outflow maximum length and velocity}
\label{s:d-max}
\begin{figure*}[t!]
    \center
     \subfloat{\includegraphics[width=0.52\hsize]{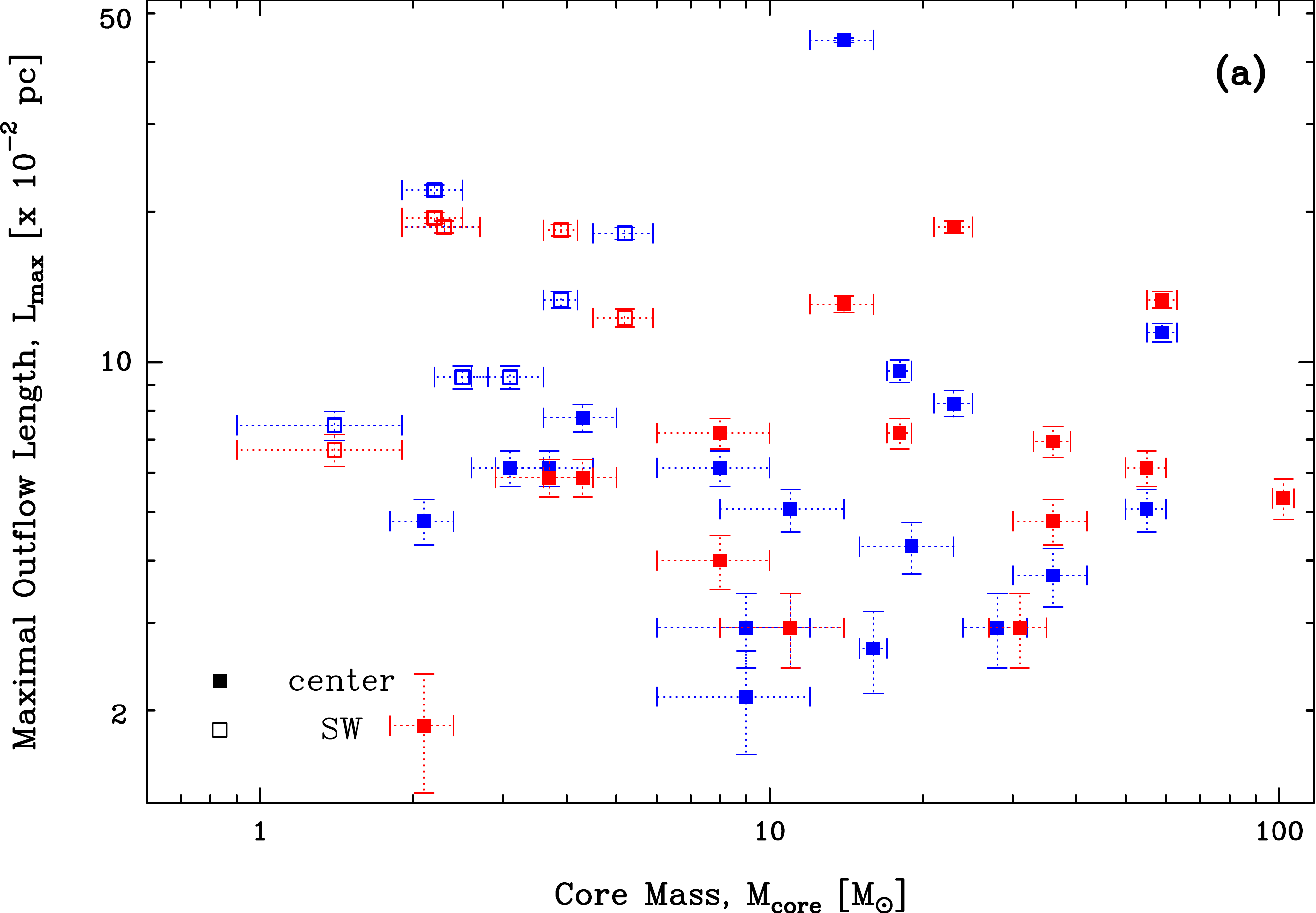}} 
     \subfloat{\includegraphics[width=0.47\hsize]{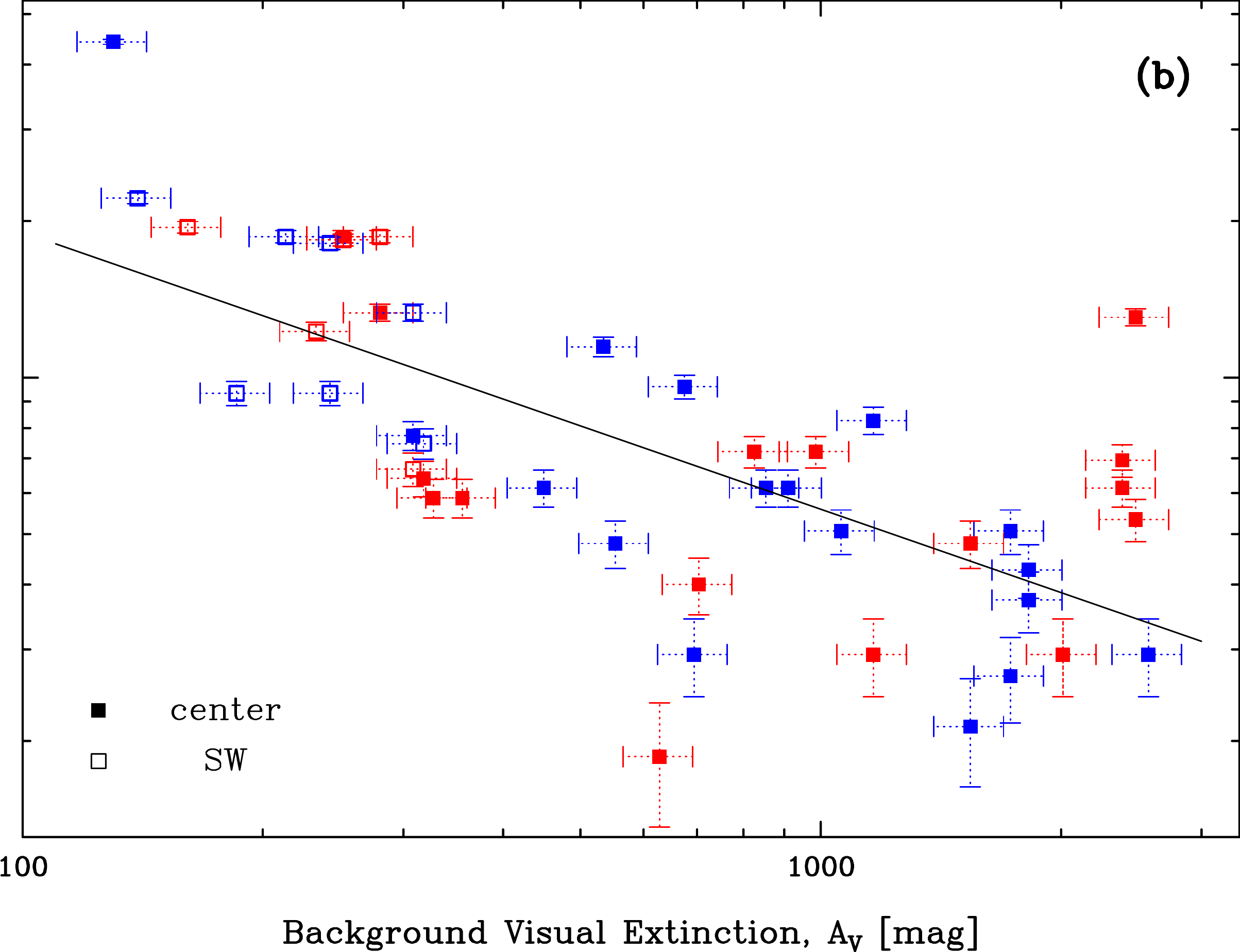}}\\
    \caption{Maximal outflow length observed projected on the plane of the sky, $L_{\rm max}$, vs. the mass of the launching core, $M_{\rm core}$, (in \textbf{a}) and the visual extinction of the cloud background crossed by the outflow, $A_{\rm V}$, (in \textbf{b}). For the longest outflows, measurements at their tip (in parenthesis in Table~\ref{tab:outflow}) are used instead of that averaged over their complete extent. Filled and empty squares pinpoint outflow lobes developing in the central and south-western parts of W43-MM1, respectively. Blue and red colors show the blue- and red-shifted lobes, respectively. 
    The anti-correlation found in \textbf{b} (best fit line $L_{\rm max}\propto A_{\rm V}^\alpha$ with $\alpha= -0.5 \pm 0.1$) suggests that molecular outflows are shorter in denser environments.
    }
    \label{f:outflow-char}
\end{figure*}

Figures~\ref{f:outflow-char}a-b compare the maximum length of molecular outflows with the mass of the launching cores and the extinction level, that is used as a proxy for the density in which the outflows propagate. 
While no relation is seen between the outflow length and the core mass in Fig.~\ref{f:outflow-char}a, a clear anti-correlation of the length of outflow lobes projected on the plane of the sky with the visual extinction appears in Fig.~\ref{f:outflow-char}b. A linear regression gives the relation $L_{\rm max}\propto A_{\rm V}^\alpha$ with $\alpha= -0.5 \pm 0.1$, with a Pearson correlation coefficient of -0.70 and a p-value of $7\times10^{-8}$. This anti-correlation is consistent with the asymmetry in size of the blue and red lobes driven by cores \#7 and \#8: 0.08~pc versus 0.18~pc and 0.44~pc versus 0.13~pc, respectively. Their longest outflow lobe indeed propagates away from the cluster and their shortest lobe toward the inner part of the cluster (see Figs.~\ref{f:co-ctr-high}b-c).

We also compared the maximal velocity, $\Delta V_{\rm max}$, of the molecular jets with both the mass of the driving core and the background cloud extinction (see Figs.~\ref{f:outflow-char-Vmax}a-b). The large scatters of the observed points do not suggest any correlation. However, we recall that the unknown inclination of the individual lobes introduces errors on the length and velocity measurements.
As shown for dynamical times in Sect.~\ref{s:d-angle}, this should introduce a scatter when plotting observed length and velocity characteristics for the complete sample of 46 outflow lobes. This scatter could hide real physical trends, like those expected between the mass of the launching core and the outflow characteristics, in particular the correlation observed with the outflow momentum flux \citep[e.g.,][]{Bontemps96,Duarte13CO, Maud15}. \\
 
Regardless the correlations or non-correlations discussed above, they are all obviously limited by our ability to detect protostellar outflows in the central part of W43-MM1. This ridge contains extremely-dense, $10^4-10^7$~cm$^{-3}$ \citep[see Sect.~\ref{s:env},][]{nguyen13,louvet14,louvet16}, material that should slow down the propagation of molecular outflows. While no global slowing down is observed in Fig.~\ref{f:outflow-char-Vmax}b, the inspection of the PV diagrams of the asymmetric outflows of cores \#8 and \#7 possibly give some insights (see Figs.~\ref{f:PV}a and \ref{f:PV}c). 
For core~\#8, the velocity of knots, when they remain observable, are not much impacted, with the velocity offsets, $\Delta V_{\rm knot}$, of R3 and R4 similar to that of R1 (see Fig.~\ref{f:PV}a) and to the mean velocity of the blue knots, $\overline{\Delta V_{\rm knot}}^{\rm blue\;lobe}$
(see Table~\ref{tab:tdyn}). In contrast, the series of nested shock structures, seen in the PV diagram of Fig.~\ref{f:PV}a as a continuous emission plateau over $\Delta V \simeq 20-40~\kms$ connecting knots, is less extended in the X axis for the red lobe (containing R1 to R3) than for its blue  counterpart (containing B0 to B3): $\sim$0.1~pc vs. $\sim$0.19~pc. It also gets more intense at low velocity, reaching $I_{\rm CO}=0.5$~Jy/beam vs. $I_{\rm CO}=0.3$~Jy/beam. These two last elements suggest quenching/containment of the outflowing gas by the denser gas of the globally infalling cloud structure inside which outflows develop. A similar trend is observed in Fig.~\ref{f:PV}c, where the complete red lobe of core~\#7 is twice less extended than its blue lobe.

Besides the 
increased pressure expected for an increased density, the complex kinematics of W43-MM1 should be considered. The W43-MM1 ridge is indeed a highly dynamical cloud with $5-10~\kms$ gradients \citep[][]{nguyen13, louvet16} and inflowing gas at several tens of $\kms$ \citep{Louvet15Phd}. 
These complex gas dynamics result in a blurring effect along the velocity axis that prevents the identification of molecular outflows with velocities lower than $\sim$10~$\kms$. Other processes, such as the global infall that was reported in W43-MM1 by \cite{motte05} and \cite{cortes10}, could also slow down or even disperse the gas contained within protostellar outflows. This global infall is probably neither spherical nor continuous in time but it would rather consist of sporadic accretion flows along filaments thus following privileged directions \citep[e.g,.][]{schneider10, peretto13, motte18a}.
The most massive protostars \#1 and \#4 are the best sites where this scenario could apply since they are located at the center of gravity of the infalling protocluster \citep{Herpin12, cortes10} and they only developed small, low-velocity monopolar (red) outflows (see Fig.~\ref{f:co-ctr-high}c). Interestingly, both their missing lobes and the largest gas inflow observed toward the clump hosting cores \#1 and \#4 are from the blue-shifted part of the spectrum (Pouteau et al. in prep.). 
Therefore one can expect this inflow of material to partly quench and disperse the outflowing gas, thus preventing the development
of well-organized blue lobes. The general lack of cavity shells developing around each molecular jet could as well be explained by the dispersal of these moderate-velocity outflow components by infalling gas. 
Numerical simulations of protostellar outflows interacting with a dense and dynamical environment are definitively needed to go beyond the qualitative hints given above. They could use the shape in PV diagrams of both the high-velocity part of the plateau and the low-velocity arch observed in, e.g., Fig.~\ref{f:PV}a, to distinguish the passive quenching of density from the active dispersion of a complex gas inflow.

\begin{figure*}
\subfloat{\includegraphics[width=0.5\hsize]{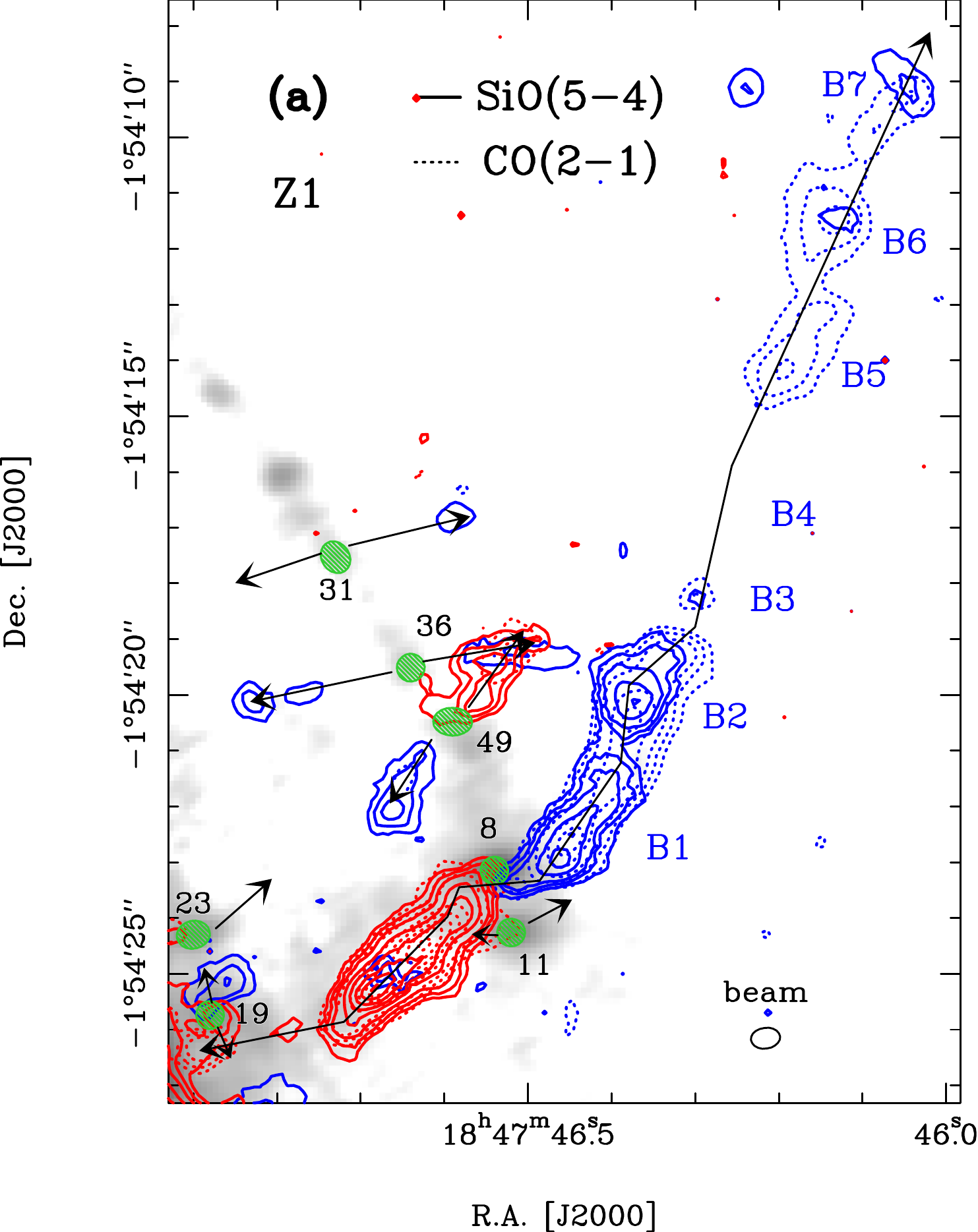}}
\hskip 2cm
\subfloat{\includegraphics[width=0.3\hsize]{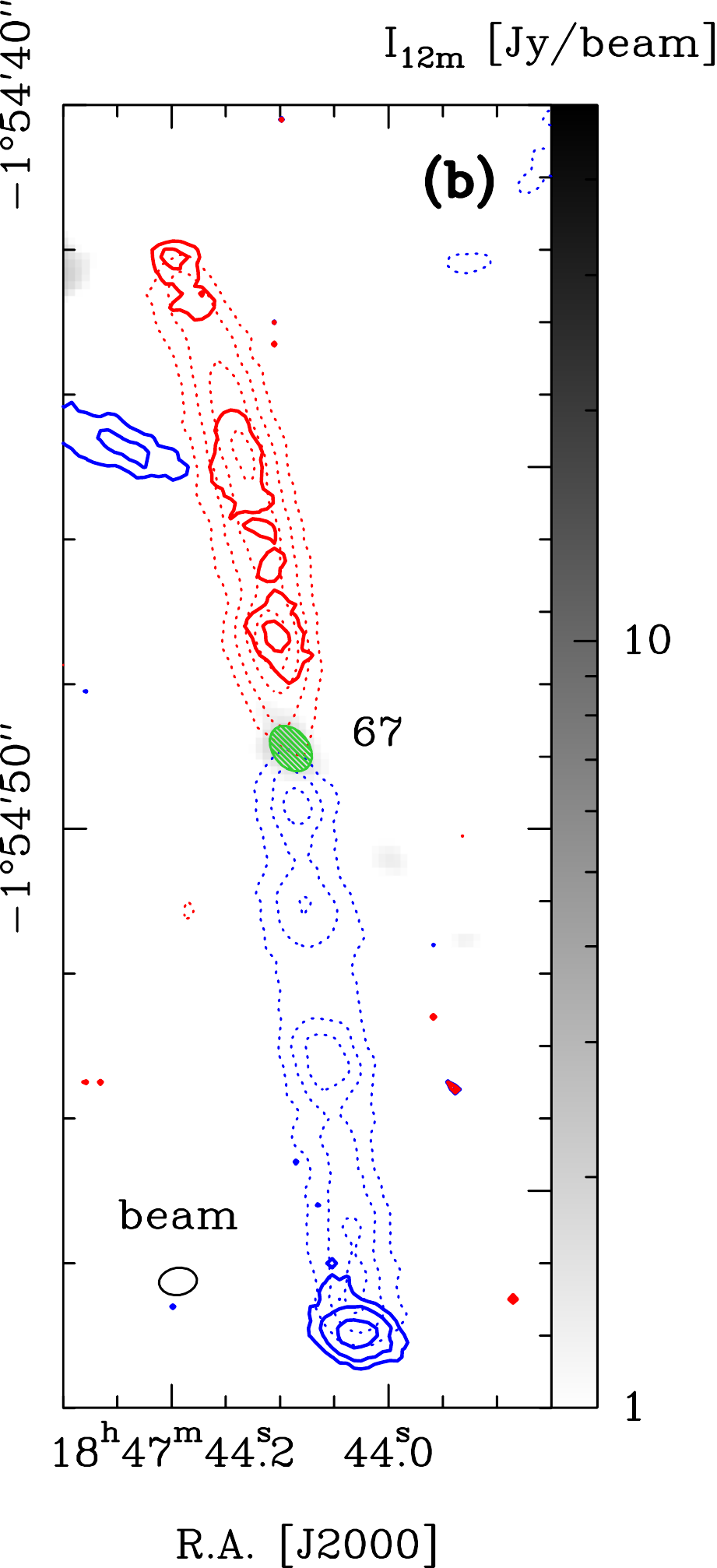}}

\caption{\label{fig:co-sio-main} Molecular outflows observed in SiO(5-4) compared with their CO(2-1) emission in the central region (in \textbf{a}, Z1 field of Fig.~\ref{f:co-ctr-high}a) and in the surrounding of core~\#67 (in \textbf{b}). 
The SiO line is integrated over $43-93~\kms$ (continuous blue contours) and $103-153~\kms$ (continuous red contours), with contours of 6, 11, 18 to 198 by steps of 15 (in \textbf{a}) and 5, 10, 20 (in \textbf{b}), in unit of $\sigma_{\rm SiO}=23\,\rm mJy\,beam^{-1}\,\kms$.
The CO line is integrated over $21-64~\kms$ (dotted blue contours), with contours 7, 15, 30 to 270 by steps of 40 (in \textbf{a}) and  5, 12, 19 (in \textbf{b}), in unit of $\sigma_{\rm CO,B}=32\,\rm mJy\,beam^{-1}\,\kms$; and $128-158~\kms$ (dotted red contours), with contours of 7, 15, 30 to 280 by steps of 50 (in \textbf{a}) and  5, 15, 30, 45 (in \textbf{b}), in unit of $\sigma_{\rm CO,R}=22\,\rm mJy\,beam^{-1}\,\kms$.
Contours are overlaid on the 1.3~mm continuum emission of the 12~m array (gray scale).
Green ellipses locate the W43-MM1 cores, arrows (sometimes broken) indicate the direction of their outflows. Knots of the molecular outflow of
core~\#8 are labelled B1 to B7.
}

\end{figure*}

\subsubsection{Outflow chemistry through CO versus SiO tracers}
\label{s:sio}

Another argument for the large impact of the cloud kinematics on W43-MM1 outflows arises from the comparison of SiO(5-4) and CO(2-1) tracers. SiO has indeed proven to give insights on the cloud chemistry and, as a result, on the cloud dynamics \citep[e.g.,][]{louvet16}.

In total 42 out of the 46 outflow lobes (i.e. 91\%) are detected both in CO(2-1) and SiO(5-4) (see Figs.~\ref{fig:co-sio-main}a-b and \ref{fig:co-sio-app}a-c). Such a high detection rate of the molecular outflows in SiO is frequent in high-mass star-forming regions \citep[see however][]{Widmann16}.
Low-angular resolution surveys of the SiO(2-1), (3-2) or (5-4) emission lines toward 50-400 high-mass molecular clumps indeed reported detection rates as high as $\sim$60\%--90\% \citep{lopez-sepulcre2011, Csengeri16, Li19}.
These rates are in marked contrast with the rare detection of the SiO(5-4) emission line toward low-mass star-forming regions. For instance,
less than 30\% of the $\sim$20 best-known outflows arising from low-mass protostars have been detected by \cite{chandler-richer97} and \cite{Gibb04}. 
This discrepancy is most likely due to the excitation conditions of the SiO(5-4) emission line, that has a high critical density of $(5-10)\times 10^6$~cm$^{-3}$. Such high densities are more easily reached in high-mass than in low-mass star-forming regions. 

Beyond the detection rate of the molecular outflows in the SiO(5-4) emission, the important new result here is that the morphologies of SiO outflows are very similar to those of the CO outflows (see Figs.~\ref{fig:co-sio-main}a and \ref{fig:co-sio-app}a-c). The high-angular resolution studies by \cite{Duarte14SiO} and \cite{zhang15} reported a similarity, but less striking, for 10--20 outflow lobes in Cygnus~X and IRDC~G28.34+0.06P1, respectively. This excellent correlation between CO and SiO probably stems from the peculiar chemistry of SiO in W43. 
We recall that the most efficient path for the SiO formation is a gas phase reaction between O$_2$ and Si \citep{lepicard01}. Getting Si in the gas phase generally requires high-velocity shocks that will erode the core of the dust grain \citep[$\geq$25~$\kms$, ][]{schilke97, Gusdorf08}.
In W43-MM1, \cite{louvet16} have however shown that a wide-spread SiO emission was associated with low-velocity shocks ($\sim$10~$\kms$). 
They explained this atypical association by the presence of a fraction of the total Si abundance ($1\%$ to 10$\%$) in the form of SiO located in the mantle of the dust grain. Such hypothesis can be explained if a first event of high-velocity shocks formed the SiO that got depleted onto the mantle of dust grain. This proposed peculiarity would permit that SiO emits over a large range of shock velocities within the outflows of W43-MM1, and not only where high-velocity ($>$25~$\kms$) shocks are present.

In this respect, the case of core \#67, the most isolated core of the W43-MM1 protocluster, is enlightening. It drives a bipolar outflow that is nicely traced in CO(2-1) whereas SiO(5-4) is only detected at the terminal bow shocks, plus a few locations along the red-shifted lobe (see Fig.~\ref{fig:co-sio-main}b). The core \#67 is located in a background cloud with visual extinction four times lower than the median $A_{\rm v}=600~$mag value. We speculate that along the outflow of core \#67 the volume density is not enough to reach the critical density of the SiO(5-4) emission line,
except at the terminal bow shocks. The blue lobe of core \#8 shows another good example of the impact of the cloud density on the SiO detection in outflows. While the first part of the jet develops in gas column densities typical of the central region, $A_{\rm v} \simeq 400~$mag, its second part is characterized by a lower background level, $A_{\rm v} \simeq 150~$mag. The first knots, labeled B0 to B2, propagating in high-density gas are strongly detected in SiO ($>$20$\sigma_{\rm SiO}$, see Fig.~\ref{fig:co-sio-main}a), while the knots further away, labeled B3 to B6, are not or barely detected ($\leq$6$\sigma_{\rm SiO}$). Finally, the furthest knot, labelled B7, is itself detected; it probably corresponds to the terminal bow shock of the outflow lobe, which presents a stronger density.

\subsubsection{Angle of ejection versus the direction of outflow lobes}
\label{s:angle}

A last argument for the influence of cloud kinematics on W43-MM1 molecular outflows comes from the study of outflow propagation directions. As shown in Figs.~\ref{f:co-ctr-high}b-d and \ref{f:co-ext}a-b, we observed changes of directions for the longest molecular jets. Most of them are consistent with `wiggling', periodic oscillation of the atomic and/or molecular jet position (see outflows ejected by cores \#8, \#3, \#9, and \#39 in Figs.~\ref{f:co-ctr-high}b-d and  Fig.~\ref{f:co-ext}a). This pattern could be a signature of jet precession, associated with the misalignment of the rotation and magnetic-field axes or orbital motion due to the presence of close multiple protostars \citep[e.g.,][]{Terquem99}.

When the separation of protostellar multiples is large, one could expect multiples to have formed from the fragmentation of the envelope rather than from that of the disk. Their outflow would then have a greater probability of being misaligned,  which would facilitate the detection of multiple outflows driven by multiple protostars hosted in a single core.
The only clear-cut case of ejections from multiple systems is observed for core \#22. It drives a first bipolar outflow along the NE-SW direction (\#22a in Table~\ref{tab:outflow}) and a single red lobe (\#22b) almost perpendicular to \#22a (see Fig.~\ref{f:co-ctr-high}c). 
This small detection number of multiple outflows (1 out of 27 cores) could be understood if secondary outflows tend to be weaker and/or lower velocity or if multiples would generally have small separations. The first interpretation would point to uneven mass binaries, which is amiss for a cluster dominated by high-mass protostars \cite[][]{Duchene13}. The second interpretation recalls results obtained for young Class~0s \citep{maury10} and agrees with the youth of the W43-MM1 protostars, inferred from their low luminosity \citep{motte03,motte18b} and strong outflow collimation (see Sect.~\ref{s:flow} and Figs.~\ref{f:co-ctr-high}-\ref{f:co-ext}).

In a few cases, long portions of the molecular jets are deflected, a pattern sometimes inconsistent with jet wiggling. In particular, the red lobes of cores \#7 and \#67 show, at their mid-length, a bend of $\sim$30$\degree$ and $\sim$10$\degree$, respectively (see Figs.~\ref{f:co-ctr-high}c and \ref{f:co-ext}b). 
For core \#67, the gradual bend of its red lobe is in marked contrast with the blue counter-jet that stays straight and narrow along its whole length. 
In the case of core \#7, one cannot directly compare the red bent lobe to its blue counterpart because the latter develops in an environment confused by other outflows. Since these two bends are obvious and occur locally, three interpretations are possible. 
These outflows could have encountered an obstacle, as for instance a dense core although none were yet detected at the location of these bends. 
These deflections could also trace density inhomogenities in the surrounding medium, assuming molecular outflow consists of entrained gas loosely associated with the jet -- as gas forming the cavity shell.
Otherwise, if outflows, and especially jet-like outflows, are constituted of both ejected and dragged molecular gas, outflow deflection would indicate that outflows propagate in a gas stream with varying velocities. This third interpretation seems to be the most likely since the W43-MM1 cloud consists of several layers of gas, which all host cores and move at several $\kms$ relative to each others and thus relative to some of the cores \citep{Louvet15Phd}.
Similar outflow bends have been observed for a few protostars, including IRAS~4A and IRAS~18162-2048 \citep{Choi05, Fernandez13}.

\begin{figure*}[htbp]
    \centering
    \subfloat{\includegraphics[width=0.505\hsize]{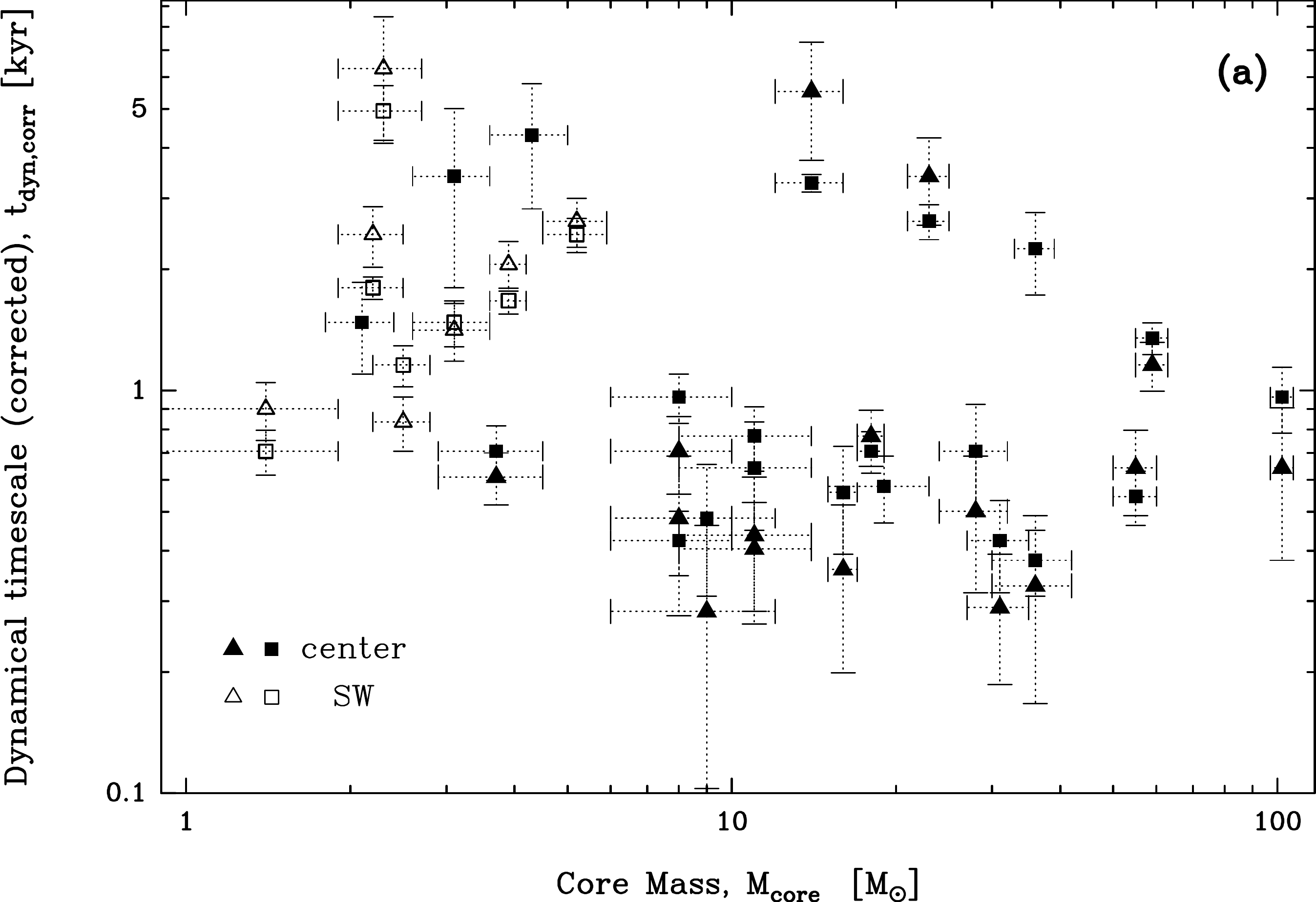}}
    \vspace{0.3cm}
    \subfloat{\includegraphics[width=0.46\hsize]{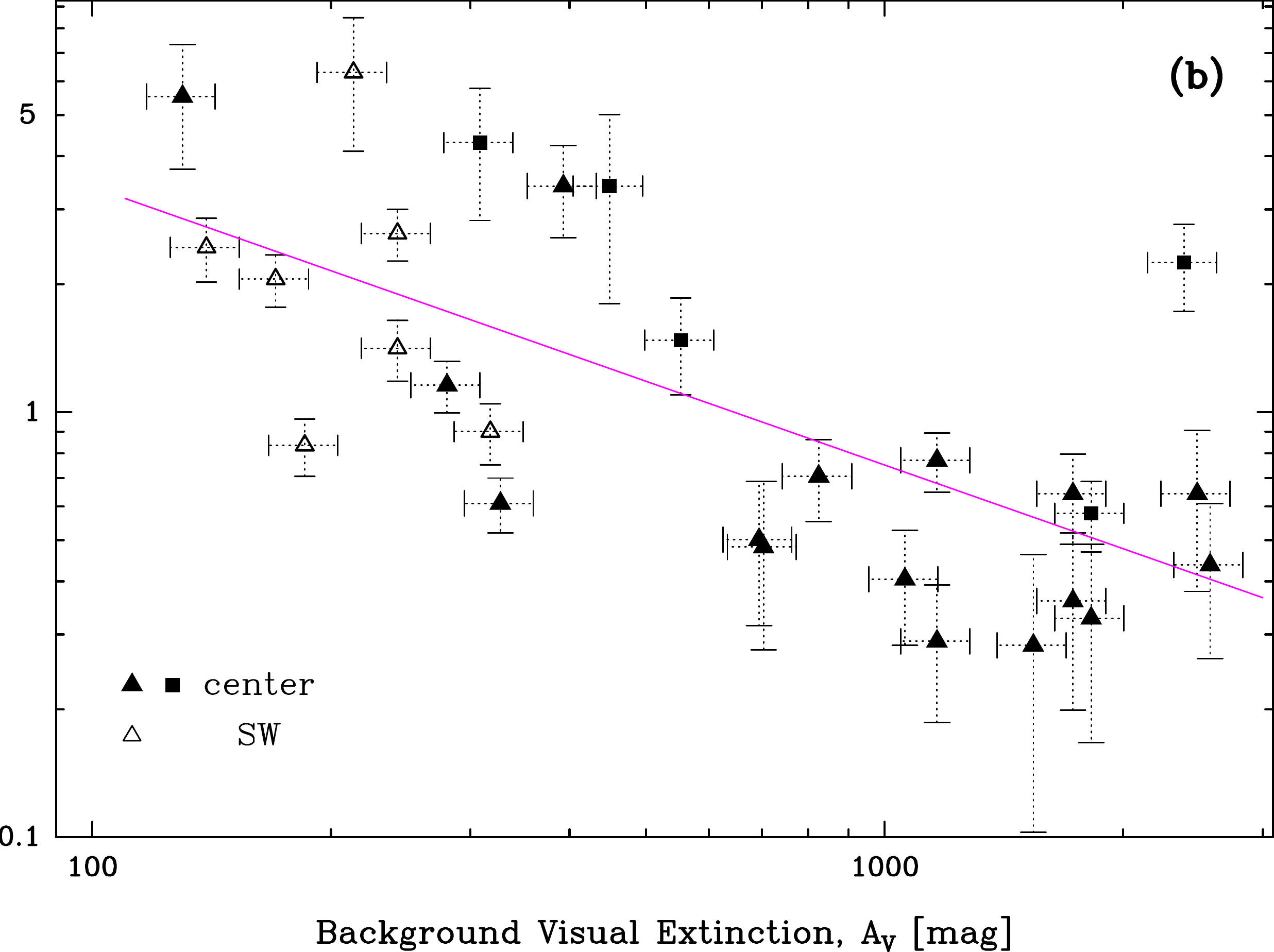}}
    \caption{\label{f:tdyn} Outflow dynamical timescale
    corrected for a mean projection angle of $57.3\degree$, $t_{\rm dyn,\,corr}$, vs. the mass of the launching core, $M_{\rm core}$, (in \textbf{a}) and the visual extinction of the cloud background crossed by the outflow, $A_{\rm V}$, (in \textbf{b}). 
    The dynamical timescale of a given core is evaluated in two ways: as the oldest ejection event (triangles) and as the $L_{\rm max}/{\Delta V}_{\rm max}$ ratio (squares).
    Both dynamical timescales are plotted in \textbf{a}, showing that they generally are consistent within errorbars. A single estimate is used in \textbf{b}, that corresponds to the oldest ejection event (triangles) when it can be measured and the $L_{\rm max}/{\Delta V}_{\rm max}$ ratio otherwise (squares).
     No significant correlation is observed in \textbf{a}. 
    The anti-correlation between $t_{\rm dyn}$ and $A_{\rm V}$ in \textbf{b}, $t_{\rm dyn} \propto A_{\rm V}^\alpha$ with $\alpha= -0.7 \pm 0.1$ (magenta line) and a Pearson correlation coefficient of -0.67, recalls that between $L_{\rm max}$ and $A_{\rm V}$ in Fig.~\ref{f:outflow-char}b.
     }
\end{figure*}

Another deflection could also explain the peculiar behavior of the blue lobe of core \#8. At high velocities ($\Delta V > 50~\kms$), its blue jet is not aligned with the core center and points toward core \#11, located $0.8\arcsec$ away. At lower velocities however, it connects with core \#8, following a direction similar to the last ejection (first knot) of the red outflow lobe. 
This strong global deflection, with an angle of $\sim$40$\degree$ and the velocity gradient ($\sim$3500~AU over $60~\kms$) observed perpendicular to the jet suggest that the blue outflow lobe of core \#8 strongly interacted with the surrounding gas, at the location of this bend, $\sim$0.08~pc east from the core. We cannot exclude that similar deflections happened in other molecular outflows (especially those of cores \#3 and \#9) since they generally are smaller in size and/or more confused by nearby outflows. Further investigations of the cloud structure and kinematics are necessary to properly conclude on the main reason for the outflow deflections observed here.

The three elements presented in Sects.~\ref{s:d-max}--\ref{s:angle} argue for a large impact of the dynamical environment of molecular outflows on their characteristics. While the maximum velocity of outflows may not strongly be influenced, their length, chemistry, and direction angle are.

\subsection{History of outflow ejections and protostellar accretion}
\label{s:d-history}

Molecular outflows have been widely used to trace the history of protostellar accretion. Here, we measure the dynamical timescale of the 28 protostellar outflows developing in W43-MM1 (Sect.~\ref{s:d-timescale}) and use PV diagrams (Sects.~\ref{s:d-PV}-\ref{s:d-angle}) to constrain the history of outflow ejection and protostellar accretion in this region (Sect.~\ref{s:d-variable}).

\subsubsection{Dynamical timescales}
\label{s:d-timescale}
Determining protostellar ages is far from trivial. Estimations are obtained from various indirect measurements such as the protostar location in envelope mass vs. luminosity diagrams, spectral energy distribution (SED) slopes, chemical ages, or the dynamical timescale of outflows. We here estimated the dynamical timescale of outflows using two different methods. The first one classically estimates it as the ratio between the maximum length and maximum velocity of outflows, $t_{\rm dyn} = L_{\rm max}/\Delta V_{\rm max}$. Since these maximum values, listed in Table~\ref{tab:outflow}, are independently measured in the CO cube, length and velocity do not necessarily correspond to the same gas element and, even less, to the furthest knot corresponding to the oldest ejection event. We therefore adopted as second estimation of $t_{\rm dyn}$ the maximum value of the dynamical timescales of knots. They were estimated as the ratio of the distance from the knot to the core and the velocity (offset) of the outflow lobe in which it formed, $t_{\rm knot}=r_{\rm knot}/\Delta V_{\rm lobe}$ (see Table~\ref{tab:tdyn} and Sect.~\ref{s:knot}). The uncertainty of this dynamical timescale depends on the dispersion of $V_{\rm knot}$ and on the angular resolution used to identify knots in each PV diagram, here half of the beam ($\sim$0.2$\arcsec$ or $\sim$0.005~pc). We consider that this second method provides the most reliable estimates of the ejection timescale. The two timescale estimates given above are consistent to within the uncertainties for 19 out of the 23 cores or multiples (see Fig.~\ref{f:tdyn}a). The dynamical timescales of outflows in W43-MM1, before any correction of projection effects, range from 400~yr to $10^4$~yr, with a median value of $\sim$1500~yr.

To estimate the real values of dynamical timescales, one needs to correct from projection effects, process which is hazardous when the inclination angle remains unknown (see however Sect.~\ref{s:d-angle}). 
Measured velocities and lengths in fact are velocities projected along the line of sight, $\Delta V_{\rm max}= \Delta V_{\rm real}\,\times\cos{(i)}$, where $i$ is the angle to the line of sight, and lengths projected on the plane of the sky, $L_{\rm max} = L_{\rm real} \times \sin{(i)}$. The dynamical timescales corrected for the projection angle would then be $t_{\rm dyn, corr}= t_{\rm dyn} / \tan{(i)}$. 
Assuming a random distribution of inclination angles, the mean value is given by $\overline{i}=\int_{0}^{\pi/2}i\sin{(i)}\,\mathrm{d}i=1\,\text{rad}= 57.3\degree$. This leads to dynamical timescales that, once statistically corrected for homogeneous projection effects, range from $t_{\rm dyn,corr}=280$~yr to 6300~yr.

Figures~\ref{f:tdyn}a--b display the dynamical timescale of protostellar outflows as a function of the mass of the launching core and as a function of the visual extinction of the background cloud. While no correlation is observed with the core's mass (see Fig.~\ref{f:tdyn}a), an anti-correlation is found between the dynamical timescale of outflows and the visual extinction of the background they cross (see Fig.~\ref{f:tdyn}b). As shown in Figs.~\ref{f:tdyn}a-b, the dynamical timescale of outflows in the central part of W43-MM1 (filled symbols) seems on average -- but not systematically -- a couple of times shorter than the dynamical timescale of protostellar outflows in the south-westernmost part of W43-MM1 (empty symbols).
At face value and, assuming that this dynamical time is representative of the complete ejection history and thus of the true protostellar lifetime, this would imply that cores in the central region are twice younger than those in the south-west.
This would suggest that cloud and subsequent star formation would have started in the south-western part of W43-MM1 first. This fits in the framework of recent cloud formation models \citep[e.g.,][]{LeeHe16, vazquez19}, where several cloud and star formation events sum up to form the final population of stars.
That said, the anti-correlation observed in Fig.~\ref{f:tdyn}b recalls the one found in Fig.~\ref{f:outflow-char}b between $L_{\rm max}$ and $A_{V}$. 
We recall that outflows in the central part of W43-MM1 generally arise from massive protostars and develop in dense environments \citep[][and Table~\ref{tab:outflow}]{motte18b}. Therefore, the trend observed here could be completely environmental (see Sect.~\ref{s:d-core+envi}). In any case we stress that outflow dynamical timescales should only be considered as 
rough estimates of the protostellar lifetimes. A first reason is that outflows very likely get missed or dispersed in the cloud, if they do not simply fall below our detection limit.
A second important reason is that our angular resolution does not allow to measure the velocity of the atomic and/or molecular jet directly ejected by the protostar, but only the velocity of the entrained molecular layer that could be a factor of a few to ten times slower.

\subsubsection{PV diagrams as open books on ejection variability}
\label{s:d-PV}

To go beyond the global characterization of outflow lobes and study the evolution of the ejection process, PV diagrams are among the most powerful tools. The PV diagrams of Figs.~\ref{f:PV}a-c, \ref{f:PV-49}a-f, and \ref{f:PV-15}a-b display fingers that generally have velocities that increase, approximately linearly, with their positional offsets, and thus with their distance, from the core. This is nicely illustrated by the handful of linear features found in the outflow lobes of core \#8 (connected to knots R2, B1, B2 in Fig.~\ref{f:PV}a), of core \#7 (connected to knots R2, R3, B1 in Fig.~\ref{f:PV}c), of core \#49 (see Fig.~\ref{f:PV-49}b), and of core \#44 (see Fig.~\ref{f:PV-49}e). 
Interestingly, the slope of these fingers tends to be steeper or even vertical close to the core and shallower afterwards (see B1, B2, and B3 in Fig.~\ref{f:PV}a and R1, R2, and R3 in Fig.~\ref{f:PV}c). In a ballistic scenario, this behavior could indicate that the molecular ejecta plus the gas layer at its immediate interface, both lying within our spatial resolution unit, are initially all located at close distances from the core with various velocities. Then after propagation, the ejecta, roughly corresponding to the knot, reaches further distances than the entrained gas layer that now draws a finger in the PV diagram. Similar changes of slopes have been predicted by models of variable ejections \cite[e.g.,][]{Rohde19}. The distance-velocity relations of these fingers are frequently referred to as `Hubble laws'. We here interpret them as
the sum of 1/ molecular gas ejected by the protostar and ballistically propagating plus 2/ ambient gas dragged and accelerated by the jet \citep[see also][]{Arce06, Rohde19}. In our case, fingers with Hubble-law gas distributions represent traces of enhanced ejection events that add to a more continuous average ejection level.
At the tip of these fingers observed in PV diagrams (see particularly Figs.~\ref{f:PV}a-c and Figs.~\ref{f:PV-49}b-c, \ref{f:PV-49}f), vertical segments appear that correspond to knots. These knots, also called internal shocks, mark enhanced velocity and density spots of the forward shock  
where the ejecta and/or its accompanying accelerated gas catches up with slower material from previous ejections \citep[e.g.,][]{Jhan16}. 
In some PV diagrams, a gas layer/structure presents a vertical or strongly declining slope that connects the knot to the $V_{\rm LSR}$, recalling the features seen in simulations \citep[e.g.,][]{Lee01}.
This gas layer could correspond to the lateral forward shocks.

In our general case, fingers with increasing velocities, knots, and layers with decreasing velocities create triangular-shape structures in the PV diagrams (see Figs.~\ref{f:PV}a, \ref{f:PV}c, Figs.~\ref{f:PV-49}a-b and \ref{f:PV-49}e-f). 
Several of these triangular shape structures add up forming nested shocks that appear like
a continuous emission plateau in the PV diagrams (see, e.g., Figs.~\ref{f:PV}a, \ref{f:PV}c).
At large offsets, most of the low- to intermediate-velocity ($<$30~\kms) emission disappears from PV diagrams, leaving the high-velocity vertical segments relatively isolated (case of B5 to B7 in Fig.~\ref{f:PV}a).
For a couple of knots, their highest velocity appears upstream, drawing a `slow-head, fast-tail' profile (see B2 and B3 of core \#67 in Fig.~\ref{f:PV-49}f). \cite{Santiago09} and more recently \cite{Jhan16} attributed these `inverse-Hubble laws' to projection effects of internal shocks, in agreement with the pulsed jet simulations of \cite{Stone93}.

Most of the PV diagrams observed for outflow lobes of W43-MM1 protostars display several knots (see, e.g., Figs.~\ref{f:PV}a-c and \ref{f:PV-49}f). This probably is the signature of episodic ejection, with variation of the velocity and/or of the mass of the gas ejected, which could be related to episodic accretion. When observed with lower spatial and/or spectral resolutions or lower sensitivity, only the high-velocity part of the PV diagram stands out. These velocity peaks, corresponding to internal shocks, are then the only structures that can clearly be identified.
When several velocity peaks appear, PV diagrams present a jagged profile sometimes called `Hubble wedge' \citep{Arce01a, Stoji06, Arce13, Plunkett15}, that should resemble to a smoothed version of, e.g., Fig.~\ref{f:PV}a.

\subsubsection{Time span between consecutive ejection events and projection effects}
\label{s:d-angle}

\begin{figure}[hb]
    \vskip -0.4cm
    \centerline{\includegraphics[width=10.5cm]{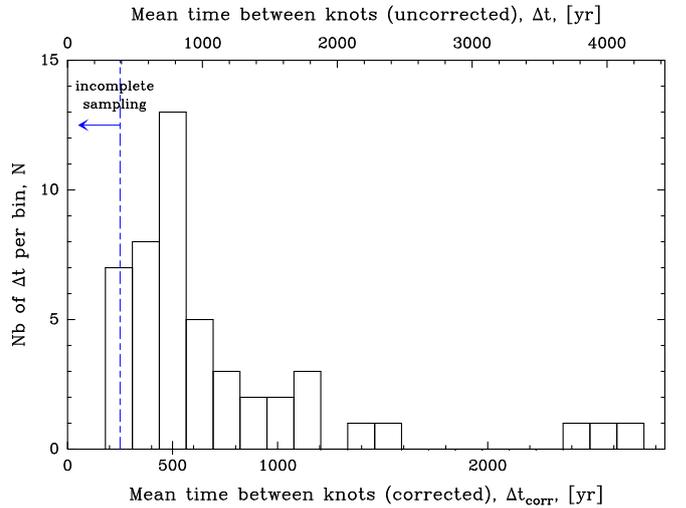}}
    \vskip -0.7cm
    \caption{\label{fig:histo-dt} Distribution of the timescale differences between successive ejection events. The top and bottom axes show the timescales before and after the correction by an homogeneous inclination angle of $57.3\degree$, $\Delta \rm t_{\rm corr}$ and $\Delta \rm t$ respectively. The $75\%$ completeness limit, $\sim$250~yr after deprojection, is represented with a dashed line. Outflows display a variability with a typical timescale, corrected for projection effects, of $\overline{\Delta \rm t_{\rm corr}} \sim 500$~yr. }
\end{figure}

Despite the fact that most of the lobes in the central region seem limited by the environment (see Sect.~\ref{s:d-core+envi}), 22 lobes from 14 cores or multiples display two knots or more. For those, we measured the timescale difference between knots, $\Delta t$ (see Table~\ref{tab:tdyn}), suggesting it could be an estimate, within the limit of projection effects, of the time between two successive ejection events. The distribution of the 48 $\Delta t$ values between these 72 knots varies from 290~yr to 4300~yr, with a median of $\sim$780~yr (before correction for projection effect). 
Since, for each lobe, we used the same velocity to calculate all $t_{\rm knot}$ values, the $\Delta t$ measurements mostly relate to the distance between consecutive knots, $\Delta r$, through the equation $\Delta t=\Delta r / \Delta V_{\rm lobe}$. Uncertainties of about 200~yr have been calculated using those set for $\Delta V_{\rm lobe}$ (see Table~\ref{tab:tdyn}) and a constant uncertainty of 0.4$\arcsec$ for $\Delta r$. 
Figure~\ref{fig:histo-dt} displays the distribution of timescale differences between knots.
Our $\Delta t$ sample should be complete as long as the referred two knots can be separated in the PV diagrams. This will be the case when their relative distance exceeds the spatial resolution of the CO cube, $\rm FWHM=0.013$~pc, and their velocity offset the first quartile from which $\Delta t$ values are measured in our sample, $\Delta V_{\rm lobe}\sim 33~\kms$ (see Sect.~\ref{s:flow} and Table~\ref{tab:tdyn}). When corrected for an average projection angle of $57.3\degree$, the median time span between ejection events and the $75\%$ completeness limit of our sample are $\sim$500~yr and $\sim$250~yr. The histogram of Fig.~\ref{fig:histo-dt} clearly peaks at $\sim$500~yr, well above the completeness level, suggesting that a typical timescale between two ejecta exists in W43-MM1. 

\begin{figure*}[]
    \includegraphics[width=0.51\hsize]{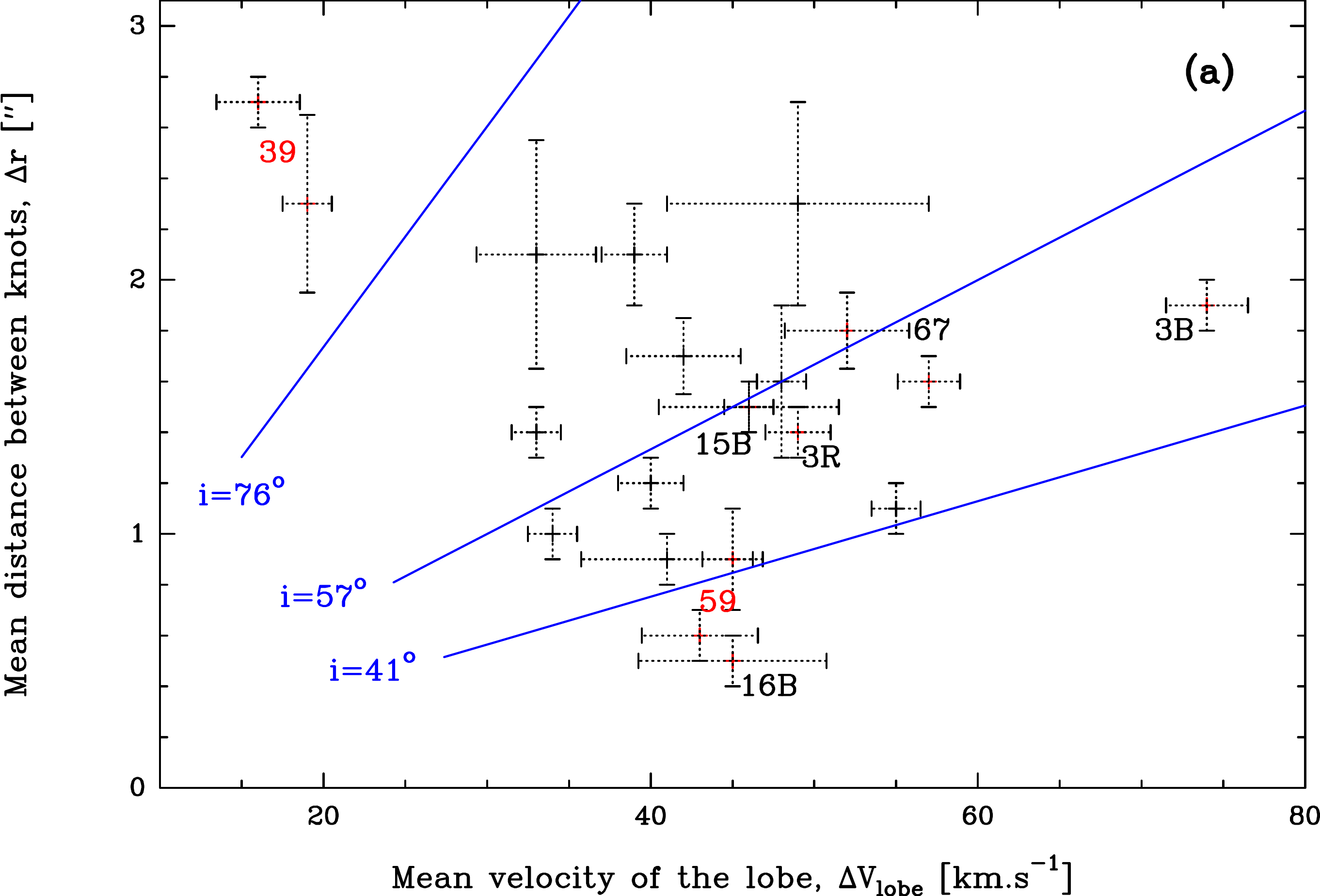} \hskip 0.5cm
    \includegraphics[width=0.46\hsize]{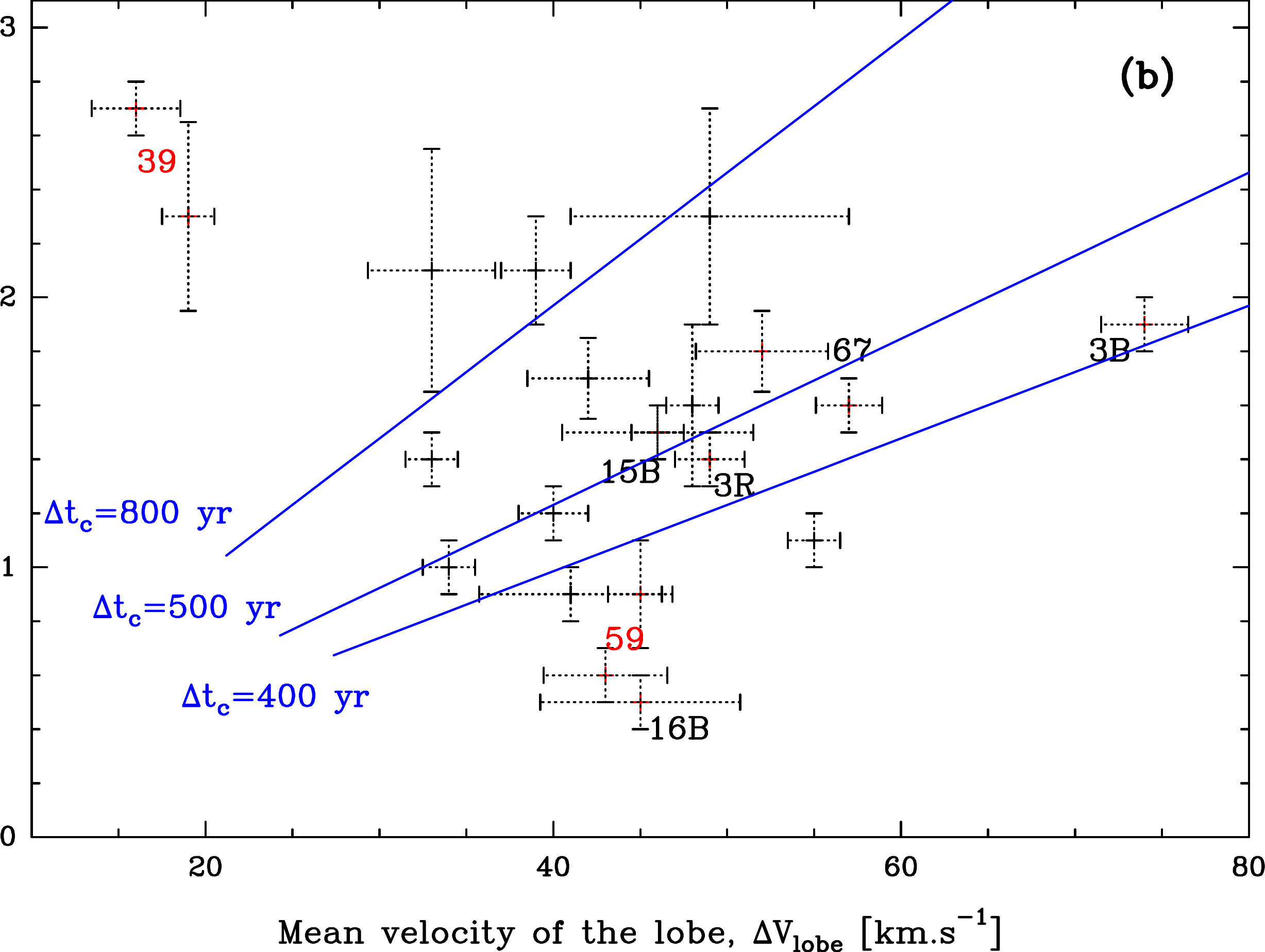}
    \caption{\label{fig:dr-V} Mean projected distance between knots, $\overline{\Delta r}^{\rm lobe}$, vs. the mean velocity of knots, $\Delta V_{\rm lobe}$, for 21 outflow lobes with at least 2 knots. Error bars represent the $\pm 1\sigma$ dispersion. 
    Measurements and core numbers are indicated in red for outflow lobes discussed below.
    In \textbf{a}, blue lines locate the measurements for the mean inclination angle of a random distribution of outflows ($57.3\degree$) and for the inclination angles of the first and last quartiles ($41\degree$, $76\degree$), assuming a constant variability timescale of $\Delta t_{\rm corr} = 500$~yr.
    In \textbf{b}, blue lines locate the measurements for the median timescale difference between knots, $\Delta t_{\rm corr}$ = 500 yr, and the first and last quartiles ($\Delta t_{\rm corr}$ = 400 and 800 yr), assuming a constant inclination angle of $57.3\degree$.}
\end{figure*}

We evaluated the projection effects of $\Delta t$ measurements by considering two extreme configurations presented in Figs.~\ref{fig:dr-V}a-b. First, we assumed that ejections in W43-MM1 have the same periodicity
throughout the whole cluster and that the observed dispersion is entirely due to projection effects. We recall that measured distances and velocities are projected on the plane of the sky and on the line of sight, respectively: $\Delta r=\Delta r_{\rm corr}\times\sin(i)$ and $\Delta V_{\rm lobe}=\Delta V_{\rm lobe,corr}\times\cos(i)$.
In order to look for strong projection effects, we plotted the mean distance between knots of a given lobe, $\overline{\Delta r}^{\rm lobe}$, against the mean velocity of knots relative to the $V_{\rm LSR}$, $\Delta V_{\rm lobe}$ (see Fig.~\ref{fig:dr-V}a).
We assumed that the most probable inclination angle, 57.3$\degree$, is associated to lobes in the median mode of our sample, which means those that present a median distance between knots of $\overline{\Delta r}^{\rm lobe} = 1.5\arcsec$ (or 8300~AU) and a median velocity of $\Delta V_{\rm lobe}=45~\kms$. A configuration with a single periodicity implies inclination angles varying from 29$\degree$ to 81$\degree$ to cover the full dispersion in our sample. In Fig.~\ref{fig:dr-V}a, we plotted the $\overline{\Delta r}^{\rm lobe}$ vs. $\Delta V_{\rm lobe}$ linear relations expected for the most probable inclination angle of $i=57.3\degree$ and for the first and last quartile angles, $i=41\degree$ and $i=75\degree$. Most lobes of our sample lie within this 50\% percentile zone, with striking outliers being lobes driven by core \#39 (see Fig.~\ref{fig:dr-V}a). The low-velocity, long outflow lobes (see Figs.~\ref{f:co-ext}a and \ref{f:PV-49}d) driven by core \#39 could be lying close to the plane of the sky, $i \gtrsim 80\degree$.
If however the low-velocity outflow is intrinsic to the core, the latter could be more evolved than the typical cores of W43-MM1. 
On the other hand, the high-velocity, short outflow lobes driven by core \#59, and to a lesser extent the blue lobe of core \#16 (see Figs.~\ref{f:co-ctr-high}c and \ref{f:co-ext}a) argue for them to lie closer to the line of sight than the average, $i \lesssim 30\degree$. 
Given that these cores lie in a confused environment, their peculiarity needs to be confirmed to reinforce this interpretation. 

In the second configuration, we once again assumed that each lobe is seen with the same inclination angle, $57.3\degree$, and thus that the observed dispersion of the timescale differences between knots, $\Delta t$, is caused by an intrinsic dispersion of time variability in outflows. In that case, the timescale difference between two knots, corrected for projection effects, directly is $\Delta t_{\rm corr}=\Delta t/\tan(57.3\degree)$ and thus has a median value of $\sim$500~yr (see Fig.~\ref{fig:histo-dt}). 
The distribution of timescale differences clearly peaks at $\sim$500~yr, but ten $\Delta t_{\rm corr}$ values are above $\sim$1000~yr. Four out of the five outliers above $\sim$1200~yr correspond to knots of the \#39 outflow,which would thus correspond to the most particular outflow of our sample. In this configuration, timescale variations $\Delta t_{\rm corr}=400-800$ yr can account for the dispersion of half of the sample (see Fig.~\ref{fig:dr-V}b). 

The other methods we used to infer the inclination angle of observed outflows use the morphology of the ejection in integrated images and/or PV diagrams. In our sample, this is most likely the case of cores~\#15, \#3 and \#67.
The bipolar outflow of core \#15 indeed displays a unique morphology in our sample, with an ejection both blue- and red-shifted, both in its eastern and western lobes (see Fig.~\ref{f:co-ext}a). The western jet and the western and eastern cavities contain both blue- and red-shifted gas, with the blue-shifted emission more pronounced east of the core and the red-shifted emission west of it. The outflow cavity, though not centered on the jet direction has a maximum opening angle from its axis of $\theta_{\rm m} = 16\degree$.  This configuration suggests that the outflow of core \#15 is observed in the third geometrical configuration of \cite{cabrit86}, a bipolar lobe seen close to the plane of the sky: $i \geq 90-\theta_{\rm m} \simeq 74\degree$. 
As for core \#3, its PV diagram presents curved fingers (see Fig.~\ref{f:PV}b) that resemble the arc-like structures observed by \cite{Lee15}. This study suggests that carved gas layers could come from sideways ejections from an internal bow shock observed with large inclination angles, which means close to the plane of the sky, $i \sim 90\degree$. 
With similar arguments and following the interpretation of \cite{Santiago09} for the IRAS~04166 protostar, core \#67, which presents internal shocks with `slow-head, fast-tail' profiles (see Sect.~\ref{s:d-PV}), would be seen with an inclination angle of about $45\degree$.

Since the three indirect methods (statistical, morphology, and PV diagrams arguments) do not converge toward the same objects that should lie on the plane of the sky or on the line of sight, we chose not to individually correct outflow characteristics but to keep the most probable angle, $57.3\degree$, for homogeneously deprojecting timescales.

\subsubsection{Episodic ejection constraining variable accretion}
\label{s:d-variable}

Our study revealed the existence of a typical timescale for the molecular outflow episodicity in W43-MM1, $\overline{\Delta \rm t_{\rm corr}} \sim 500\pm^{300}_{100}$~yr, with a dispersion that can partly be attributed to projection effects (see Figs.\ref{fig:histo-dt}-\ref{fig:dr-V} and Sect.~\ref{s:d-angle}). If this timescale is not induced by propagation instabilities along the outflow, but is rather associated with ejection events, it could characterize episodic accretion.

Constraining the origin, within the close environment of the outflowing protostar, of the 
instabilities creating this episocity requires the comparison with dedicated models.
Unfortunately, very few models have so far attempted to link the ejection and accretion processes using their temporal variability. Among the most recent, the model of \cite{Vorobyov18} describes the accretion instabilities in the disk of low-mass protostars, like the one whose outflow variability has been studied by \cite{Plunkett15}. 
This model revealed luminosity bursts with a bi-modal distribution  constituted by one major and one minor mode with $10^3-10^4$~yr and $10-100$~yr timescales, respectively. The first mode would be associated with the accretion of disk cloudlets that remain unique entities, the second one would correspond to cloudlets that undergo tidal torques and fragment into a series of smaller cloudlets. 
The \cite{Vorobyov18} model is likely not suitable for describing the intermittent accretion expected for intermediate- to high-mass protostars in W43-MM1. 
Indeed, this model does not consider 
irregular feeding of the disk from the envelope and possibly its environment, while intermittent inflows are expected from a dynamical cloud like W43-MM1 \citep[e.g.,][]{louvet16}.
Moreover, as mentioned by \cite{Vorobyov18}, accretion from a more massive disk, like the ones surrounding more massive protostars, would lead to more intense luminosity bursts and possibly longer delays between bursts. We therefore refrain from comparing the episodicity timescale we obtained in W43-MM1 with the timescales found for the minor and major modes of accretion variability modeled by \cite{Vorobyov18}.

Discrete emission peaks have, for long, been observed along the axis of outflows driven by low-mass protostars \citep[see the reviews of, e.g.,][]{Arce06, Bally16}. Very few published studies have however quantified the time span  between knots, even taking into account studies done 20 years ago \citep{Bachiller90, Cernicharo96, Arce01a} and more recent studies \citep{Arce13, Plunkett15, Chen16}.
These handful of measurements also have low statistical value since they were obtained for a single source and they are, except for \cite{Plunkett15}, limited by the small number of observed ejecta, typically 3--5 (22 in \citealt{Plunkett15}).
In marked contrast, present study measured a statistically-robust characteristic timescale using 72 knots, detected in a cluster of outflows driven by low- to high-mass protostars. 
The timescale measured here, $\sim$500$\pm^{300}_{100}$~yr, is broadly consistent with the $300-1000$~yr range of values obtained by the handful of studies mentioned above. If we limit our comparison to the statistically-robust study of \cite{Plunkett15}, our estimated timescale is slightly larger: $\sim$500$\pm^{300}_{100}$~yr vs. $\sim$310$\pm 150$~yr\footnote{This timescale would become smaller if corrected for the inclination angle proposed by \cite{Vorobyov18}.}.
A word of caution needs to be added on the relevance of direct comparisons with the previously-published timescales. 
While some measurements are computed using velocities typical of atomic jets \citep[200--300~$\kms$,][]{Arce01a, Arce13}, our study and others assume the measured, probably lower, velocity of knots \citep[10--50$~\kms$,][]{Cernicharo96, Plunkett15}. As for the inclination angles, \cite{Cernicharo96,Arce13,Chen16} had a precise measurement derived from proper motions but others \citep{Bachiller90, Arce01a, Plunkett15} only roughly estimated it based on strong, and often unverifiable, assumptions.
Our study itself uses a sample of 22 lobes to circumvent the bias associated with unknown inclination angles.
Interestingly, outflows observed at increasing angular resolutions generally present several spatial -- and thus temporal -- characteristic scales. L1448 has for instance been studied with 3000~AU and 250~AU resolutions by \cite{Bachiller90} and \cite{Hirano10}, revealing timescales of $\sim$400~yr\footnote{The published timescale of 1000~yr is here corrected for the inclination angle of $i\simeq 21\degree$ with respect to the plane of the sky, which was later determined by proper motions \citep{Girart01}.} and $\sim$20~yr, that could correspond to several modes of accretion episodicity.
Our timescale estimate also recalls the 100--200~yr timescales found for jet wiggling \citep[e.g.,][]{Choi01, Lee15} but it is not obvious that similar instabilities would be at the origin of both accretion episodicity and jet wiggling.
In addition to the absolute value of the measured variability timescales, one can compare the number of molecular knots found in various studies. We have discovered more knots along the molecular outflows of W43-MM1 protostars than those reported for low-mass protostars, even though the latter have been studied for already three decades.
If ejection episodicity reflects variability in the accretion process, protostellar accretion could be more variable or episodicity is simply easier to detect in W43-MM1 than in nearby clouds.

To conclude, we argue that studying protostellar clusters is an efficient method to statistically constrain the episodicity of the ejection, and possibly the accretion, processes. It allows to circumvent the difficulty to measure inclination angles in very embedded protostars where proper motions are difficult to study.
Interpreting the outflow time variability, measured to have a $\overline{\Delta \rm t_{\rm corr}} \sim 500\pm^{300}_{100}$~yr typical timescale, will require dedicated models of the accretion episodicity and its link to outflow variability. We emphasize that complete physical ingredients should be implemented into these models to fully represent the complex environment of massive disk  surrounding the intermediate- to high-mass protostars of W43-MM1.

\section{Conclusion}
\label{s:conc}

We used ALMA to investigate, at high-spatial, 2600~AU, resolution and with CO(2-1) and SiO(5-4) molecular line tracers, the molecular outflows developing in W43-MM1. Our main conclusions can be summarized as follows:

\begin{enumerate}
    \item We discovered a rich cluster of 46 outflow lobes driven by 27 protostars (see Fig.~\ref{f:general} and Figs.~\ref{f:co-ctr-high}-\ref{f:co-ext}). They are typically $\sim$0.1~pc long, strongly collimated jet-like structures with high-velocity components reaching typical velocities of $\Delta V \sim 50~\kms$ (see Table~\ref{tab:outflow}).
    \item The protostellar cores driving these molecular outflows span a large mass range, from 1 to 100\,\Msol, yet no significant variation of the maximal length and maximal velocity of outflows with the cores' mass is observed. 18 out of the 28 molecular outflows are bipolar and only one core is associated with two outflows.
    \item We showed a clear anti-correlation between the outflow maximal length and the visual extinction of the crossed background cloud (see Fig.~\ref{f:outflow-char}b). We propose that the high-density, dynamical cloud of the central region (up to $A_{\rm V}=3000$~mag and $\sim$10~$\kms$) limits the propagation of protostellar outflows.
    \item The SiO(5-4) emission of protostellar outflows presents a remarkable morphological and kinematical coincidence with the CO(2-1) emission (see Figs.~\ref{fig:co-sio-main} and \ref{fig:co-sio-app}).
    This excellent correspondence, which remains unusual, is explained by the fact that the SiO(5-4) critical density is easily reached in the background crossed by the W43-MM1 outflows (see Sect.~\ref{s:sio}).
    \item The jet-like component of molecular outflows consists of a continuous emission plus a series of knots (see Figs.~\ref{f:co-ctr-high}-\ref{f:co-ext}). Observed in the CO(2-1) and SiO(5-4) integrated maps as local maxima, knots are typically separated by 1.5$\arcsec$, or $\sim$8000~AU. These knots appear in PV diagrams at constant distances from the core (see Figs.~\ref{f:PV}-\ref{f:PV-15}) and are interpreted as internal bow shocks (see Sect.~\ref{s:d-PV}). They could be created by the temporal variations of the velocity and/or the mass flux of the protostellar ejection. The large number of knots detected, 86 along 38 outflow lobes, suggests that W43-MM1 protostars undergo accretion bursts.
    %
    \item The detailed study made for numerous PV diagrams revealed complex kinematic structures, including fingers with velocities increasing approximately linearly with the distance from the core (see, e.g., Figs.~\ref{f:PV}a and \ref{f:PV}c). These structures, sometimes referred as Hubble laws, could be due to CO gas of the envelope dragged and accelerated 
    by the fast atomic and/or molecular jet. 
    \item We estimated the timescales associated to each knot and discovered, thanks to the large statistics achieved in W43-MM1, a typical timescale between ejecta. Once corrected with a uniform inclination angle of 57.3$\degree$, the ejection episodicity timescale in W43-MM1 is $\sim$500$\pm^{300}_{100}$~yr. This timescale is consistent with the few variability timescales observed for low-mass protostars. The physical interpretation of this typical timescale, in terms of instabilities in the close environment of the protostars, remains unclear. 
   
\end{enumerate}

\begin{acknowledgements}
We are grateful to Paolo Cortes, Catherine Dougados, and Jonathan Ferreira for useful discussions.
This paper makes use of the following ALMA data: \#2013.1.01365.S. ALMA is a partnership of ESO (representing its member states), NSF (USA) and NINS (Japan), together with NRC (Canada), MOST and ASIAA (Taiwan), and KASI (Republic of Korea), in cooperation with the Republic of Chile. The Joint ALMA Observatory is operated by ESO, AUI/NRAO and NAOJ.
This work was supported by the Programme National de Physique Stellaire and Physique et Chimie du Milieu Interstellaire (PNPS and PCMI) of CNRS/INSU (with INC/INP/IN2P3) co-funded by CEA and CNES.
This project has received funding from the European Union's Horizon 2020 research and innovation programme StarFormMapper under grant agreement No 687528.
\end{acknowledgements}

\bibliographystyle{aa}
\bibliography{biblio-article}	

\begin{thebibliography}{80}
\expandafter\ifx\csname natexlab\endcsname\relax\def\natexlab#1{#1}\fi

\bibitem[{{Alves} {et~al.}(2019){Alves}, {Caselli}, {Girart}, {Segura-Cox},
  {Franco}, {Schmiedeke}, \& {Zhao}}]{Alves19}
{Alves}, F.~O., {Caselli}, P., {Girart}, J.~M., {et~al.} 2019, Science, 366, 90

\bibitem[{{Alves} {et~al.}(2007){Alves}, {Lombardi}, \& {Lada}}]{Alves07}
{Alves}, J., {Lombardi}, M., \& {Lada}, C.~J. 2007, \aap, 462, L17

\bibitem[{{Arce} \& {Goodman}(2001)}]{Arce01a}
{Arce}, H.~G. \& {Goodman}, A.~A. 2001, \apj, 551, L171

\bibitem[{{Arce} {et~al.}(2013){Arce}, {Mardones}, {Corder}, {Garay},
  {Noriega-Crespo}, \& {Raga}}]{Arce13}
{Arce}, H.~G., {Mardones}, D., {Corder}, S.~A., {et~al.} 2013, \apj, 774, 39

\bibitem[{{Arce} {et~al.}(2007){Arce}, {Shepherd}, {Gueth}, {Lee}, {Bachiller},
  {Rosen}, \& {Beuther}}]{Arce06}
{Arce}, H.~G., {Shepherd}, D., {Gueth}, F., {et~al.} 2007, in Protostars and
  Planets V, ed. B.~{Reipurth}, D.~{Jewitt}, \& K.~{Keil}, 245

\bibitem[{{Audard} {et~al.}(2014){Audard}, {{\'A}brah{\'a}m}, {Dunham},
  {Green}, {Grosso}, {Hamaguchi}, {Kastner}, {K{\'o}sp{\'a}l}, {Lodato},
  {Romanova}, {Skinner}, {Vorobyov}, \& {Zhu}}]{Audart14}
{Audard}, M., {{\'A}brah{\'a}m}, P., {Dunham}, M.~M., {et~al.} 2014, in
  Protostars and Planets VI, 387

\bibitem[{{Bachiller} {et~al.}(1990){Bachiller}, {Cernicharo},
  {Martin-Pintado}, {Tafalla}, \& {Lazareff}}]{Bachiller90}
{Bachiller}, R., {Cernicharo}, J., {Martin-Pintado}, J., {Tafalla}, M., \&
  {Lazareff}, B. 1990, \aap, 231, 174

\bibitem[{{Bachiller} {et~al.}(1994){Bachiller}, {Tafalla}, \&
  {Cernicharo}}]{Bachiller94}
{Bachiller}, R., {Tafalla}, M., \& {Cernicharo}, J. 1994, \apjl, 425, L93

\bibitem[{{Bally}(2016)}]{Bally16}
{Bally}, J. 2016, Annual Review of Astronomy and Astrophysics, 54, 491

\bibitem[{{Beuther} {et~al.}(2007){Beuther}, {Churchwell}, {McKee}, \&
  {Tan}}]{Beuther07}
{Beuther}, H., {Churchwell}, E.~B., {McKee}, C.~F., \& {Tan}, J.~C. 2007, in
  Protostars and Planets V, ed. B.~{Reipurth}, D.~{Jewitt}, \& K.~{Keil}, 165

\bibitem[{Beuther {et~al.}(2002)Beuther, Schilke, Sridharan, Menten, Walmsley,
  \& Wyrowski}]{Beuther02}
Beuther, H., Schilke, P., Sridharan, T.~K., {et~al.} 2002, Astronomy \&
  Astrophysics, 383, 892

\bibitem[{{Bontemps} {et~al.}(1996){Bontemps}, {Andre}, {Terebey}, \&
  {Cabrit}}]{Bontemps96}
{Bontemps}, S., {Andre}, P., {Terebey}, S., \& {Cabrit}, S. 1996, \aap, 311,
  858

\bibitem[{{Bontemps} {et~al.}(2010){Bontemps}, {Motte}, {Csengeri}, \&
  {Schneider}}]{Bontemps2010}
{Bontemps}, S., {Motte}, F., {Csengeri}, T., \& {Schneider}, N. 2010, \aap, 524

\bibitem[{Cabrit \& Bertout(1986)}]{cabrit86}
Cabrit, S. \& Bertout, C. 1986, The Astrophysical Journal, 307, 313

\bibitem[{{Carlhoff} {et~al.}(2013){Carlhoff}, {Nguyen Luong}, {Schilke},
  {Motte}, {Schneider}, {Beuther}, {Bontemps}, {Heitsch}, {Hill}, {Kramer},
  {Ossenkopf}, {Schuller}, {Simon}, \& {Wyrowski}}]{Carlhoff13}
{Carlhoff}, P., {Nguyen Luong}, Q., {Schilke}, P., {et~al.} 2013, \aap, 560,
  A24

\bibitem[{{Cernicharo} \& {Reipurth}(1996)}]{Cernicharo96}
{Cernicharo}, J. \& {Reipurth}, B. 1996, \apjl, 460, L57

\bibitem[{{Chandler} \& {Richer}(1997)}]{chandler-richer97}
{Chandler}, C.~J. \& {Richer}, J.~S. 1997, in IAU Symposium, Vol. 182,
  Herbig-Haro Flows and the Birth of Stars, ed. B.~{Reipurth} \& C.~{Bertout},
  76

\bibitem[{Chen {et~al.}(2016)Chen, Arce, Zhang, Launhardt, \& Henning}]{Chen16}
Chen, X., Arce, H.~G., Zhang, Q., Launhardt, R., \& Henning, T. 2016, \apj,
  824, 72

\bibitem[{{Cheng} {et~al.}(2019){Cheng}, {Qiu}, {Zhang}, {Wyrowski}, {Menten},
  \& {G{\"u}sten}}]{Cheng19}
{Cheng}, Y., {Qiu}, K., {Zhang}, Q., {et~al.} 2019, \apj, 877, 112

\bibitem[{{Choi}(2001)}]{Choi01}
{Choi}, M. 2001, \apj, 553, 219

\bibitem[{{Choi}(2005)}]{Choi05}
{Choi}, M. 2005, \apj, 630, 976

\bibitem[{{Choi} {et~al.}(2017){Choi}, {Kang}, {Lee}, {Tatematsu}, {Kang},
  {Sayers}, {Evans}, {Cho}, {Kwon}, \& {Park}}]{Choi17}
{Choi}, M., {Kang}, M., {Lee}, J.-E., {et~al.} 2017, \apjs, 232, 24

\bibitem[{{Cody} \& {Hillenbrand}(2018)}]{Cody18}
{Cody}, A.~M. \& {Hillenbrand}, L.~A. 2018, \aj, 156, 71

\bibitem[{{Cortes} {et~al.}(2010){Cortes}, {Parra}, {Cortes}, \&
  {Hardy}}]{cortes10}
{Cortes}, P.~C., {Parra}, R., {Cortes}, J.~R., \& {Hardy}, E. 2010, \aap, 519,
  A35

\bibitem[{{Csengeri} {et~al.}(2016){Csengeri}, {Leurini}, {Wyrowski},
  {Urquhart}, {Menten}, {Walmsley}, {Bontemps}, {Wienen}, {Beuther}, {Motte},
  {Nguyen-Luong}, {Schilke}, {Schuller}, {Zavagno}, \& {Sanna}}]{Csengeri16}
{Csengeri}, T., {Leurini}, S., {Wyrowski}, F., {et~al.} 2016, \aap, 586, A149

\bibitem[{Cunningham {et~al.}(2016)Cunningham, Lumsden, Cyganowski, Maud, \&
  Purcell}]{Cunningham16}
Cunningham, N., Lumsden, S.~L., Cyganowski, C.~J., Maud, L.~T., \& Purcell, C.
  2016, Monthly Notices of the Royal Astronomical Society, 458

\bibitem[{{Duarte-Cabral} {et~al.}(2014){Duarte-Cabral}, {Bontemps}, {Motte},
  {Gusdorf}, {Csengeri}, {Schneider}, \& {Louvet}}]{Duarte14SiO}
{Duarte-Cabral}, A., {Bontemps}, S., {Motte}, F., {et~al.} 2014, \aap, 570, A1

\bibitem[{Duarte-Cabral {et~al.}(2013)Duarte-Cabral, Bontemps, Motte,
  Hennemann, Schneider, \& Andr{\'e}}]{Duarte13CO}
Duarte-Cabral, A., Bontemps, S., Motte, F., {et~al.} 2013, \aap, 558, A125

\bibitem[{{Duch{\^e}ne} \& {Kraus}(2013)}]{Duchene13}
{Duch{\^e}ne}, G. \& {Kraus}, A. 2013, \araa, 51, 269

\bibitem[{{Fern{\'a}ndez-L{\'o}pez} {et~al.}(2013){Fern{\'a}ndez-L{\'o}pez},
  {Girart}, {Curiel}, {Zapata}, {Fonfr{\'\i}a}, \& {Qiu}}]{Fernandez13}
{Fern{\'a}ndez-L{\'o}pez}, M., {Girart}, J.~M., {Curiel}, S., {et~al.} 2013,
  \apj, 778, 72

\bibitem[{{Frank} {et~al.}(2014){Frank}, {Ray}, {Cabrit}, {Hartigan}, {Arce},
  {Bacciotti}, {Bally}, {Benisty}, {Eisl{\"o}ffel}, {G{\"u}del}, {Lebedev},
  {Nisini}, \& {Raga}}]{Frank14}
{Frank}, A., {Ray}, T.~P., {Cabrit}, S., {et~al.} 2014, in Protostars and
  Planets VI, ed. H.~{Beuther}, R.~S. {Klessen}, C.~P. {Dullemond}, \&
  T.~{Henning}, 451

\bibitem[{{Gibb} {et~al.}(2004){Gibb}, {Richer}, {Chandler}, \&
  {Davis}}]{Gibb04}
{Gibb}, A.~G., {Richer}, J.~S., {Chandler}, C.~J., \& {Davis}, C.~J. 2004,
  \apj, 603, 198

\bibitem[{{Girart} \& {Acord}(2001)}]{Girart01}
{Girart}, J.~M. \& {Acord}, J. M.~P. 2001, \apjl, 552, L63

\bibitem[{{Gueth} \& {Guilloteau}(1999)}]{Gueth99}
{Gueth}, F. \& {Guilloteau}, S. 1999, \aap, 343, 571

\bibitem[{{Gueth} {et~al.}(1998){Gueth}, {Guilloteau}, \&
  {Bachiller}}]{Gueth98}
{Gueth}, F., {Guilloteau}, S., \& {Bachiller}, R. 1998, \aap, 333, 287

\bibitem[{{Gusdorf} {et~al.}(2008){Gusdorf}, {Cabrit}, {Flower}, \& {Pineau Des
  For{\^e}ts}}]{Gusdorf08}
{Gusdorf}, A., {Cabrit}, S., {Flower}, D.~R., \& {Pineau Des For{\^e}ts}, G.
  2008, \aap, 482, 809

\bibitem[{{Herpin} {et~al.}(2012){Herpin}, {Chavarr{\'\i}a}, {van der Tak},
  {Wyrowski}, {van Dishoeck}, {Jacq}, {Braine}, {Baudry}, {Bontemps}, \&
  {Kristensen}}]{Herpin12}
{Herpin}, F., {Chavarr{\'\i}a}, L., {van der Tak}, F., {et~al.} 2012, \aap,
  542, A76

\bibitem[{{Hirano} {et~al.}(2010){Hirano}, {Ho}, {Liu}, {Shang}, {Lee}, \&
  {Bourke}}]{Hirano10}
{Hirano}, N., {Ho}, P. P.~T., {Liu}, S.-Y., {et~al.} 2010, \apj, 717, 58

\bibitem[{{Jhan} \& {Lee}(2016)}]{Jhan16}
{Jhan}, K.-S. \& {Lee}, C.-F. 2016, \apj, 816, 32

\bibitem[{{Konigl} \& {Pudritz}(2000)}]{Konigl00}
{Konigl}, A. \& {Pudritz}, R.~E. 2000, in Protostars and Planets IV, ed.
  V.~{Mannings}, A.~P. {Boss}, \& S.~S. {Russell}, 759

\bibitem[{{Le Picard} {et~al.}(2001){Le Picard}, {Canosa}, {Pineau des
  For{\^e}ts}, {Rebrion-Rowe}, \& {Rowe}}]{lepicard01}
{Le Picard}, S.~D., {Canosa}, A., {Pineau des For{\^e}ts}, G., {Rebrion-Rowe},
  C., \& {Rowe}, B.~R. 2001, \aap, 372, 1064

\bibitem[{{Lee} {et~al.}(2015){Lee}, {Hirano}, {Zhang}, {Shang}, {Ho}, \&
  {Mizuno}}]{Lee15}
{Lee}, C.-F., {Hirano}, N., {Zhang}, Q., {et~al.} 2015, \apj, 805, 186

\bibitem[{{Lee} {et~al.}(2006){Lee}, {Ho}, {Beuther}, {Bourke}, {Zhang},
  {Hirano}, \& {Shang}}]{Lee06}
{Lee}, C.-F., {Ho}, P. T.~P., {Beuther}, H., {et~al.} 2006, \apj, 639, 292

\bibitem[{{Lee} {et~al.}(2001){Lee}, {Stone}, {Ostriker}, \& {Mundy}}]{Lee01}
{Lee}, C.-F., {Stone}, J.~M., {Ostriker}, E.~C., \& {Mundy}, L.~G. 2001, \apj,
  557, 429

\bibitem[{{Lee} \& {Hennebelle}(2016)}]{LeeHe16}
{Lee}, Y.-N. \& {Hennebelle}, P. 2016, \aap, 591, A30

\bibitem[{{Lefloch} {et~al.}(2015){Lefloch}, {Gusdorf}, {Codella},
  {Eisl{\"o}ffel}, {Neri}, {G{\'o}mez-Ruiz}, {G{\"u}sten}, {Leurini},
  {Risacher}, \& {Benedettini}}]{Lefloch15}
{Lefloch}, B., {Gusdorf}, A., {Codella}, C., {et~al.} 2015, \aap, 581, A4

\bibitem[{{Li} {et~al.}(2019){Li}, {Wang}, {Fang}, {Zhang}, {Li}, {Zhang},
  {Li}, {Zhu}, \& {Zeng}}]{Li19}
{Li}, S., {Wang}, J., {Fang}, M., {et~al.} 2019, \apj, 878, 29

\bibitem[{{Lodato} \& {Clarke}(2004)}]{Lodato04}
{Lodato}, G. \& {Clarke}, C.~J. 2004, \mnras, 353, 841

\bibitem[{{L{\'o}pez-Sepulcre} {et~al.}(2011){L{\'o}pez-Sepulcre}, {Walmsley},
  {Cesaroni}, {Codella}, {Schuller}, {Bronfman}, {Carey}, {Menten}, {Molinari},
  \& {Noriega-Crespo}}]{lopez-sepulcre2011}
{L{\'o}pez-Sepulcre}, A., {Walmsley}, C.~M., {Cesaroni}, R., {et~al.} 2011,
  \aap, 526, L2

\bibitem[{{Louvet}(2015)}]{Louvet15Phd}
{Louvet}, F. 2015, PhD thesis, Université Paris Sud

\bibitem[{{Louvet} {et~al.}(2016){Louvet}, {Motte}, {Gusdorf}, {Nguy{\^e}n
  Luong}, {Lesaffre}, {Duarte-Cabral}, {Maury}, {Schneider}, {Hill}, {Schilke},
  \& {Gueth}}]{louvet16}
{Louvet}, F., {Motte}, F., {Gusdorf}, A., {et~al.} 2016, \aap, 595, A122

\bibitem[{{Louvet} {et~al.}(2014){Louvet}, {Motte}, {Hennebelle}, {Maury},
  {Bonnell}, {Bontemps}, {Gusdorf}, {Hill}, {Gueth}, {Peretto},
  {Duarte-Cabral}, {Stephan}, {Schilke}, {Csengeri}, {Nguyen Luong}, \&
  {Lis}}]{louvet14}
{Louvet}, F., {Motte}, F., {Hennebelle}, P., {et~al.} 2014, \aap, 570, A15

\bibitem[{{Maud} {et~al.}(2015){Maud}, {Moore}, {Lumsden}, {Mottram},
  {Urquhart}, \& {Hoare}}]{Maud15}
{Maud}, L.~T., {Moore}, T.~J.~T., {Lumsden}, S.~L., {et~al.} 2015, \mnras, 453,
  645

\bibitem[{{Maury} {et~al.}(2010){Maury}, {Andr{\'e}}, {Hennebelle}, {Motte},
  {Stamatellos}, {Bate}, {Belloche}, {Duch{\^e}ne}, \& {Whitworth}}]{maury10}
{Maury}, A.~J., {Andr{\'e}}, P., {Hennebelle}, P., {et~al.} 2010, \aap, 512,
  A40

\bibitem[{{Motte} {et~al.}(2018{\natexlab{a}}){Motte}, {Bontemps}, \&
  {Louvet}}]{motte18a}
{Motte}, F., {Bontemps}, S., \& {Louvet}, F. 2018{\natexlab{a}}, \araa, 56, 41

\bibitem[{{Motte} {et~al.}(2005){Motte}, {Bontemps}, {Schilke}, {Lis},
  {Schneider}, \& {Menten}}]{motte05}
{Motte}, F., {Bontemps}, S., {Schilke}, P., {et~al.} 2005, in IAU Symposium,
  Vol. 227, Massive Star Birth: A Crossroads of Astrophysics, ed.
  R.~{Cesaroni}, M.~{Felli}, E.~{Churchwell}, \& M.~{Walmsley}, 151--156

\bibitem[{{Motte} {et~al.}(2018{\natexlab{b}}){Motte}, {Nony}, {Louvet},
  {Marsh}, {Bontemps}, {Whitworth}, {Men'shchikov}, {Nguyen Luong}, {Csengeri},
  {Maury}, {Gusdorf}, {Chapillon}, {K{\"o}nyves}, {Schilke}, {Duarte-Cabral},
  {Didelon}, \& {Gaudel}}]{motte18b}
{Motte}, F., {Nony}, T., {Louvet}, F., {et~al.} 2018{\natexlab{b}}, Nature
  Astronomy, 2, 478

\bibitem[{{Motte} {et~al.}(2003){Motte}, {Schilke}, \& {Lis}}]{motte03}
{Motte}, F., {Schilke}, P., \& {Lis}, D.~C. 2003, \apj, 582, 277

\bibitem[{{Nguyen Luong} {et~al.}(2013){Nguyen Luong}, {Motte}, {Carlhoff},
  {Louvet}, {Lesaffre}, {Schilke}, {Hill}, {Hennemann}, {Gusdorf}, {Didelon},
  {Schneider}, {Bontemps}, {Duarte-Cabral}, {Menten}, {Martin}, {Wyrowski},
  {Bendo}, {Roussel}, {Bernard}, {Bronfman}, {Henning}, {Kramer}, \&
  {Heitsch}}]{nguyen13}
{Nguyen Luong}, Q., {Motte}, F., {Carlhoff}, P., {et~al.} 2013, \apj, 775, 88

\bibitem[{{Nguyen Luong} {et~al.}(2011){Nguyen Luong}, {Motte}, {Schuller},
  {Schneider}, {Bontemps}, {Schilke}, {Menten}, {Heitsch}, {Wyrowski},
  {Carlhoff}, {Bronfman}, \& {Henning}}]{nguyen11b}
{Nguyen Luong}, Q., {Motte}, F., {Schuller}, F., {et~al.} 2011, \aap, 529, A41

\bibitem[{{Nony} {et~al.}(2018){Nony}, {Louvet}, {Motte}, {Molet}, {Marsh},
  {Chapillon}, {Gusdorf}, {Brouillet}, {Bontemps}, {Csengeri}, {Despois},
  {Nguyen Luong}, {Duarte-Cabral}, \& {Maury}}]{Nony18}
{Nony}, T., {Louvet}, F., {Motte}, F., {et~al.} 2018, \aap, 618, L5

\bibitem[{{Parks} {et~al.}(2014){Parks}, {Plavchan}, {White}, \&
  {Gee}}]{Parks14}
{Parks}, J.~R., {Plavchan}, P., {White}, R.~J., \& {Gee}, A.~H. 2014, The
  Astrophysical Journal Supplement Series, 211, 3

\bibitem[{{Peretto} {et~al.}(2013){Peretto}, {Fuller}, {Duarte-Cabral},
  {Avison}, {Hennebelle}, {Pineda}, {Andr{\'e}}, {Bontemps}, {Motte},
  {Schneider}, \& {Molinari}}]{peretto13}
{Peretto}, N., {Fuller}, G.~A., {Duarte-Cabral}, A., {et~al.} 2013, \aap, 555,
  A112

\bibitem[{{Plunkett} {et~al.}(2013){Plunkett}, {Arce}, {Corder}, {Mardones},
  {Sargent}, \& {Schnee}}]{Plunkett13}
{Plunkett}, A.~L., {Arce}, H.~G., {Corder}, S.~A., {et~al.} 2013, \apj, 774, 22

\bibitem[{{Plunkett} {et~al.}(2015){Plunkett}, {Arce}, {Mardones}, {van
  Dokkum}, {Dunham}, {Fern{\'a}ndez-L{\'o}pez}, {Gallardo}, \&
  {Corder}}]{Plunkett15}
{Plunkett}, A.~L., {Arce}, H.~G., {Mardones}, D., {et~al.} 2015, \nat, 527, 70

\bibitem[{{Rohde} {et~al.}(2019){Rohde}, {Walch}, {Seifried}, {Whitworth},
  {Clarke}, \& {Hubber}}]{Rohde19}
{Rohde}, P.~F., {Walch}, S., {Seifried}, D., {et~al.} 2019, \mnras, 483, 2563

\bibitem[{Santiago-Garcia {et~al.}(2009)Santiago-Garcia, Tafalla, Johnstone, \&
  Bachiller}]{Santiago09}
Santiago-Garcia, J., Tafalla, M., Johnstone, D., \& Bachiller, R. 2009,
  Astronomy \& Astrophysics, 495, 169

\bibitem[{{Schilke} {et~al.}(1997){Schilke}, {Walmsley}, {Pineau des Forets},
  \& {Flower}}]{schilke97}
{Schilke}, P., {Walmsley}, C.~M., {Pineau des Forets}, G., \& {Flower}, D.~R.
  1997, \aap, 321, 293

\bibitem[{{Schneider} {et~al.}(2010){Schneider}, {Csengeri}, {Bontemps},
  {Motte}, {Simon}, {Hennebelle}, {Federrath}, \& {Klessen}}]{schneider10}
{Schneider}, N., {Csengeri}, T., {Bontemps}, S., {et~al.} 2010, \aap, 520, A49+

\bibitem[{{Shu} {et~al.}(2000){Shu}, {Najita}, {Shang}, \& {Li}}]{Shu00}
{Shu}, F.~H., {Najita}, J.~R., {Shang}, H., \& {Li}, Z.~Y. 2000, in Protostars
  and Planets IV, ed. V.~{Mannings}, A.~P. {Boss}, \& S.~S. {Russell}, 789--814

\bibitem[{{Stojimirovi{\'c}} {et~al.}(2006){Stojimirovi{\'c}}, {Narayanan},
  {Snell}, \& {Bally}}]{Stoji06}
{Stojimirovi{\'c}}, I., {Narayanan}, G., {Snell}, R.~L., \& {Bally}, J. 2006,
  \apj, 649, 280

\bibitem[{{Stone} \& {Norman}(1993)}]{Stone93}
{Stone}, J.~M. \& {Norman}, M.~L. 1993, \apj, 413, 210

\bibitem[{{Tafalla} {et~al.}(2004){Tafalla}, {Santiago}, {Johnstone}, \&
  {Bachiller}}]{Tafalla04}
{Tafalla}, M., {Santiago}, J., {Johnstone}, D., \& {Bachiller}, R. 2004, \aap,
  423, L21

\bibitem[{{Terquem} {et~al.}(1999){Terquem}, {Eisl{\"o}ffel}, {Papaloizou}, \&
  {Nelson}}]{Terquem99}
{Terquem}, C., {Eisl{\"o}ffel}, J., {Papaloizou}, J.~C.~B., \& {Nelson}, R.~P.
  1999, \apjl, 512, L131

\bibitem[{{V{\'a}zquez-Semadeni} {et~al.}(2019){V{\'a}zquez-Semadeni}, {Palau},
  {Ballesteros-Paredes}, {G{\'o}mez}, \& {Zamora-Avil{\'e}s}}]{vazquez19}
{V{\'a}zquez-Semadeni}, E., {Palau}, A., {Ballesteros-Paredes}, J.,
  {G{\'o}mez}, G.~C., \& {Zamora-Avil{\'e}s}, M. 2019, \mnras, 2348

\bibitem[{{Vorobyov} {et~al.}(2018){Vorobyov}, {Elbakyan}, {Plunkett},
  {Dunham}, {Audard}, {Guedel}, \& {Dionatos}}]{Vorobyov18}
{Vorobyov}, E.~I., {Elbakyan}, V.~G., {Plunkett}, A.~L., {et~al.} 2018, \aap,
  613, A18

\bibitem[{{Widmann} {et~al.}(2016){Widmann}, {Beuther}, {Schilke}, \&
  {Stanke}}]{Widmann16}
{Widmann}, F., {Beuther}, H., {Schilke}, P., \& {Stanke}, T. 2016, \aap, 589,
  A29

\bibitem[{{Zhang} {et~al.}(2014){Zhang}, {Moscadelli}, {Sato}, {Reid},
  {Menten}, {Zheng}, {Brunthaler}, {Dame}, {Xu}, \& {Immer}}]{zhang14}
{Zhang}, B., {Moscadelli}, L., {Sato}, M., {et~al.} 2014, \apj, 781, 89

\bibitem[{{Zhang} {et~al.}(2015){Zhang}, {Wang}, {Lu}, \&
  {Jim{\'e}nez-Serra}}]{zhang15}
{Zhang}, Q., {Wang}, K., {Lu}, X., \& {Jim{\'e}nez-Serra}, I. 2015, \apj, 804,
  141

\bibitem[{{Zhu} {et~al.}(2009){Zhu}, {Hartmann}, \& {Gammie}}]{Zhu09}
{Zhu}, Z., {Hartmann}, L., \& {Gammie}, C. 2009, \apj, 694, 1045

\end{thebibliography}
   
\begin{appendix}

\renewcommand{\thefigure}{A\arabic{figure}}

\begin{figure*}
\centering
\vspace{2cm}
\includegraphics[width=0.7\hsize]{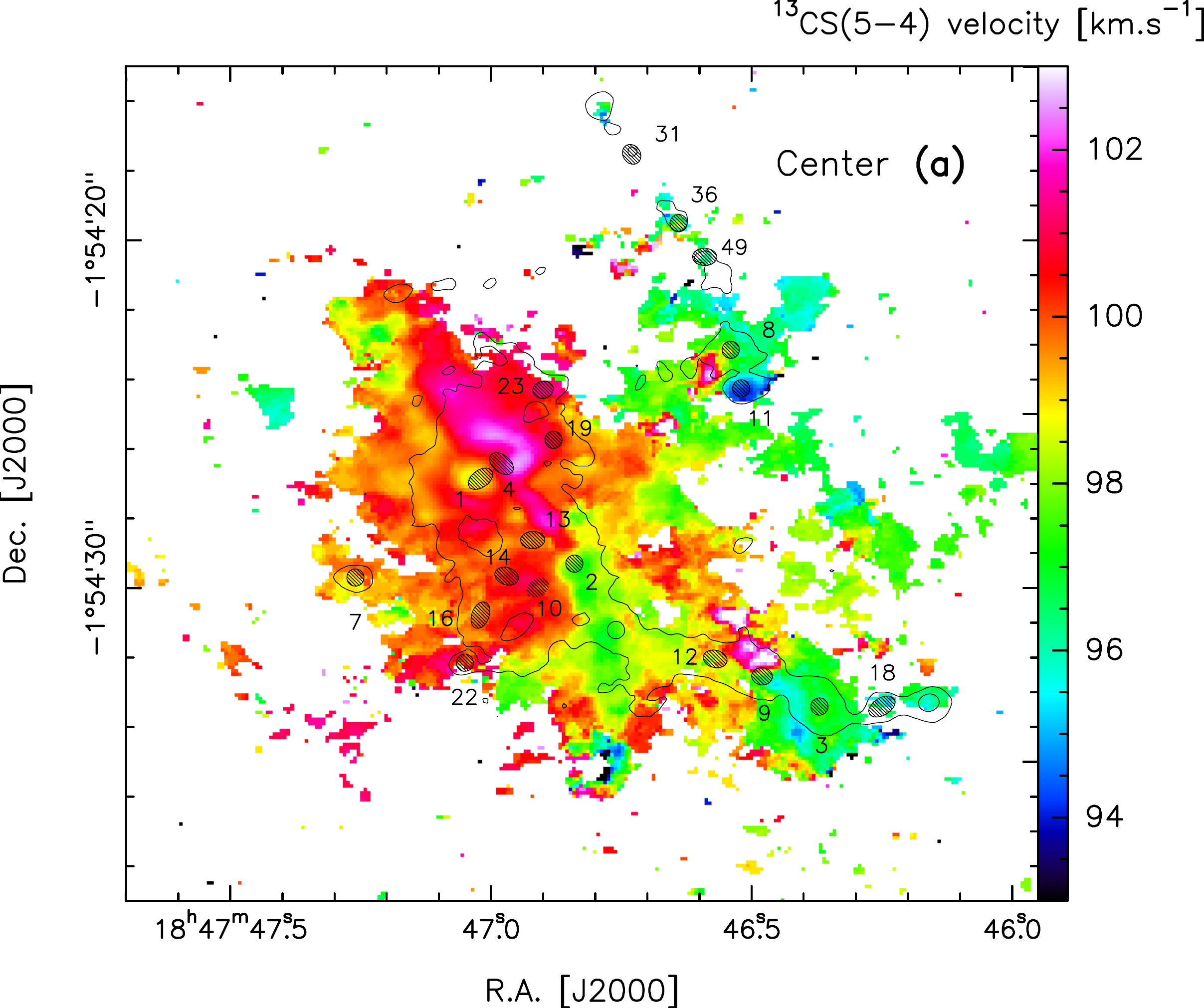}
\includegraphics[width=0.7\hsize]{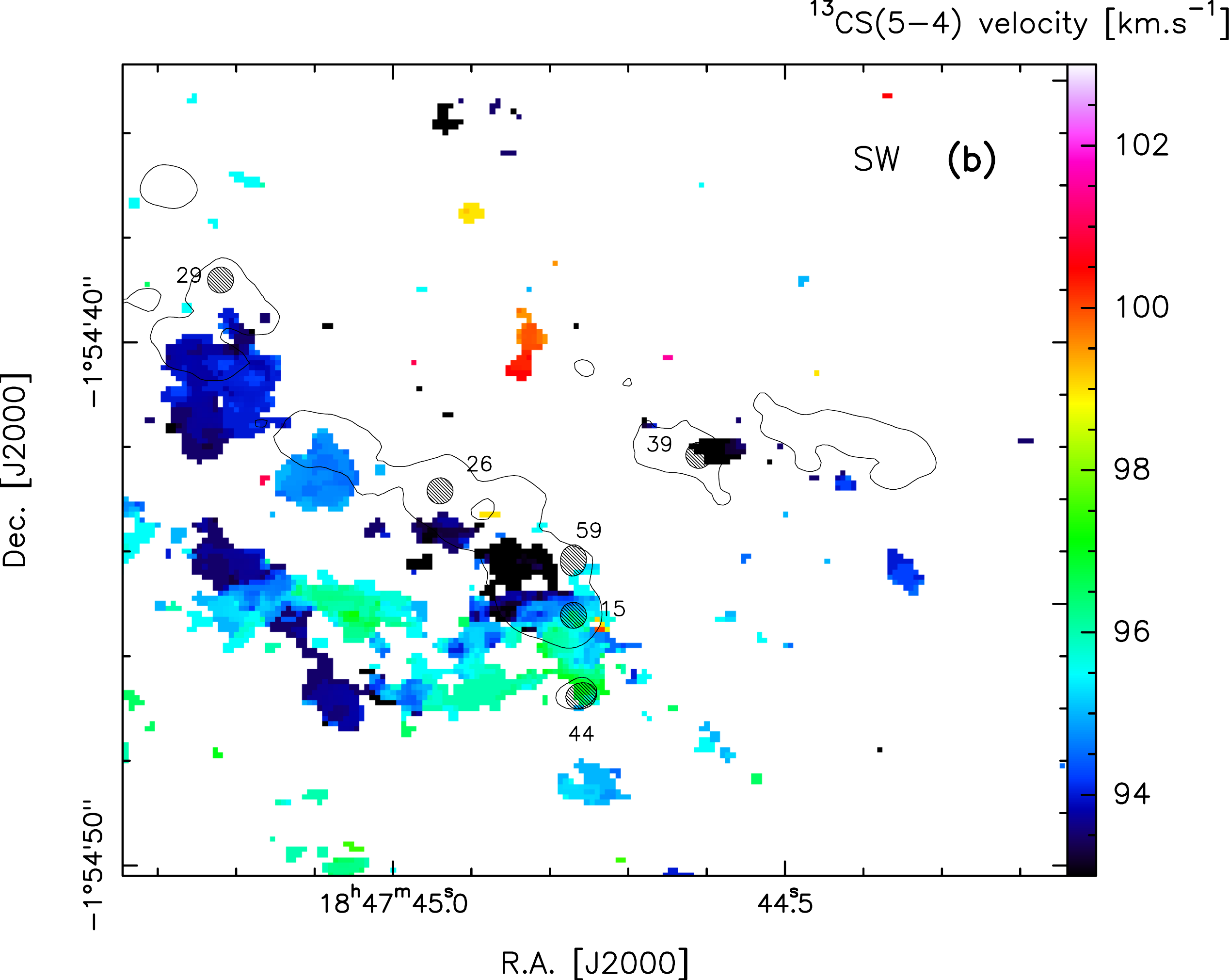}

\caption{\label{f:cs-velo} Moment 1 -- velocity field -- map of the $^{13}$CS(5-4) line in the central region  (in \textbf{a}) and the south-western region (in \textbf{b}). Hatched ellipses locate the cores with detected molecular outflows, black contours at 3 mJy\,beam$^{-1}$ (in \textbf{a}) and 7 mJy\,beam$^{-1}$ (in \textbf{b}) outline the continuum 1.3~mm emission 
of the 12~m array. }

\end{figure*}

\begin{figure*}[t!]
\centering
\vskip 1cm
\includegraphics[width=0.65\textwidth]{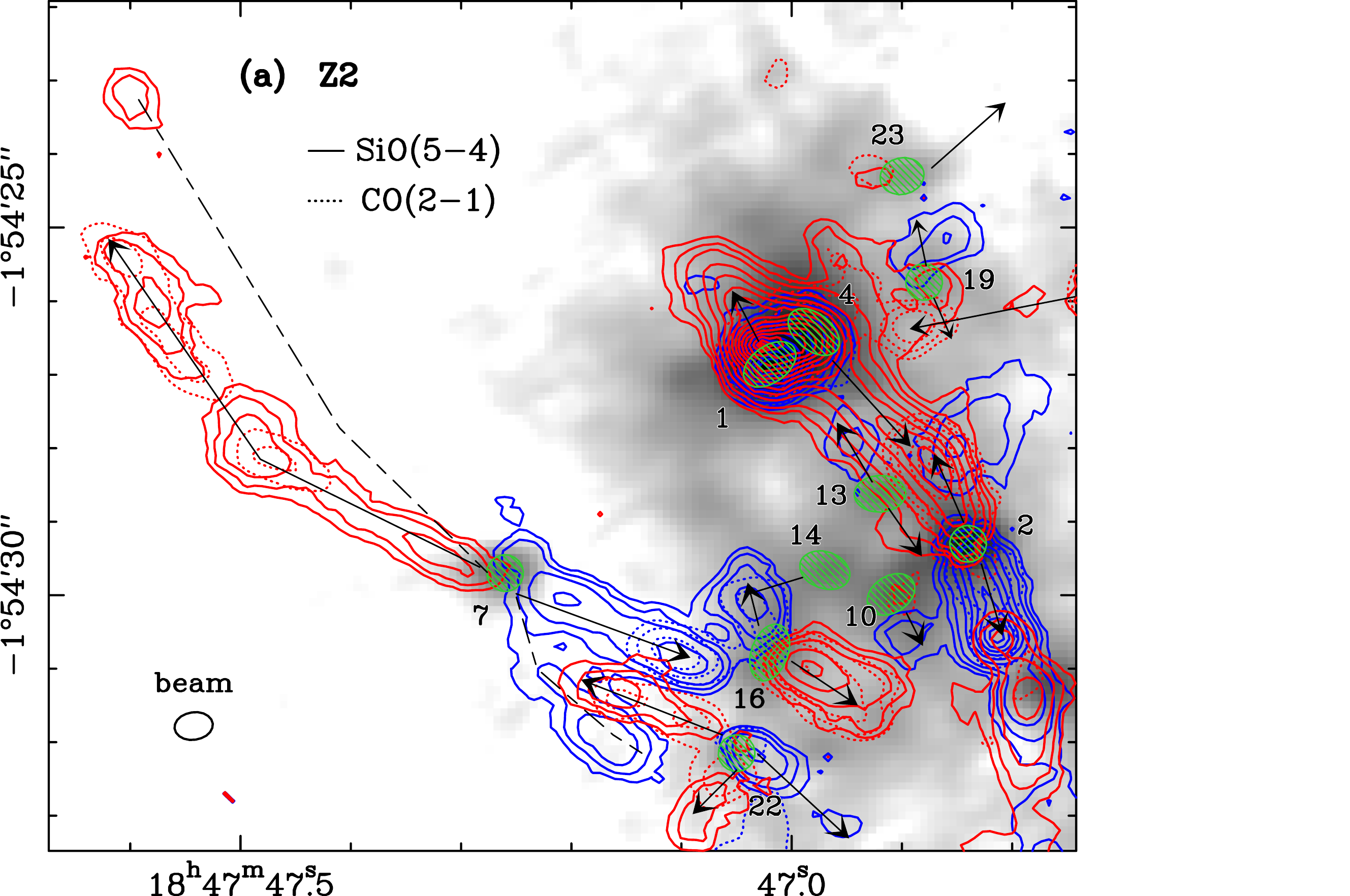} \hskip -0.3cm \includegraphics[width=0.35\textwidth]{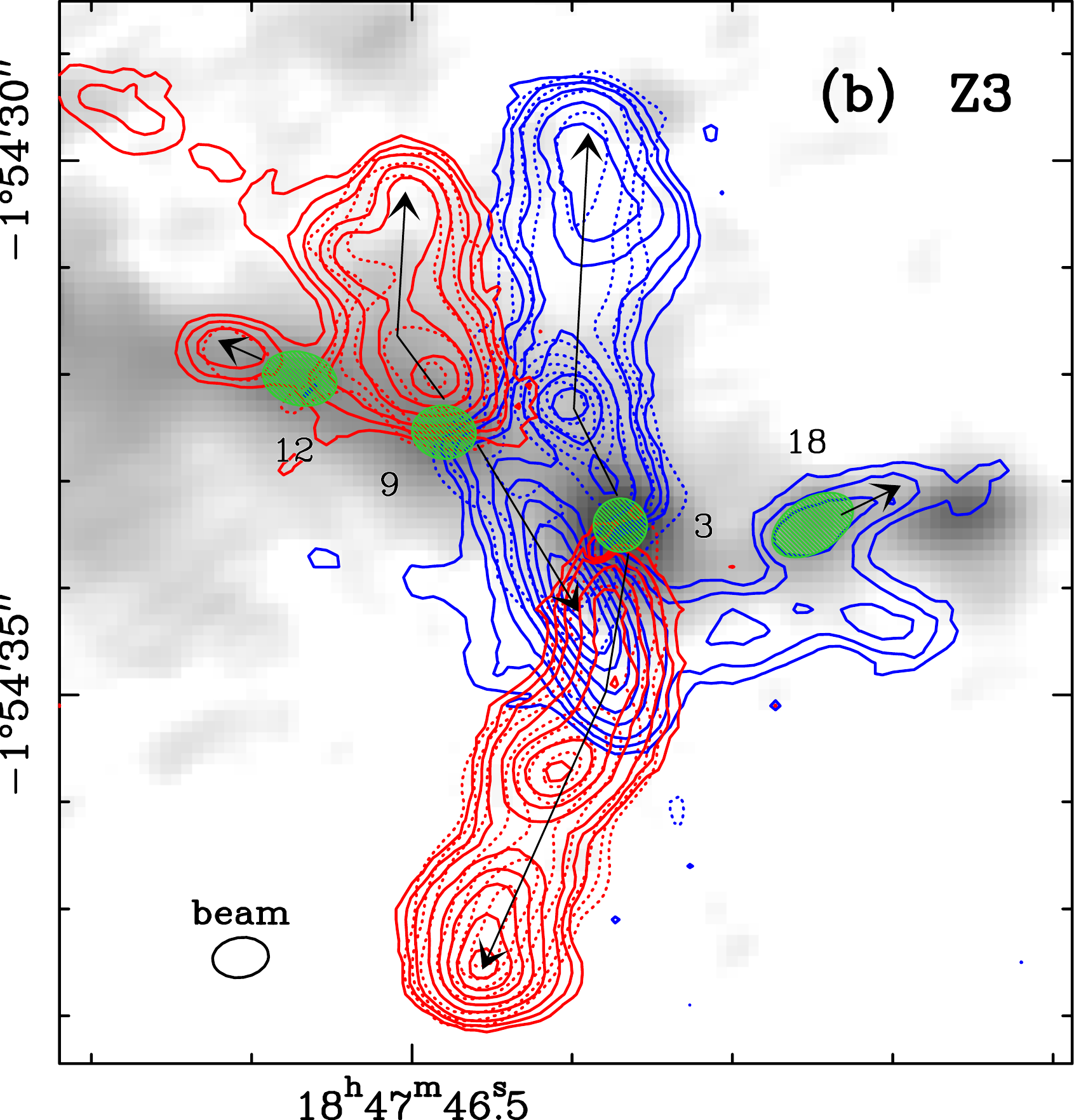}
\vskip 0.5cm
\includegraphics[width=0.65\hsize]{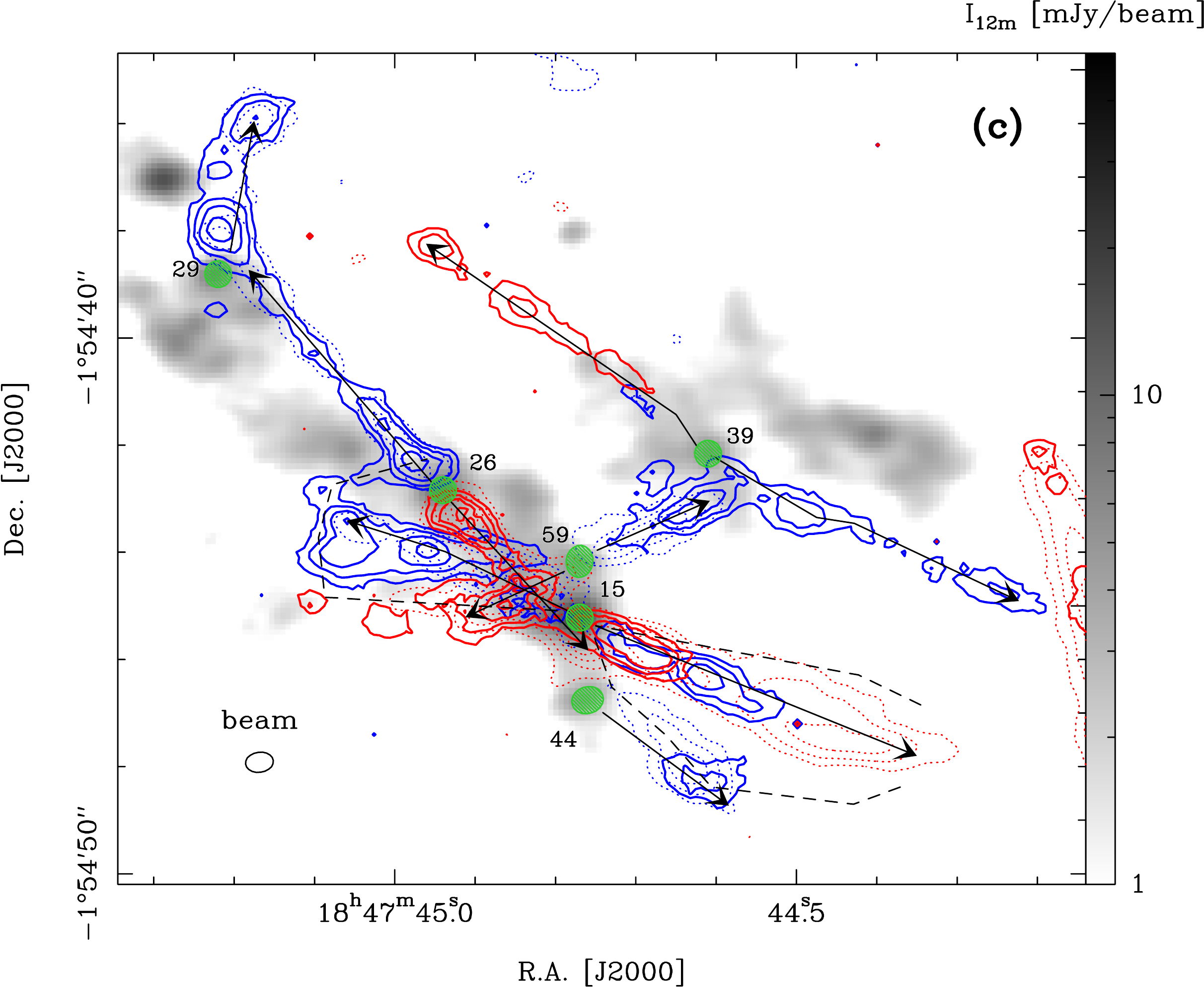}

\caption{\label{fig:co-sio-app} Molecular outflows observed in SiO(5-4) compared with their CO(2-1) emission in the central region (in \textbf{a} and \textbf{b}, Z2 and Z3 fields of Fig.~\ref{f:co-ctr-high}a) and in the south-western region (in \textbf{c}). 
The SiO line is integrated over $43-93~\kms$ (continuous blue contours) and $103-153~\kms$ (continuous red contours), with contours of 6, 11, 18 to 198 by steps of 15 (in \textbf{a} and \textbf{b}) and 6, 11, 18, 28 (in \textbf{c}), in unit of $\sigma_{\rm SiO}=23\,\rm mJy\,beam^{-1}\,\kms$.
The CO line is integrated over $21-64~\kms$ (dotted blue contours), with contours 7, 15, 30 to 270 by steps of 40 (in \textbf{a} and \textbf{b}) and  5, 12, 20, 30 (in \textbf{c}), in unit of $\sigma_{\rm CO,B}=32\,\rm mJy\,beam^{-1}\,\kms$.
It is integrated over $128-158~\kms$ (dotted red contours), with contours of 7, 15, 30 to 280 by steps of 50 (in \textbf{a} and \textbf{b}) and  5, 15, 30, 45 (in \textbf{c}), in unit of $\sigma_{\rm CO,R}=22\,\rm mJy\,beam^{-1}\,\kms$.
Contours are overlaid on the 1.3~mm continuum emission of the 12~m array (gray scale). Green ellipses locate the W43-MM1 cores, arrows (sometimes broken) indicate the direction of their outflows,
dashed lines outline the best-developed outflow cavities.
}

\end{figure*}


\newpage

\begin{figure*}
\vspace{2cm}
\centering
\subfloat{\includegraphics[width=0.52\hsize]{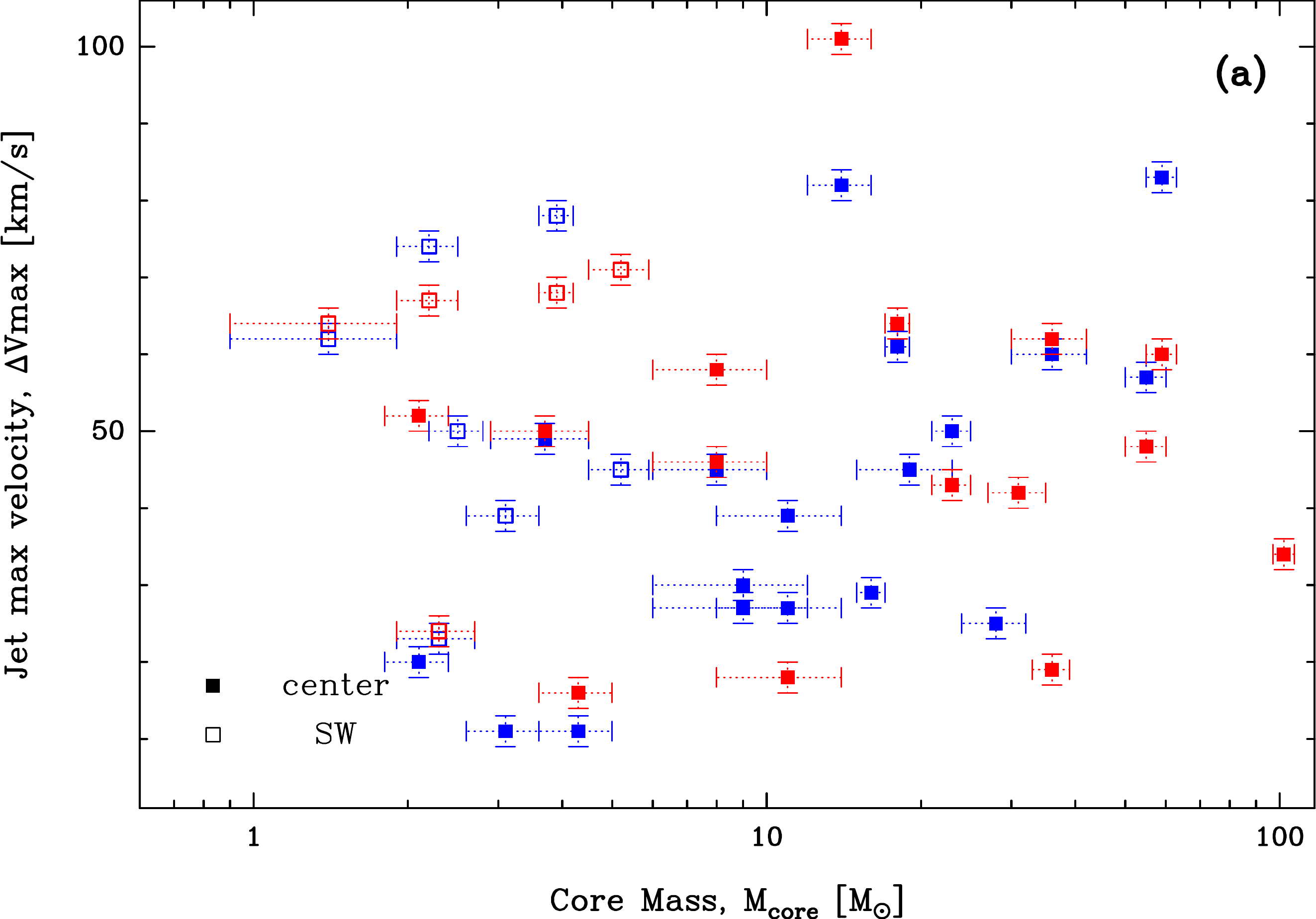}}
\subfloat{\includegraphics[width=0.47\hsize]{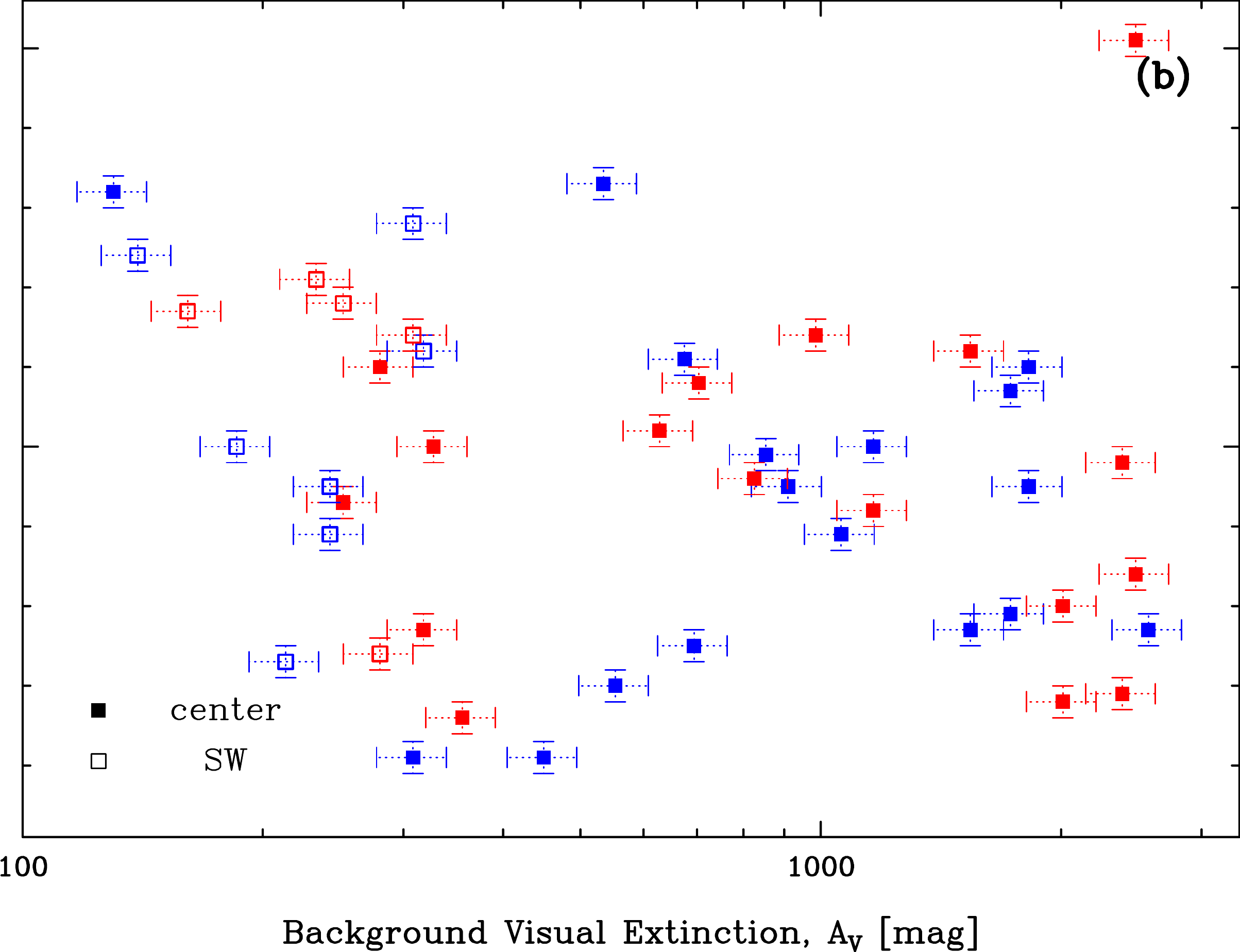}}

\caption{\label{f:outflow-char-Vmax} Jet maximal velocity, $\Delta V_{\rm max}$ vs. the mass of the launching core, $M_{\rm core}$, (in \textbf{a}) and the visual extinction of the cloud background crossed by the outflow, $A_{\rm V}$, (in \textbf{b}). Filled and empty squares pinpoint outflow lobes developing in the central and south-western part of W43-MM1, respectively. Blue and red colors show the blue- and red-shifted lobes, respectively. No clear correlation is revealed.}

\end{figure*}

\end{appendix}

\end{document}